\newif\ifmnras
\mnrasfalse
\ifmnras
	\documentclass[a4paper,fleqn,usenatbib]{mnras}
\else
    \documentclass[twocolumn]{aastex631}
\fi

\usepackage[caption=false]{subfig}

\def \octo{{\sc Octo-Tiger}}
\def \mesa{{\sc MESA}}

\ifmnras

\fi

\shorttitle{Betelgeuse as a merger}
\shortauthors{Shiber et al.}

\begin{document}
\label{firstpage}

\title{Betelgeuse as a Merger of a Massive Star with a Companion}

\correspondingauthor{Sagiv Shiber}
\email{sshiber1@lsu.edu}

\author[0000-0001-6107-0887]{Sagiv Shiber}
\affiliation{Department of Physics and Astronomy, Louisiana State University, Baton Rouge, LA 70803, USA}

\author[0000-0002-8179-1654]{Emmanouil Chatzopoulos}
\affiliation{Department of Physics and Astronomy, Louisiana State University, Baton Rouge, LA 70803, USA}
\affiliation{Institute of Astrophysics, Foundation for Research and Technology-Hellas (FORTH), Heraklion, 70013, Greece}

\author[0000-0003-4034-4900]{Bradley Munson}
\affiliation{Department of Physics and Astronomy, Louisiana State University, Baton Rouge, LA 70803, USA}

\author[0000-0003-4467-5301]{Juhan Frank}
\affiliation{Department of Physics and Astronomy, Louisiana State University, Baton Rouge, LA 70803, USA}

\begin{abstract}
We investigate the merger between a $16M_{\rm \odot}$ star, on its way to becoming a red supergiant (RSG), and a $4M_{\rm \odot}$ main-sequence companion. Our study employs three-dimensional hydrodynamic simulations using the state-of-the-art adaptive mesh refinement code \octo. The initially corotating binary undergoes interaction and mass transfer, resulting in the accumulation of mass around the companion and its subsequent loss through the second Lagrangian point (L2). The companion eventually plunges into the envelope of the primary, leading to its spin-up and subsequent merger with the helium core. We examine the internal structural properties of the post-merger star, as well as the merger environment and the outflow driven by the merger. Our findings reveal the ejection of approximately $\sim 0.6~M_{\rm \odot}$ of material in an asymmetric and somewhat bipolar outflow. We import the post-merger stellar structure into the MESA stellar evolution code to model its long-term nuclear evolution. In certain cases, the post-merger star exhibits persistent rapid equatorial surface rotation as it evolves in the H-R diagram towards the observed location of Betelgeuse. These cases demonstrate surface rotation velocities of a similar magnitude to those observed in Betelgeuse, along with a chemical composition resembling that of Betelgeuse. In other cases, efficient rotationally-induced mixing leads to slower surface rotation. This pioneering study aims to model stellar mergers across critical timescales, encompassing dynamical, thermal, and nuclear evolutionary stages.
\end{abstract}

\keywords{binaries: close --- stars: evolution --- hydrodynamics --- methods: numerical }

\section{INTRODUCTION}
\label{sec:intro}
It has been firmly established that the majority of massive stars exist within binary systems \citep{2009AJ....137.3358M,2015A&A...580A..93D}. Furthermore, a significant portion of these systems undergo binary interactions at some point during their evolution \citep{2012Sci...337..444S,2013ApJ...764..166D,2019A&A...624A..66R}. In extreme cases, these interactions can result in the complete merging of the two stars \citep{2005MNRAS.362..915B,2014ApJ...782....7D}. Depending on the initial conditions that trigger a stellar merger, there can be various outcomes \citep{2006ASPC..355..259P,2016MNRAS.457.2355S}. The merger event may be accompanied by a short-lived, relatively faint electromagnetic transient referred to as a ``merge-burst" \citep{2006MNRAS.373..733S}, as well as mass loss occurring along the orbital plane of the binary progenitor system \citep{1984ApJ...280..771B,1991ApJ...373..246T,1994ApJ...422..729T}.

If the secondary star gets past the initial common-envelope (CE) phase and spirals inward into the extended envelope of a primary star that is evolved past the main--sequence (MS) with a helium (He) core, the outcome will depend on which component has the highest central density. If mass transfer started in Case B, when the primary had already developed a compact core, the secondary may experience tidal disruption outside the core of the primary and subsequently merge with it. If no hydrogen-rich material is mixed into the primary core, we
have what we may classify as a Case B-{\emph{quiet}} merger. 
The extent to which the disrupted secondary material mixes with the core of the primary, known as the penetration depth of the stream \citep{2002MNRAS.334..819I,2002Ap&SS.281..191I}, determines whether there is sufficient fresh fuel to rejuvenate the nuclear evolution of the secondary star, leading to a new phase of core hydrogen burning \citep{2002PhDT........25I,2003fthp.conf...19I}. Rejuvenated post-merger stars are likely to evolve into compact, hot blue supergiant (BSG) supernova (SN) progenitors, similar to Sk – 69\textdegree~202, the progenitor of SN1987A \citep{2017MNRAS.469.4649M,2019MNRAS.482..438M}.  {{Recent studies have demonstrated that the degree of core penetration and rejuvenation of the primary star strongly depends on the original binary mass ratio, denoted as $q = M_{\rm 2}/M_{\rm 1}$, where $M_{\rm 1}$ and $M_{\rm 2}$ are the masses of the primary star (initially the donor) and the secondary star (initially the accretor) respectively}}. Mergers following Case B mass transfer, in systems where $q \lesssim$~0.25 tend to yield a quiet merger and evolve towards the red supergiant (RSG) phase \citep{1990A&A...227L...9P,2002MNRAS.334..819I}.

Even in the case of a Case B-quiet merger scenario, the post-merger star's evolution is unlikely to resemble that of a single massive star. Depending on the initial period (or binary separation) of the system when the CE phase begins, the secondary star may deposit significant amounts of angular momentum into the envelope of the primary star, resulting in spin-up and enhanced rotation-induced chemical mixing \citep{2017MNRAS.465.2654W}. These effects can give rise to a rapidly rotating RSG with elevated surface nitrogen ($^{14}$N) abundances. While a few observations of evolved, lower-mass red giant stars support this notion \citep{2015ApJ...807L..21C}, the most intriguing case is that of Betelgeuse ($\alpha$ Orionis), a massive, evolved RSG star exhibiting evidence of fast surface equatorial rotation \citep{1998AJ....116.2501U,1996ApJ...463L..29G,2023A&G....64.3.11W}. Recent measurements from the \textit{Atacama Large Millimeter Array} (\textit{ALMA}) indicate a surface equatorial rotational velocity of 5--15~km~s$^{-1}$ for Betelgeuse \citep{2018A&A...609A..67K}, while measurements of its surface abundances reveal nitrogen-to-carbon (N/C) and nitrogen-to-oxygen (N/O) ratios of 2.9 and 0.6, respectively \citep{1984ApJ...284..223L} far in excess of standard solar values. Recent theoretical work has shown that early Case B mergers with mass ratio in the range 0.07~$<q<$~0.25, primary mass 15--16$M_{\odot}$, secondary masses 1--4~$M_{\odot}$ where the CE phase is triggered during the expansion of the primary to the RSG stage (the ``Hertzsprung gap'' crossing phase), at primary radii 50--300~$R_{\odot}$ can reproduce the current observed state of Betelgeuse, indicating that a past merger is a viable explanation for this extraordinary star \citep{Chatzopoulos2020}.

While \citet{Chatzopoulos2020} present a compelling argument for explaining some of the observed characteristics of Betelgeuse, their approach is constrained by several simplifying assumptions that rely on basic approaches, including a spherically-symmetric 1D treatment and an analytical expression for the specific angular momentum deposition into the envelope of the primary star. These assumptions assume that the structure of the secondary star remains unaffected during the common-envelope (CE) in-spiral phase and that the in--spiral time-scale of the secondary is significantly shorter than the thermal adjustment time--scale of the primary's envelope. Additionally, there is some disagreement regarding whether the high measured rotational velocity of Betelgeuse reflects actual surface rotation or other phenomena, such as large-scale convective plume motion \citep{2018A&A...620A.199L,2023NewA...9901962J}. Furthermore, certain properties of the star, such as the enhanced surface abundance of $^{14}$N, remain puzzling within the context of the evolution of single massive stars.

The objective of this research is to comprehensively and consistently investigate the merger scenario for Betelgeuse and, more broadly, rapidly--spinning red supergiant (RSG) stars, using three-dimensional (3D) hydrodynamics merger simulations. In particular, we make use of the updated and bench-marked 3D adaptive mesh refinement (AMR) \octo\ \citep{2018MNRAS.481.3683K,Marcello2021} 
code to model the CE in--spiral and final secondary--primary core merger of a $q=$~0.25, 16~$M_{\odot}$ primary right after the end of its core H--burning phase and expanded at a radius of $\sim$~50~$R_{\odot}$ with a 4~$M_{\odot}$ main--sequence (MS) secondary. Additionally, we track the 3D evolution of the post--merger star for several orbits after the merger to obtain a realistic structure of the post--merger star, including the distribution of its internal angular momentum. We also monitor the expansion of dynamically unbound mass loss caused by the merger. The resulting 3D post--merger structure is then spherically averaged and incorporated into the 1D stellar evolution code known as \textit{Modules for Experiments in Stellar Astrophysics} ({\sc MESA}; \citealt{Paxton2011,Paxton2013,Paxton2015,Paxton2018,Paxton2019}). This integration enables us to explore the long-term nuclear evolution of the post-merger star until it reaches the RSG phase and compare our findings with observations of Betelgeuse. 

To the best of our knowledge, this is the first study in the literature that self--consistently simulates the 3D dynamics of the CE and merger phase, as well as the long--term nuclear evolution of the post--merger star in 1D. Given the extensive simulation box used to track the expansion of the mass loss induced by the merger, and the high resolution required to accurately depict the structure of the secondary star and ensure its structural stability, this simulation is the first in its kind. It took nearly an entire calendar year to complete and involved multiple successive compute time allocations. The investment of time was justified, as we were able to confirm that, under certain conditions, ``quiet" high mass-ratio mergers of this nature can indeed produce RSGs with properties similar to those observed in Betelgeuse. Moreover, the significant mass loss that can occur during the merger process can lead to non-spherically symmetric circumstellar (CS) environments with properties and geometries reminiscent of those observed in other systems, such as SN1987A \citep{1995ApJ...439..730P}.

In Section \ref{sec:numerics}, we provide a comprehensive description of our numerical setup and the initial conditions of the binary system in detail. Subsequently, in Section \ref{sec:results}, we concisely outline the outcomes derived from our three-dimensional (3D) dynamical merger simulations employing the \octo\ model, as well as the long-term nuclear evolution simulations conducted in one dimension (1D) using the {\sc MESA} code. Lastly, in Section \ref{sec:summary}, we succinctly summarize the findings of our study and engage in a thorough discussion of the principal conclusions drawn, along with their implications for the evolution of massive stars following a merger event.

\section{NUMERICAL SET-UP}
\label{sec:numerics}
Our primary and secondary models are constructed based on one-dimensional profiles that were simultaneously evolved using the \mesa\ stellar evolution code from the Zero Age Main Sequence (ZAMS). The primary star originates with a ZAMS mass of $16M_{\rm \odot}$, while the secondary star begins with a ZAMS mass of $4M_{\rm \odot}$. These masses align with the $q=0.25$ binary mergers examined in \cite{Chatzopoulos2020}, which have been identified as fitting Betelgeuse's inferred surface rotation with surface velocities ranging from $4-12~{\rm km/sec}$. These velocities persist for time spans of $7-8\times 10^5~{\rm yr}$. Both stellar models were initialized assuming solar metallicity and utilized standard mass-loss prescriptions appropriate for their respective masses and evolutionary phases \citep{deJager88,Vink01}.

Following the methodology employed in \cite{Chatzopoulos2020}, our models were terminated after the primary star completed its evolution off the main sequence and traversed the Hertzsprung gap (HG) phase. During this transitional stage, the primary's envelope expands on a thermal time scale, progressing towards the RSG state. The primary star reaches a radius of $R_1 = 50R_{\rm \odot}$, which is smaller in comparison to the primary radius ranges investigated in \cite{Chatzopoulos2020} (typically $> 200R_{\rm \odot}$). This deliberate choice was made to minimize computational time, as a larger envelope would result in longer dynamical timescales (proportional to $\sim R^{3/2}$) and require a higher cell count for adequate resolution.
At this stage, the age of the binary system is $1.11\times 10^7~{\rm yr}$, and the primary star's mass has decreased to $M_1 = 15.48M_{\rm \odot}$ due to wind-driven mass loss. The primary star exhibits a luminosity of $L_1 = 55,000L_{\rm \odot}$ and an effective temperature of $T_{\rm eff,1}=12,600~{\rm K}$. The secondary star, on the other hand, remains in the main sequence phase at this age, with a radius of $R_2 = 2.43~R_{\rm \odot}$.

To incorporate the {\sc MESA} structures into \octo's grid, we first approximate the internal structures of the two stars by fitting them to single polytropes.
We determine that polytropic indices of $n_1=4.1$ and $n_2=3.4$ provide the best fit for the primary star (referred to as star 1) and the secondary star (referred to as star 2) respectively. Subsequently, we utilize \octo's self-consistent field (SCF) technique \citep{2018MNRAS.481.3683K} to construct an equilibrium binary system with synchronous rotation. The SCF method initializes each star as a polytrope and iterates until the desired level of equilibrium is attained (for further details, refer to \citealt{Marcello2021}). The level of equilibrium is quantified by deviations from the virial relation. Specifically, the SCF produces models in which the virial error, defined as ${\rm VE}=\left|\left(2E_k + E_g + 2E_i\right)/E_g\right|$, is as small as $10^{-3}$. Here $E_k$, $E_g$, and $E_i$, are the total kinetic, gravitational and internal (thermal) energies of the binary, respectively.
The SCF process ensures the conservation of mass for each star, the ratio of core mass to the total star mass (particularly relevant in cases where there is a difference in mean molecular weight or composition between the core and the envelope, as seen in star 1), as well as the initial Roche Lobe (RL) filling factor. Additionally, the structures generated by the SCF technique incorporate the effects of tidal distortion and rotational flattening of the stars as intentional design features.

In our simulation, we assign the masses to be identical to those specified in the original {\sc MESA} 1D profiles. Additionally, we opt for a RL filling factor of 1, indicating that the primary star precisely occupies its RL. These choices establish the initial separation at $99.6R_{\rm \odot}$.  
Furthermore, to thoroughly investigate the structure, kinematics, and energetics of merger-driven outflows, we utilize an expansive simulation box with dimensions approximately 40 times the initial orbital separation, resulting in a box size of $\left(4000R_{\rm \odot}\right)^3$.

Our AMR grid has 10 levels of refinement that are controlled by a density refinement criterion, and two extra refinement levels for the secondary star. These two extra refinements are done by exploiting \octo's ability to define different species, which we use to differentiate between primary and secondary star densities. Specifically, we refine by two extra levels cells in which the secondary mass fraction exceeds the value of half. The two extra refinement levels are required for the correct computation of the gravitational force onto the secondary star's cells (see subsection~\ref{ssec:numerics_sec} for more details). This yields a minimal cell size of $(\Delta x)_{\rm min}^{\rm primary} = 0.49~R_{\rm \odot}$ and $(\Delta x)_{\rm min}^{\rm secondary} = 0.12~R_{\rm \odot}$, for the primary and secondary stars, respectively.

\begin{figure*}
    \centering
    \includegraphics[scale=0.61, trim={0 6.8cm 0 7.0cm}, clip]{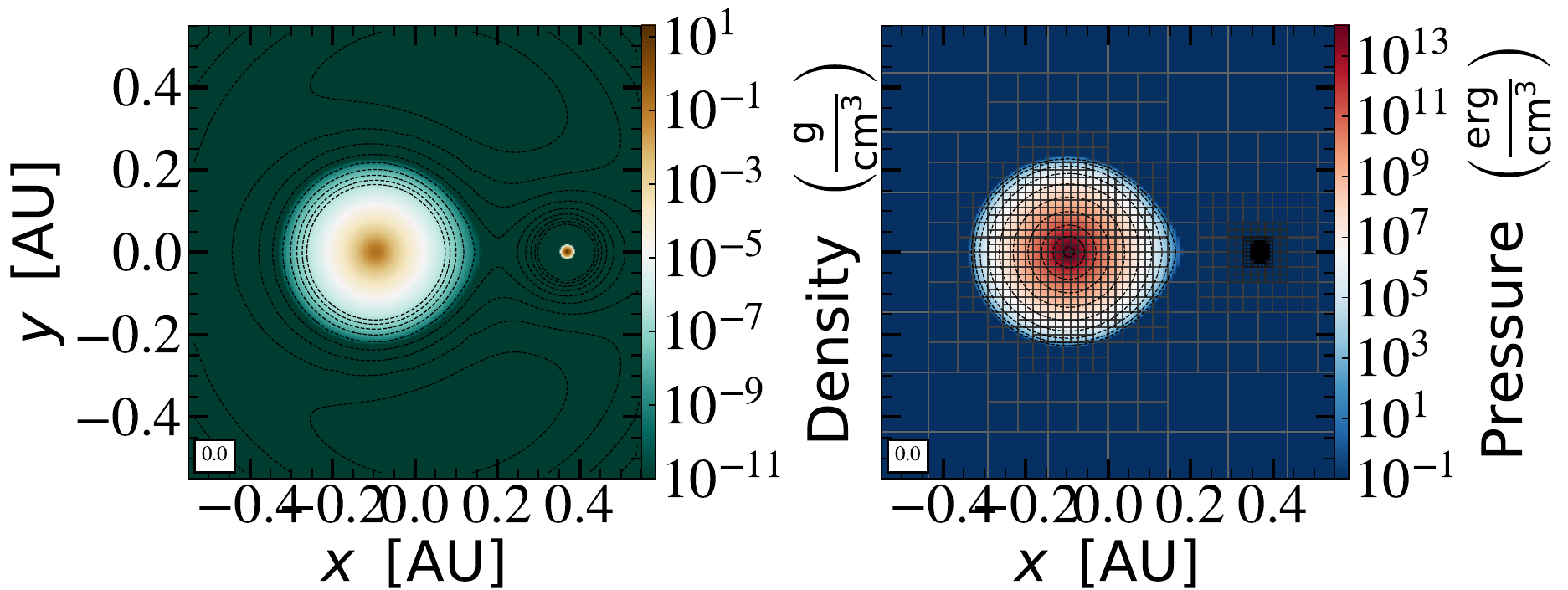}
    \caption{Density (left) and pressure (right) slices at the equatorial plane at $t=0$ of our \octo\ simulation. The right panel also shows the AMR grid and the extra refinement placed on the secondary star. The left panel also presents contours of effective potential (gravitational plus rotational). Both panels zoom-in on the central $1~{\rm AU}$ of the grid, ($\sim 1/20$ of the simulation domain size)}
    \label{fig:binary_ini}
\end{figure*}

In Figure~\ref{fig:binary_ini} we show the initial density (left) and pressure (right) maps of the binary that were produced by the SCF.
We plot density and pressure at the equatorial plane. The right panel also illustrates the adaptive mesh in our simulation (gray and black squares). Each such a square represents a subgrid consisting of $8^3=512$ cells. We resolve $\sim 200$ cells across the primary star's diameter (along the x axis), and $\sim 70$ across the secondary star's diameter. The left panel presents contours of effective potential which is the sum of the gravitational and rotational potentials, demonstrating the teardrop shape of the primary star occupying its RL. The overall initial cell count in the grid is $6.6 \times 10^6$.

During the SCF iterations, a polytropic equation of state (EoS) is assumed, with each star having a different polytropic index. However, once the simulation enters the evolution phase after initialization, an ideal gas equation of state is utilized. Radiation pressure is disregarded, and the temperature is calculated using the ideal gas law, taking into consideration the mean molecular weight of each cell. This approach may result in an overestimation of the temperature, as the inclusion of radiation pressure in the EoS would have likely resulted in lower temperatures. In the future, a revised version of \octo\ will be employed, incorporating a more comprehensive stellar equation of state (EoS) to effectively address temperature inconsistencies.

It is crucial to emphasize that \octo\ handles the evolution of inertial frame quantities within a rotating grid (for a detailed and comprehensive code description, refer to \citealt{Marcello2021}). This implementation involves a grid that constantly rotates at the same frequency as the initial binary frequency. In the rotating frame, the stars begin without initial velocities, while the dilute gas surrounding the stars does possess velocities. This deliberate approach significantly diminishes the impact of viscosity effects that could potentially hinder the accurate progression of the binary system.

The upcoming two subsections focus on individual examinations of the initial structures and dynamical stability of each star. It is demonstrated that, despite inherent limitations in resolution and computational resources, we successfully replicate the structure of both stars with a commendable level of accuracy. Notably, our simulations closely replicate the primary star's envelope as per the original {\sc MESA} structure. This enables us to investigate the impact of envelope spin-up, which stands as a significant objective of this study.

\subsection{Primary star's structure and stability}
\label{ssec:numerics_pri}

As previously mentioned, our primary star initiates the \octo\ binary simulation with a mass of $15.48M_{\rm \odot}$. Leveraging \octo's capability to track different species, we initialize the primary star with two distinct species: one for the core and one for the envelope. To determine the core mass, we locate the mass coordinate in the \mesa\ model where the abundance of Helium falls below that of Hydrogen, which corresponds to $M_{1,{\rm core}}=6.04M_{\rm \odot}$. Subsequently, we set the molecular weight of the core to the average molecular weight computed over the mass coordinate $m\leq M_{1,{\rm core}}$, resulting in $\mu_{1,{\rm core}}=1.21$. For the envelope, we calculate the average molecular weight over the mass coordinate $m>M_{1,{\rm core}}$, yielding $\mu_{1,{\rm env}}=0.63$. It is important to note that various definitions exist for the core of a star. However, we adopt a definition that primarily captures the characteristics of the envelope rather than the core, as our primary focus lies in investigating the envelope of the primary star and its acquisition of angular momentum through the merger process.

\begin{figure}
    \centering
    \includegraphics[scale=0.4]{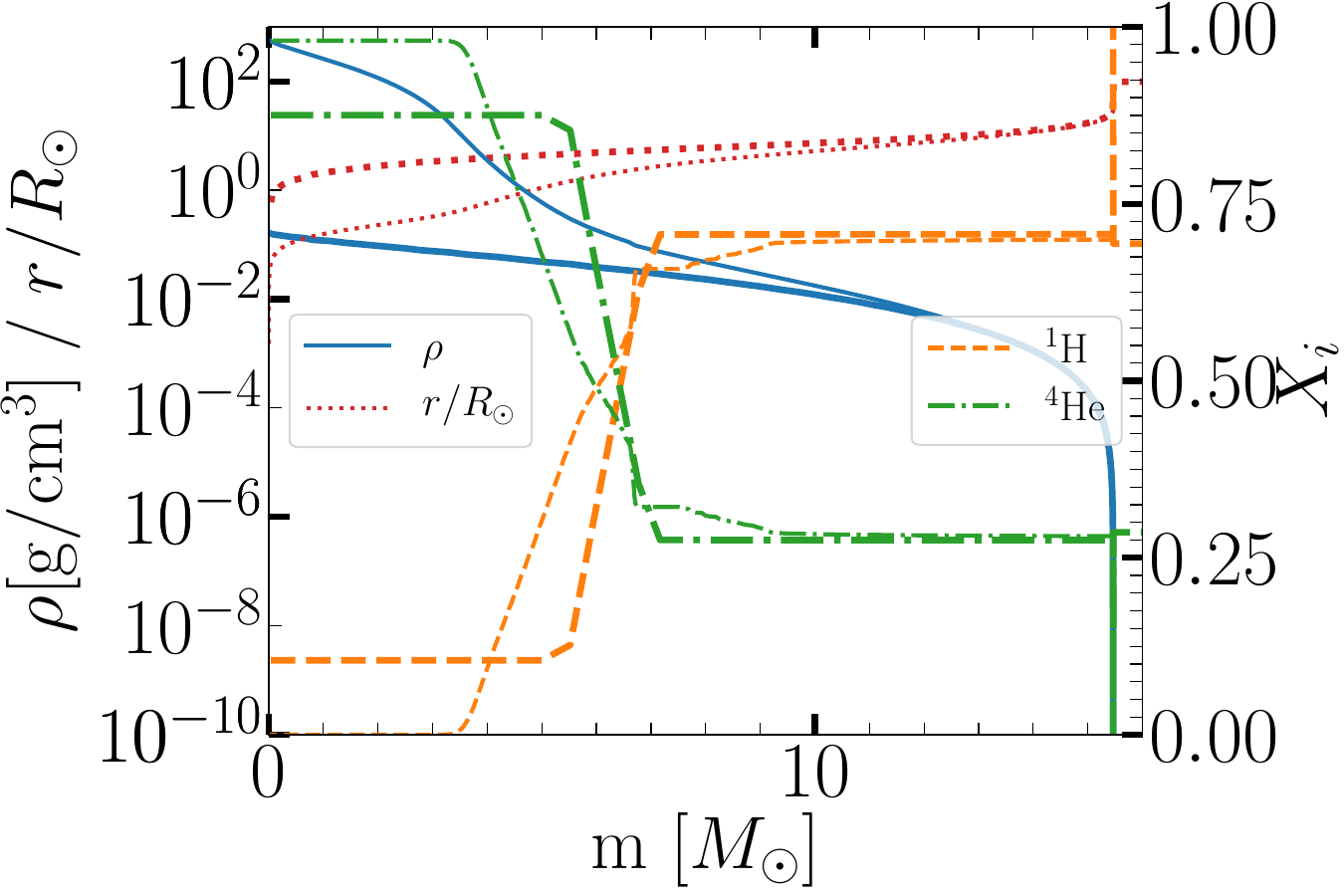}
    \caption{Comparison of the primary star structure when imported to the 3D grid (thick lines) with the original MESA model (thin lines). Density (blue solid lines), radius (red dotted lines), Hydrogen mass fraction (orange dashed lines), and Helium mass fraction (green dashed-dotted lines) are plotted as a function of the mass coordinate $m$. The adequate representation of the envelope ($m=6-15.5~M_{\rm \odot}$) enables us to study the post-merger spin-up and envelope composition}
    \label{fig:pri_prof}
\end{figure}

In Figure~\ref{fig:pri_prof} we compare the primary's structure in \octo\ (thick lines) with the \mesa\ model (thin lines). This figure is similar to Figure~1 of \cite{Lau2022a}, where they plot composition (dashed and dash-dotted lines) and density (blue solid lines) as a function of mass coordinate, but we also plot the initial composition of our 3D model as well as the radius vs mass coordinates (red dotted lines). 

\begin{figure*}
    \centering
    \includegraphics[scale=0.28]{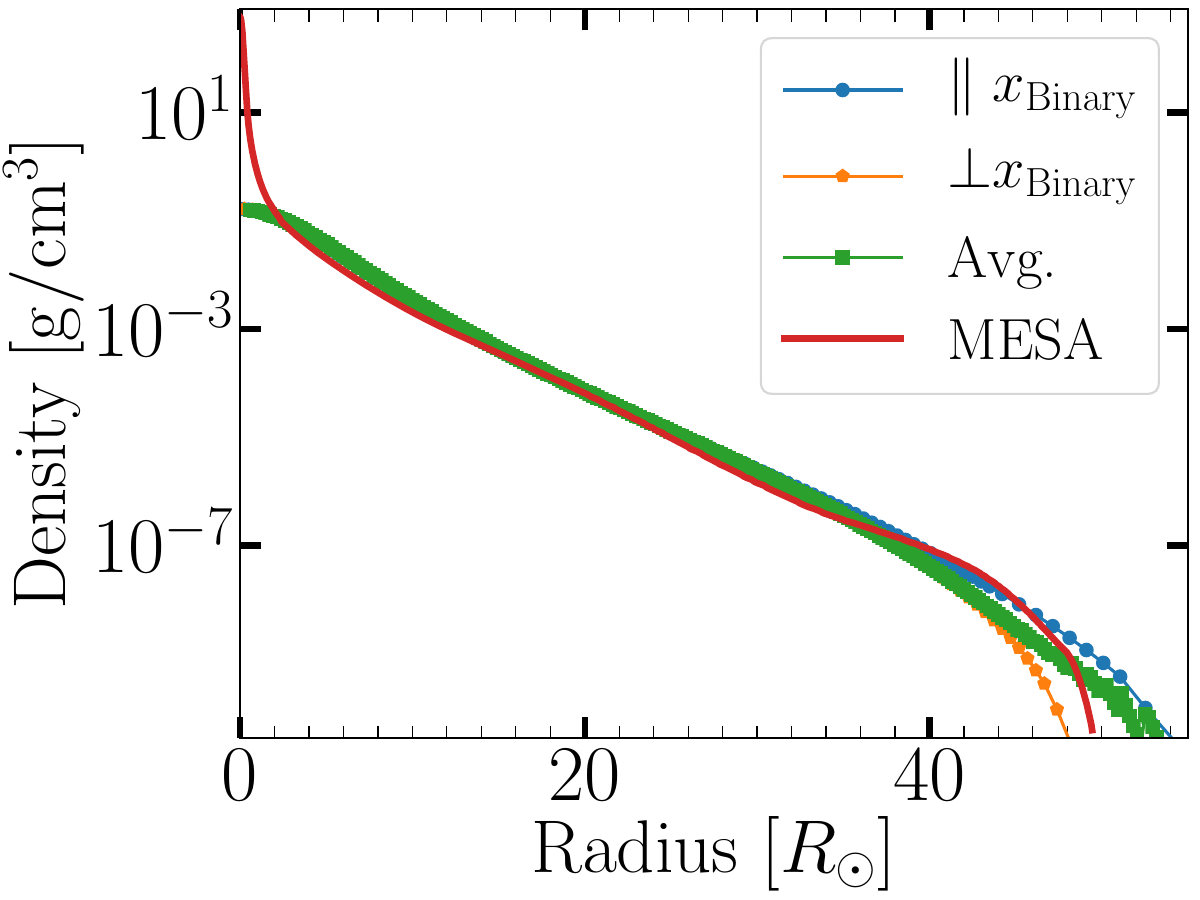}
    \includegraphics[scale=0.28]{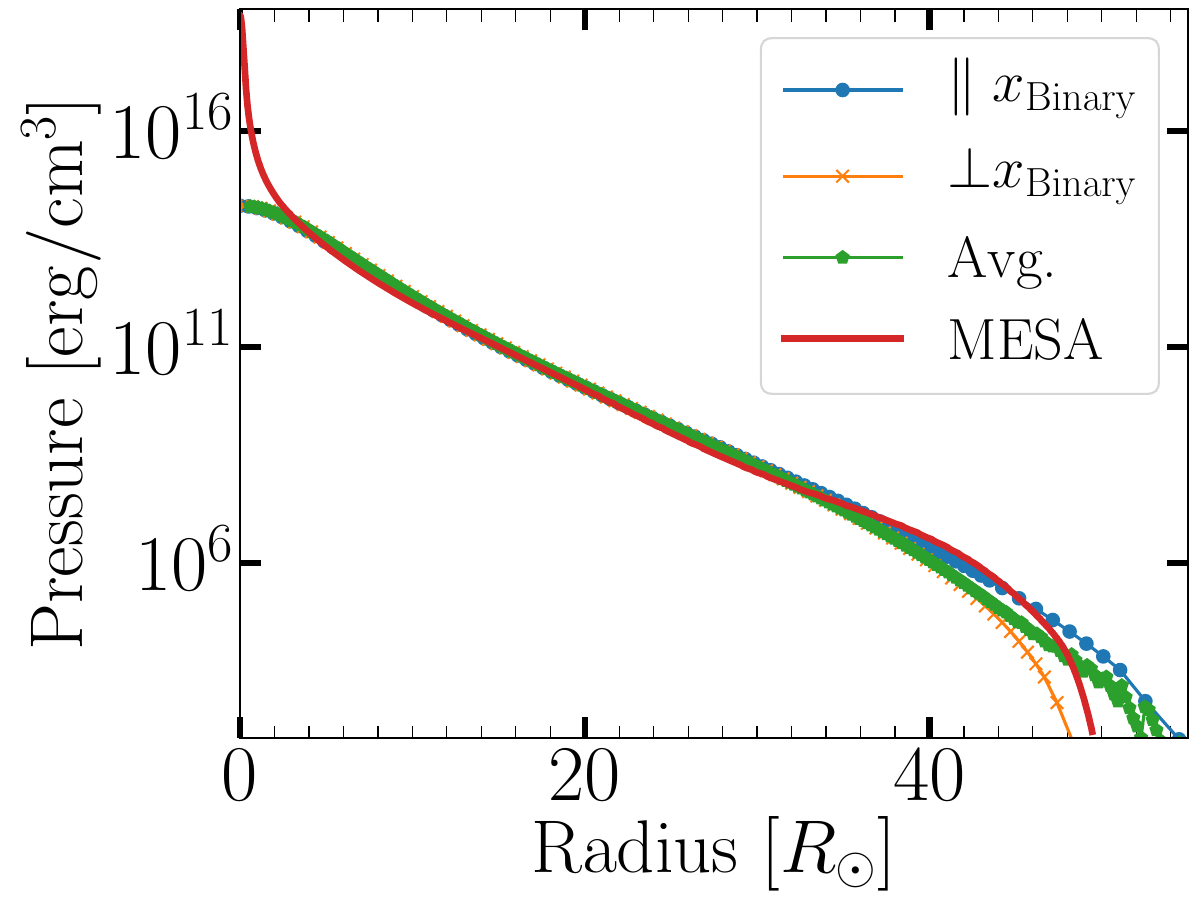}
    \includegraphics[scale=0.28]{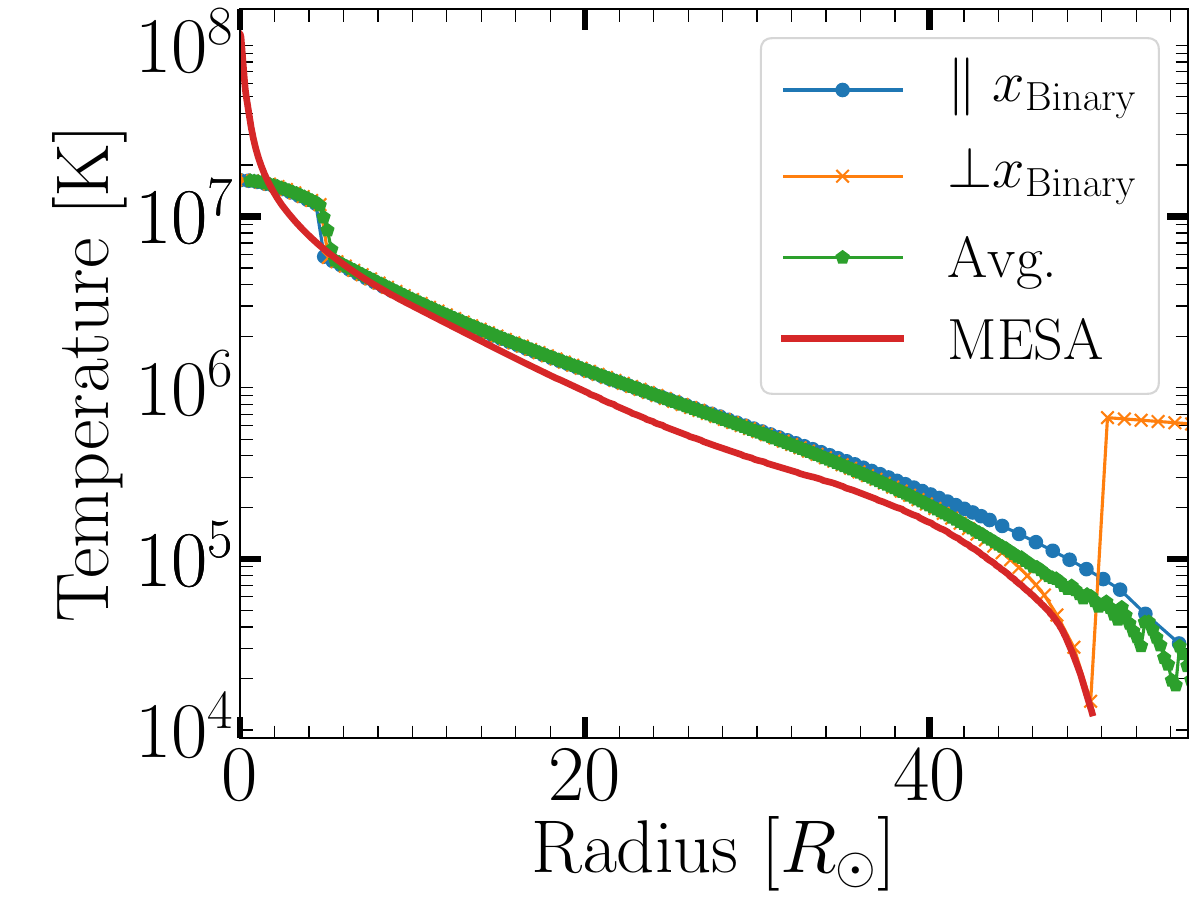}
    \caption{Density (left), pressure, (center), and temperature (right) as a function of radius of the initial primary star in our simulation. As the primary star fills its Roche Lobe, it is not completely spherically symmetric, and we therefore show lines that goes along the binary axis, $x_{\rm Binary}$ (lines that go from the center of the primary to the direction of the secondary star center; blue circles), and in perpendicular to the binary axis (orange x's) as well as mass averaged profiles (green pentagons). We also plot the 1D MESA profile (solid red line), showing again that the envelope structure in our simulation is well matched to the MESA model}
    \label{fig:prim_comp_mesa}
\end{figure*}

The density and composition in our simulation align well with the \mesa\ values in the outermost 5 solar masses of the star.
There is also a good agreement in an intermediate region spanning from $6M_{\rm \odot}$ to $10M_{\rm \odot}$. However, due to our approximation of the star as a polytrope, we face a limitation in resolving the high-density regions near the core. This limitation arises from numerical constraints and the considerable computational cost associated with our 3D simulation. The dense and compact core cannot be adequately resolved without encountering small time steps that would render the merger simulation unfeasible within a reasonable timeframe. Consequently, in \octo, the ``core" extends to approximately $5~R_{\rm \odot}$, which is approximately 2.7 times larger than the core in the \mesa\ model. In alternative studies (such as in \citealt{Lau2022a}), the dense core is replaced with a point particle to circumvent these challenges. However, the use of a point particle introduces a softening length that may influence the evolution when the secondary star approaches the core particle closely, and it remains uncertain whether a merger could occur under such circumstances. Nevertheless, ongoing work is being conducted to implement point particle treatment within the \octo\ code, which will be beneficial for future investigations. In summary, the structure of our primary star adequately represents the corresponding \mesa\ structure in the regions of primary interest for this study. This enables us to examine the post-merger spin-up and envelope composition. Figure~\ref{fig:prim_comp_mesa} illustrates the comparison between our simulation and the \mesa\ model, depicting the initial primary star's density (left), pressure (center), and temperature (right) as functions of radius. Because the primary star fills its RL, it has a teardrop shape and is more extended towards the binary axis, $x_{\rm Binary}$. Therefore we show three line-plots in each panel: along the line that go from the center of the primary to the direction of the secondary star center ($\parallel x_{\rm binary}$; blue circles), in perpendicular to this binary axis ($\bot x_{\rm binary}$; orange x's) as well as mass averaged profile ('Avg.'; green pentagons). The figure demonstrates a close match between the two models in most spatial regions.

In order to assess the stability of the primary star, we conducted single star simulations using polytropic models with an index of $n=4.1$. Our investigations revealed that resolution plays a crucial role in maintaining stability, and we determined that a minimum resolution of 128 cells across the diameter of the primary star is necessary to ensure stability over an extended period (see Appendix~\ref{app:pri_stability} for further details). At this resolution, the star experiences only a slight collapse after approximately 140 dynamical orbits. Although this collapse may potentially reduce mass transfer during the binary simulation and consequently prolong the time to merger, we mitigate this effect by employing a higher resolution in the binary simulation. In fact, we use $3/2$ more cells across the diameter of the primary star in the binary simulation, thereby minimizing the impact of this minor collapse and maintaining the star's stability throughout the binary evolution.

Furthermore, due to the limited lifetime of a star with coarser grid resolution, it is not feasible to run a binary simulation using a coarser grid as the star would collapse before the merger could take place. Conversely, a finer grid simulation would incur a prohibitively high computational cost. Consequently, we were unable to conduct a comprehensive resolution study for the specific binary merger simulation presented in this paper.

\subsection{Secondary star's structure}
\label{ssec:numerics_sec}

\begin{figure*}
    \centering
    \includegraphics[scale=0.28]{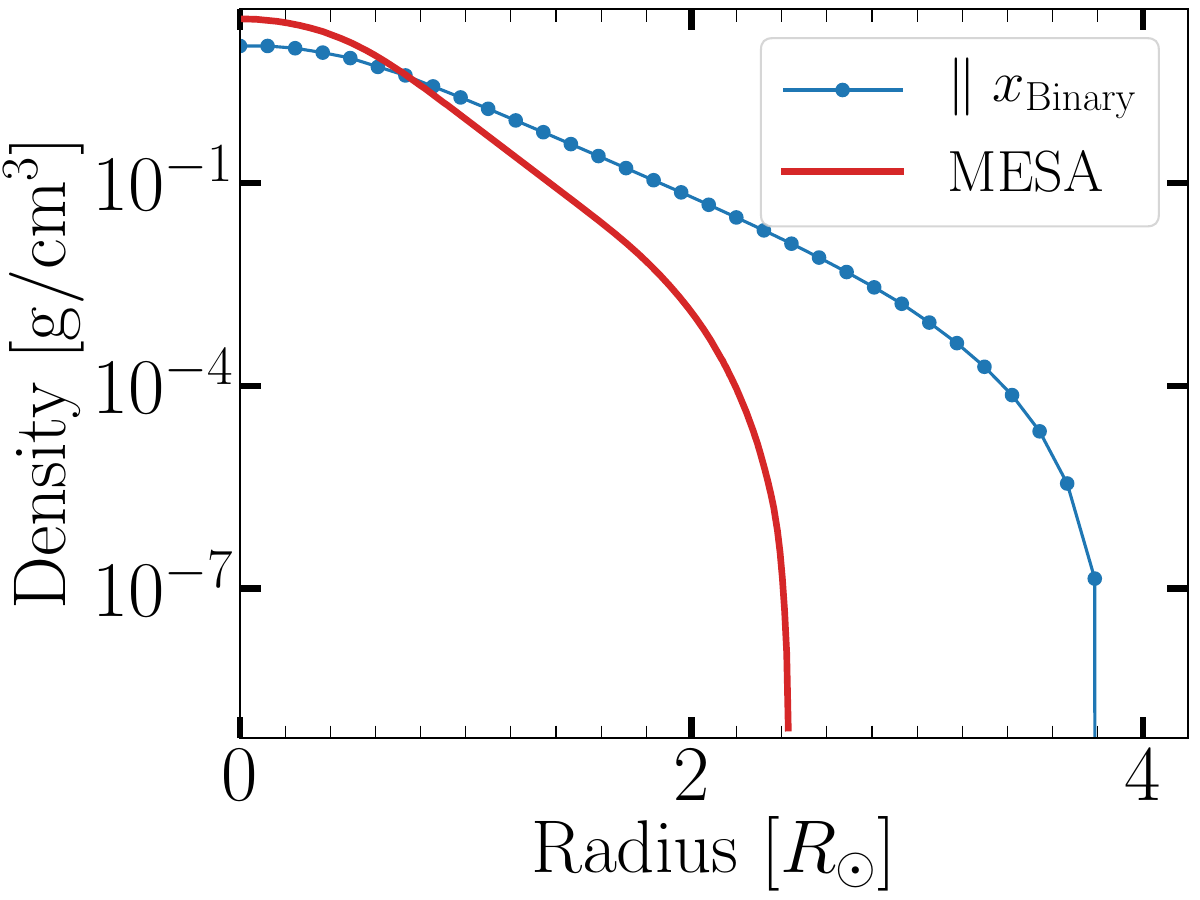}
    \includegraphics[scale=0.28]{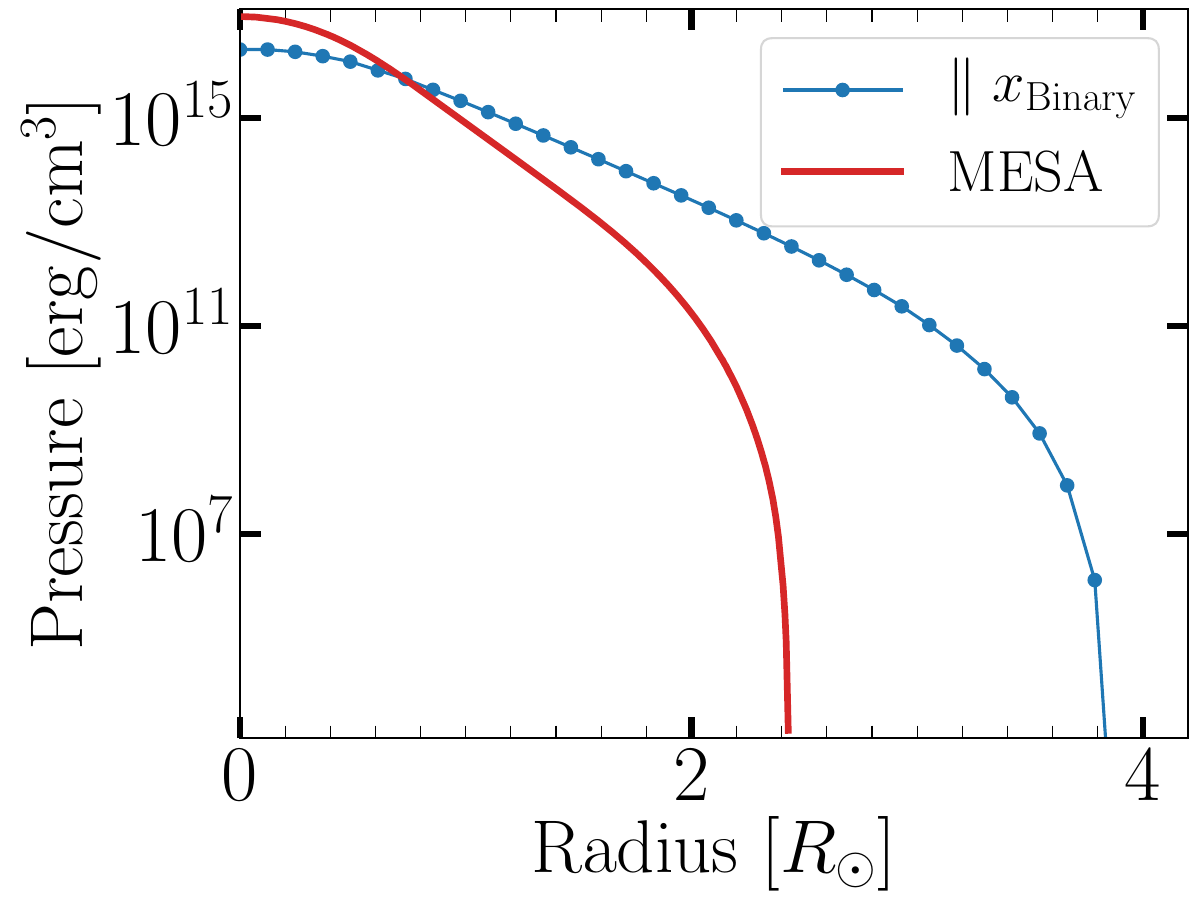}
    \includegraphics[scale=0.28]{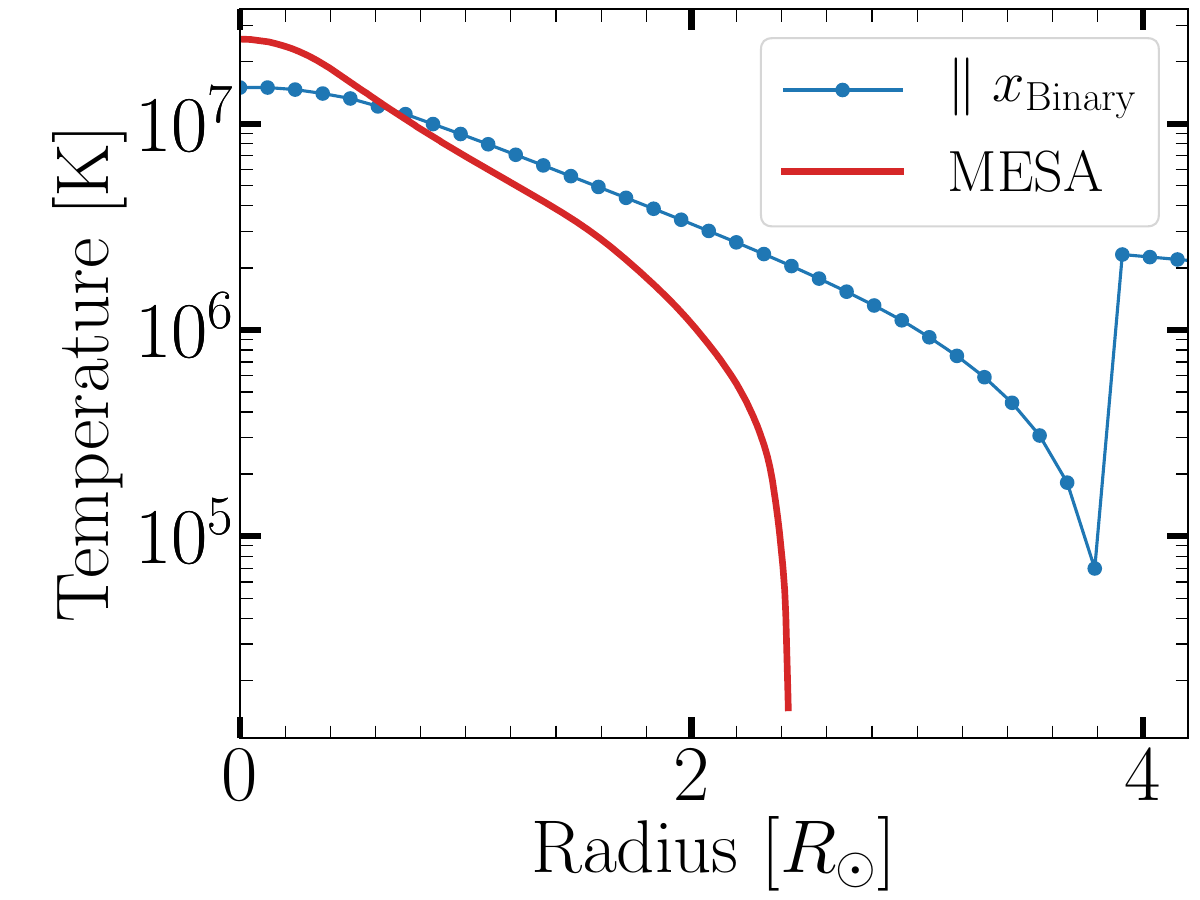}
    \caption{Density (left), pressure, (center), and temperature (right) as a function of radius of the initial secondary star in our simulation. We plot lines along the binary axis, $x_{\rm Binary}$ (lines that go from the center of the secondary star to the direction of the primary star center; blue circles), as well as the 1D MESA profile (solid red line) for comparison}
    \label{fig:sec_comp_mesa}
\end{figure*}

In accordance with the SCF technique, the size of the secondary star is determined by its RL filling factor. However, it is important to note that the resolution around the secondary star can impact its size and the SCF may not converge for certain small filling factor values if the resolution is insufficient. Insufficient resolution around the secondary star can also lead to significant fluctuations in the calculated gravitational forces, which can greatly hinder the conservation of angular momentum, and can cause the stars to wiggle around the center of mass. To address this issue, we apply two additional refinement levels around the secondary star. Additionally, we select the smallest filling factor that still converges at this resolution. 
In Figure~\ref{fig:sec_comp_mesa} we plot density (left), pressure (center), and temperature (right) as a function of radius of the resulted secondary star structure and we compare it to the \mesa\\ model. 

The secondary star obtained by the SCF is larger than the corresponding \mesa\ model but remains within a reasonable range for a main sequence star.
Increasing the resolution around the secondary star could have resulted in a smaller star size. However, it would also slow-down the calculation by at least a factor two, causing the simulation to become too expensive to run. 
{{Since the drag force depends on the accretion radius, it will not be affected by the actual size of the secondary star unless the orbital separation becomes very small. However, by that stage, ablation may have already changed the effective size of the secondary star.}}
Nonetheless, a smaller star might accrete less mass and more artificial driving by extraction of angular momentum would be required for the dynamical plunge-in to occur. Overall, resolving the secondary star is already an improvement from previous studies which utilize the secondary star as a point-particle that does not have an actual size (except maybe a numerical size like a softening length).

While the secondary star's central density is slightly smaller than in the \mesa\ model, the overall structure, characterized by an $n=3.4$ polytrope, is similar. Importantly, this structure exhibits greater stability compared to that of the primary star. Lastly, we assign a molecular weight of $\mu_2 = 0.62$ to the secondary star, based on the average molecular weight of the \mesa\ model.

\section{RESULTS}
\label{sec:results}
In this section, we present the findings from our simulations of the merger and post-merger evolution of a 16+4 $M_{\odot}$ binary system. Our analysis focuses on three key aspects.
First, in subsection \ref{ssec:evol}, we examine the binary's evolution leading up to the merger. This includes the behavior of the binary system as it approaches the merging stage. Second, in subsection \ref{ssec:outflows}, we investigate the amount of mass that becomes unbound due to the binary interaction. We analyze the properties of the mass loss resulting from the merger, including the morphology, geometry, and dynamics of the outflows. Finally, in subsection \ref{ssec:merger}, we explore the characteristics of the post-merger phase. We perform long-term nuclear evolution simulations using the {\sc MESA} code, utilizing the averaged three-dimensional post-merger structure from \octo\ to generate a one-dimensional profile. This allows us to study the properties of the post-merger system in terms of its nuclear evolution. By examining these three aspects, we gain a comprehensive understanding of the merger and post-merger processes in the 16+4 $M_{\odot}$ merger model.

\subsection{Orbital Evolution}
\label{ssec:evol}
To accelerate the merger process and minimize computational time, we employ a driving mechanism in the binary system. Following the approach outlined in \cite{Marcello2021}, we extract angular momentum from the grid at a constant rate of 1\% per initial orbit. This manipulation leads to a reduction in the orbital separation and facilitates a more significant mass transfer between the stars. Without this driving mechanism, the system would undergo numerous orbits, potentially several hundreds, before a merger occurs. Given our current resources and time constraints, simulating such a prolonged evolution would be impractical.

\begin{figure*}
    \centering
    \includegraphics[scale=0.28]{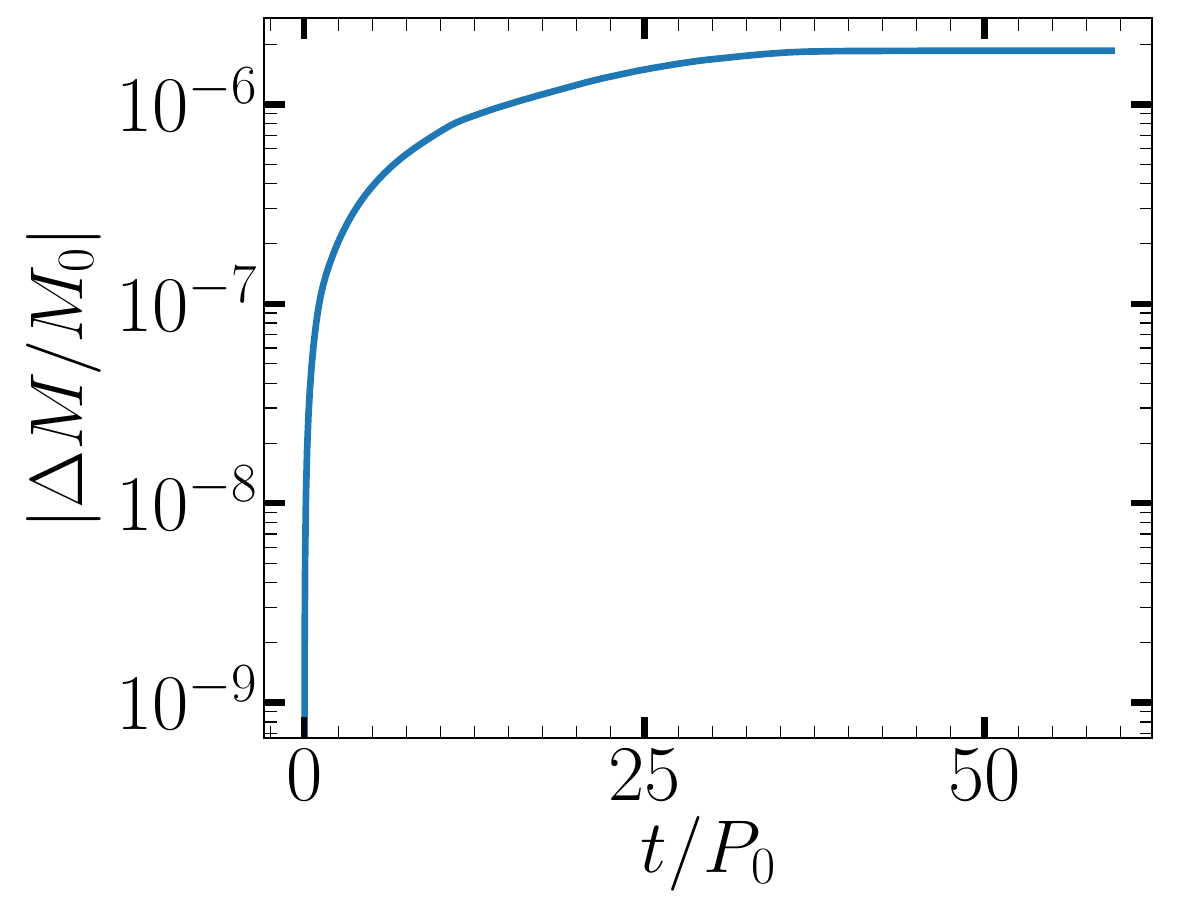}
    \includegraphics[scale=0.28]{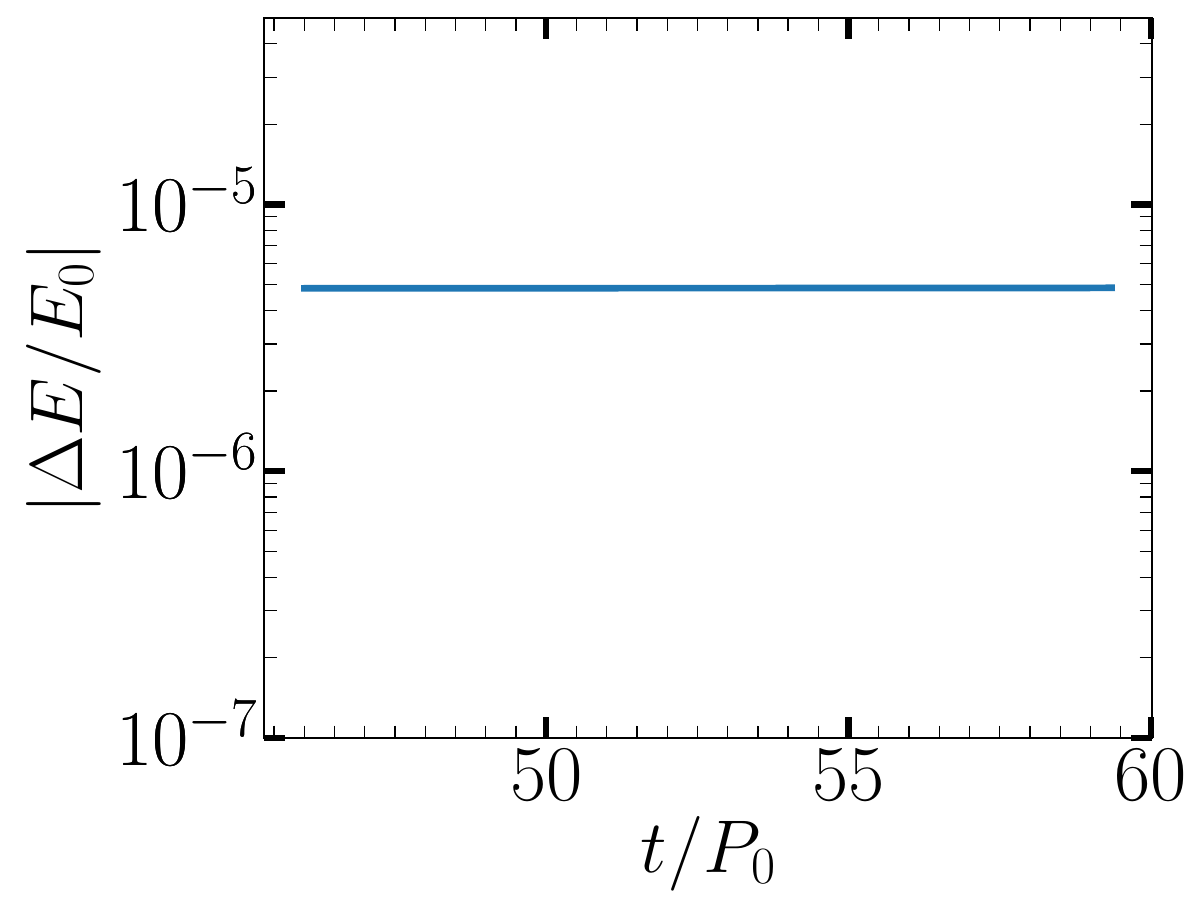}
    \includegraphics[scale=0.28]{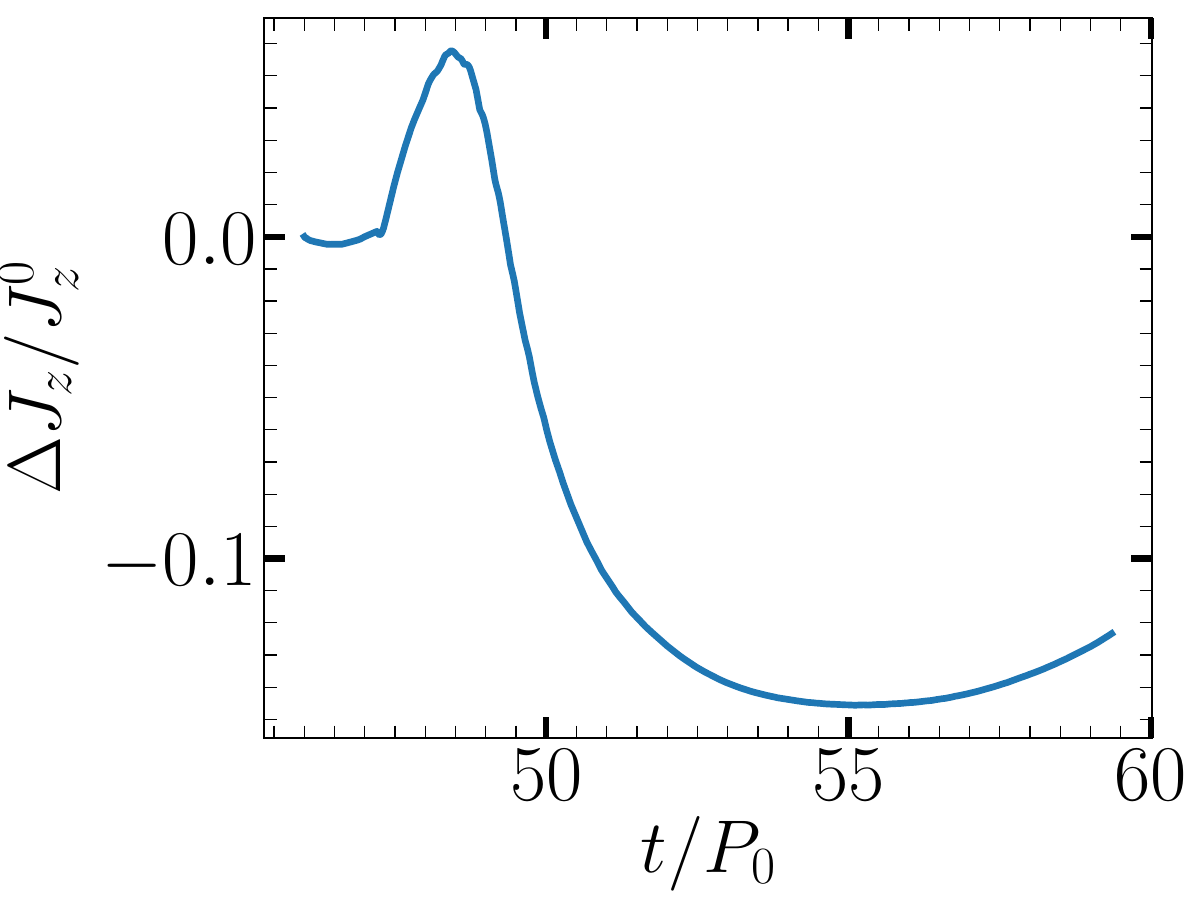}
    \caption{Conservation of mass, energy, and angular momentum. Note that the in the angular momentum plot we do not take into account angular momentum losses due to gas that flows outside of the simulation domain.}
    \label{fig:conserved_quants}
\end{figure*}

In Figure~\ref{fig:conserved_quants} we plot the conservation of mass (left), energy (center), and angular momentum (right) in the \octo\ merger simulation. 
In this plot, we examine the changes in total mass, total energy, and total z-angular momentum over time, denoted as $\Delta M$, $\Delta E$, and $\Delta J_z$ respectively. These quantities are calculated by comparing the current values to their initial values in the grid: $\Delta M = M(t) + M_{\rm out}(t) - M_0$, $\Delta E = E(t) + E_{\rm out}(t) - E_0$, and $\Delta J_z = J_z(t) - J_z^0$. Here, $M_0$, $E_0$, and $J_z^0$ represent the initial total mass, total energy, and total z-angular momentum in the grid. Additionally, $M_{\rm out}(t)$ and $E_{\rm out}(t)$ account for the mass and energy that have left the grid, respectively. For energy and angular momentum, we assess their conservation only after discontinuing the driving mechanism at $t=46P_0$. Our findings reveal that mass and energy are conserved at a precision level of $10^{-6}$. However, due to numerical viscosity-induced torques during and after the merger, the total z-angular momentum increases by 5\%. Furthermore, as mass leaves the grid, the angular momentum decreases by 20\%. Although quantifying the exact amount of z-angular momentum leaving the grid is challenging, our results indicate that angular momentum is still conserved at a level of approximately 5\%.

As described in \cite{Marcello2021} we use an iterative post-processing method to identify the cells that belong to each star. This diagnostics scheme allows us to calculate the system's orbital properties, like orbital separation and orbital angular momentum, as well as the mass, energy and spin of each star. We improved the technique and instead of just iterating over a fixed number of iterations we iterate until the stars' masses ($M_1$ and $M_2$) and the orbital separation ($a$) converged, i.e, when the values in the current iteration are at most within $10^{-5}$ of the values in the previous one. Note that when the system approaches the final merger, it becomes increasingly difficult to differentiate between the two stars and the diagnostics method cannot be reliably used. We typically find that this occurs when the binary frequency, as calculated by the center of mass, $(x^{1,2}, y^{1,2}, z^{1,2})$ and center of mass velocity $(v_x^{1,2}, v_y^{1,2}, v_z^{1,2})$ of each star, $\Omega_{\rm orb} = j_{z, {\rm orb}}/a^2 = [(x^2-x^1)(v_y^{2} - v_y^{1}) - (y^2 - y^1)(v_x^2 - v_x^1)]/a^2$, falls below half of the nominal Keplerian frequency, $\Omega_{\rm kep}=\sqrt{G(M_1 + M_2)/a^3}$. This is to be expected as the donor's envelope is tidally disrupted, and much of its mass increasingly lags behind its center of mass. Therefore we may define the binary merging time as when the binary frequency falls below half of the nominal Keplerian frequency.

\begin{figure*}
    \centering
    \subfloat[]{\includegraphics[scale=0.32]{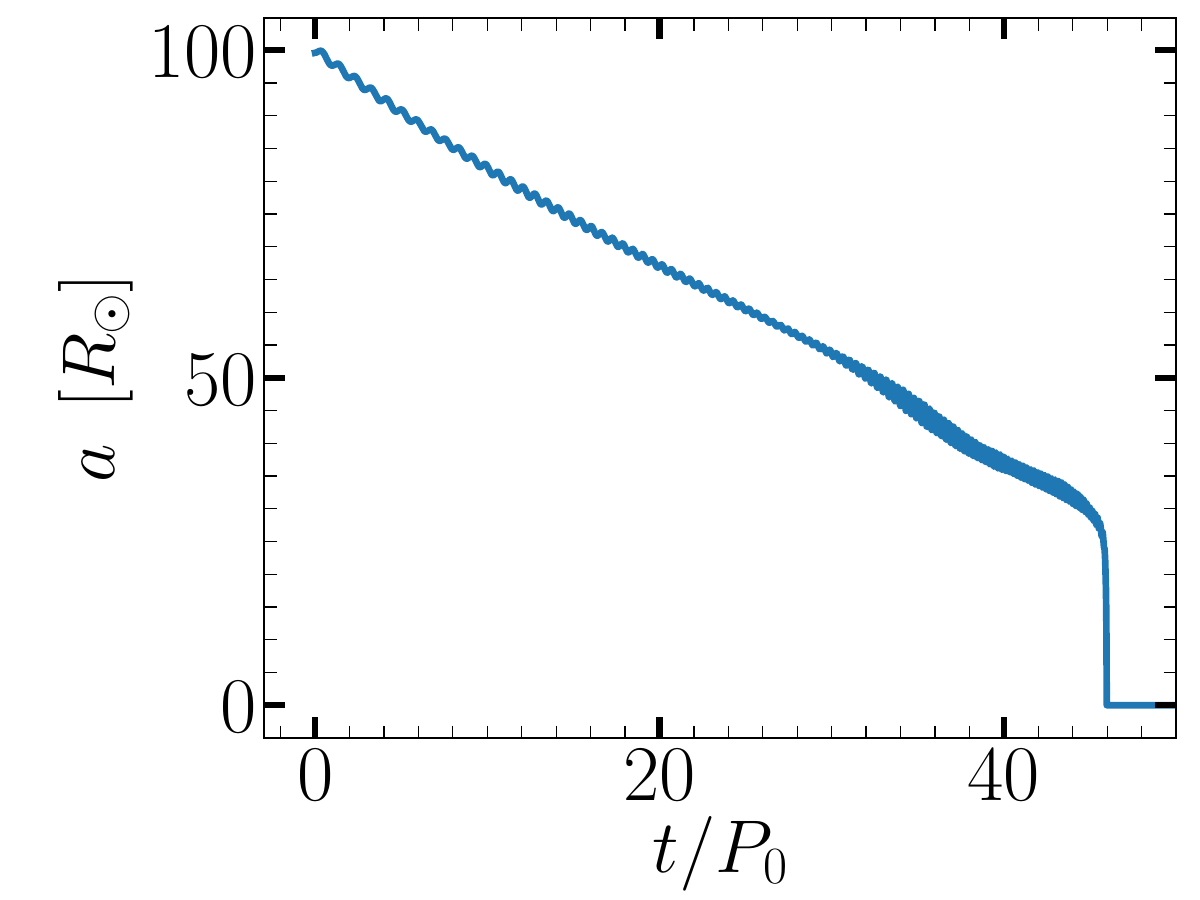}}
    \subfloat[]{\includegraphics[scale=0.32]{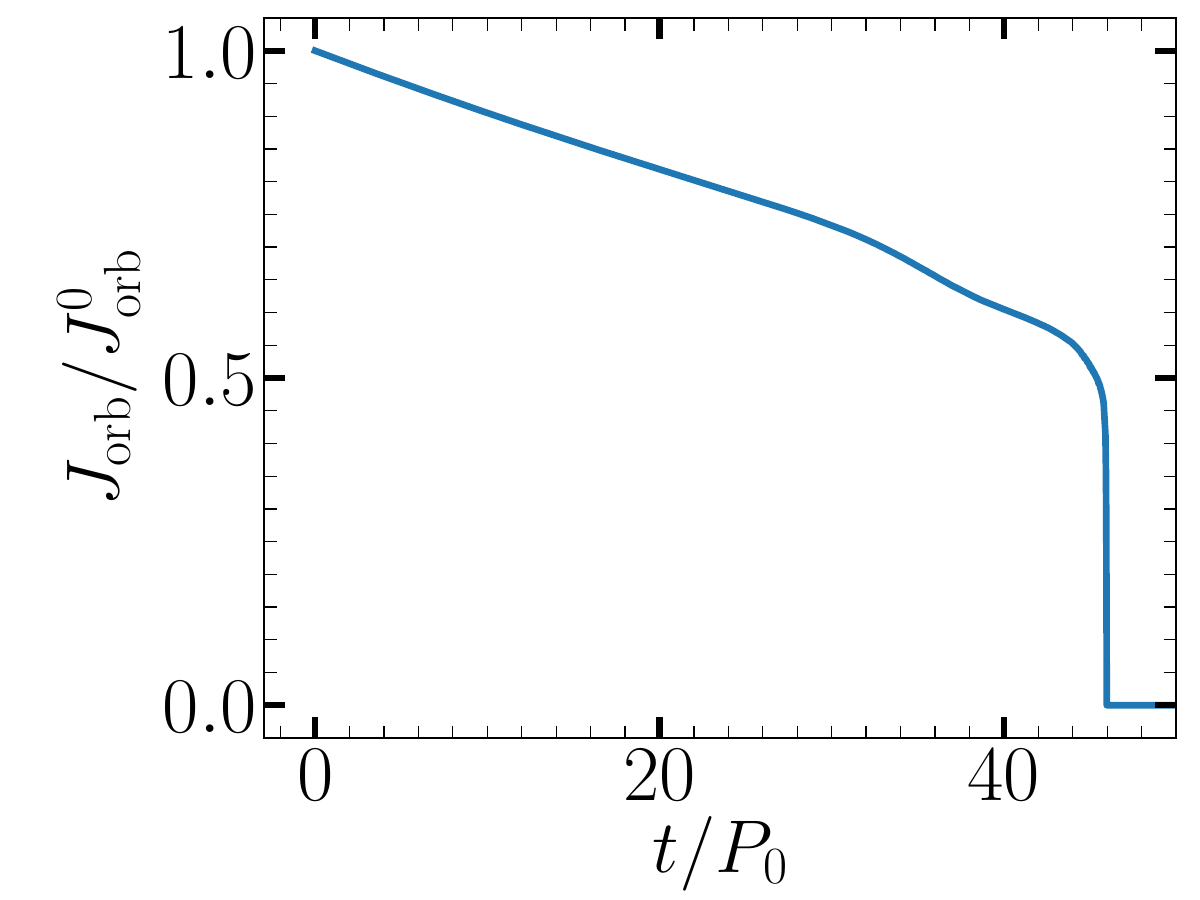}}
    \qquad
    \subfloat[]{\includegraphics[scale=0.32]{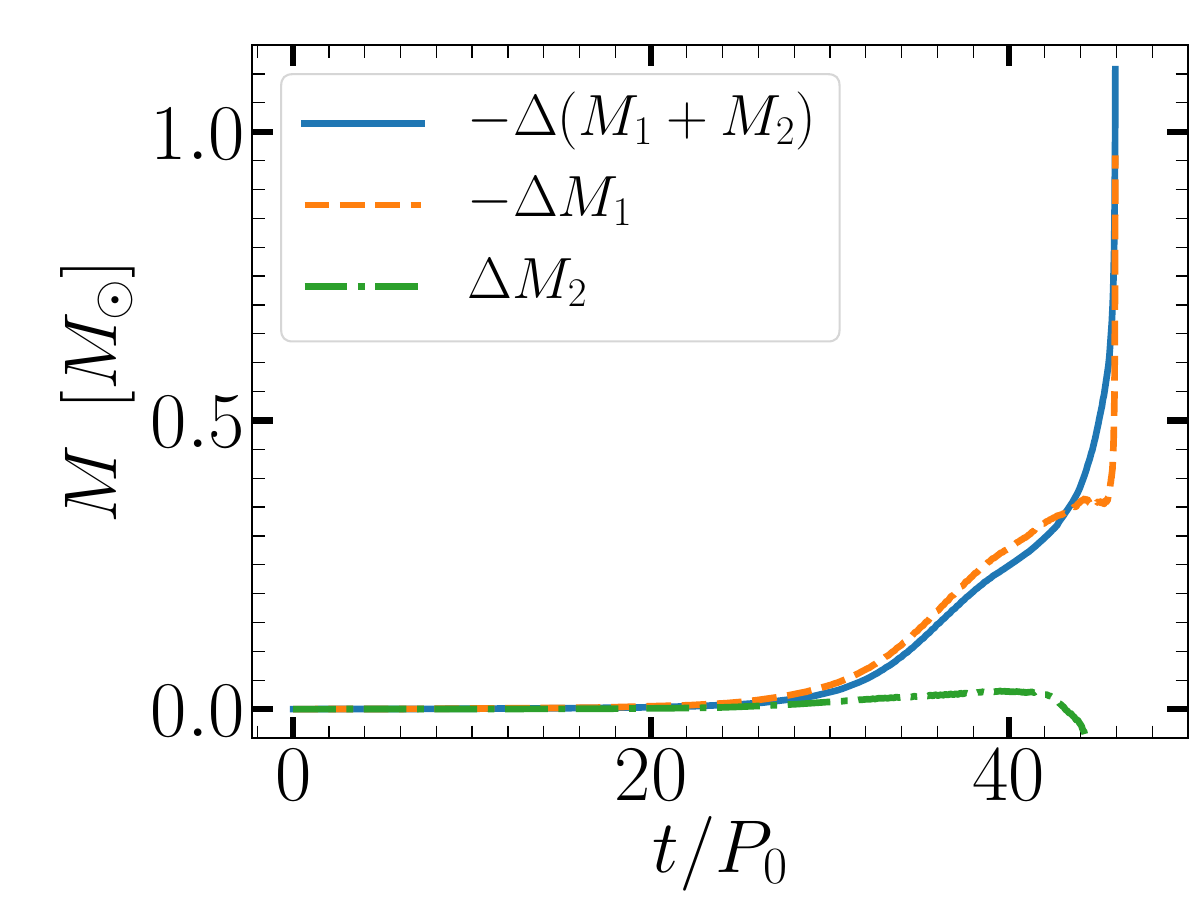}}
    \subfloat[]{\includegraphics[scale=0.32]{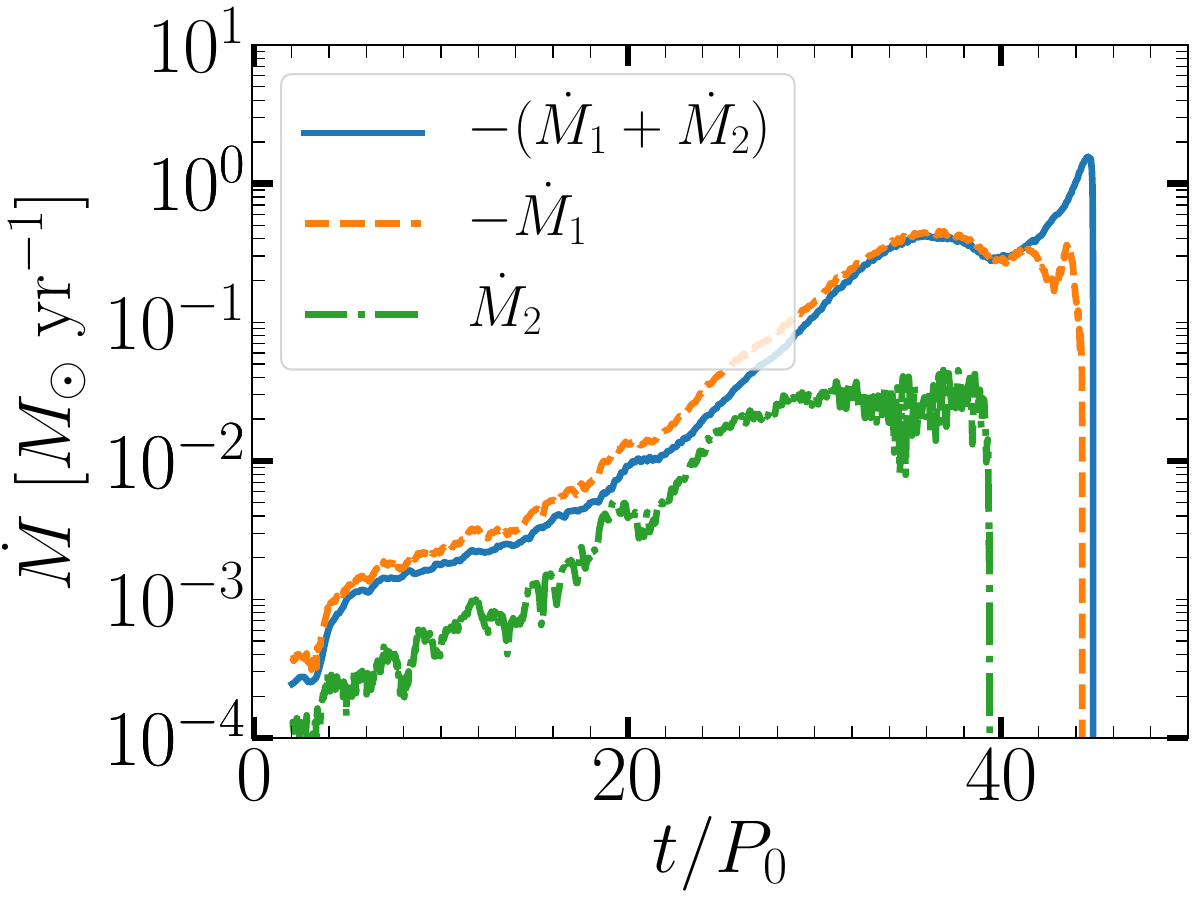}}
    
    \caption{Binary diagnostics plots as a function of time. (a) Orbital separation; (b) Orbital angular momentum; (c) mass losses, where $\Delta M = M(t) - M(0)$; and (d) mass transfer rates. The primary star (star 1) looses mass (dashed orange lines), however, the secondary star only accretes a fraction of it (dashed-dotted green lines) and the rest leaks through L2 and leaves the system (blue solid lines). As the mass transfer is otherwise stable we continuously extract angular momentum at a rate of 1\% per initial orbit to expedite the merger (see discussion on the stability of mass transfer in appendix~\ref{app:mt_stability}). The system merges after 45.96 initial periods ($P_0 = 26~{\rm days}$)}
    \label{fig:diag_plots}
\end{figure*}

In Figure~\ref{fig:diag_plots} we plot orbital separation, orbital angular momentum, mass losses ($\Delta M = M(t) - M(0)$), and mass transfer rates as a function of time (in initial orbital period units, $P_0=26~{\rm days}$). All these quantities were calculated using the diagnostics scheme discussed in the previous paragraph. The mass transfer rates are smoothed by the Savitzky-Golay filter \citep{savitzky1964smoothing} with a window size of $2P_0$.

As anticipated, the presence of driving in the system leads to an initial decrease in the orbital separation (panel a) and orbital angular momentum (panel b) that follows a nearly linear trend. At this stage, the primary star overflows its RL, resulting in a high mass loss rate of $10^{-4}-5\times 10^{-1}~M_{\rm \odot}~{\rm yr^{-1}}$ (panel d, indicated by the dashed orange line). However, only a small portion of this lost mass is accreted by the secondary star (green dash-dotted line). Surprisingly, the system experiences a net loss of mass (blue solid line) rather than a net gain by the secondary. As a result, the mass transfer rate does not increase by much, necessitating continuous driving to bring the system into contact and trigger the merger. {{This continuous driving causes the system's center of mass to spiral-out of the center of the grid to a distance of $\sim 15 R_{\rm \odot}$ by the time of the merger. In the vertical direction, the system's center of mass is not affected by this driving and it moves less than the size of one grid cell.}}

The mass loss from the primary star intensifies as the orbital separation decreases and its RL shrinks. The accretion rate onto the secondary star increases over time as well, however, it settles down on a level of $\sim 10^{-2}~M_{\rm \odot}~{\rm yr}^{-1}$, probably because the secondary star cannot accommodate a faster accretion, and the mass lost by the primary star escapes through L2 rather than accreted. These accretion rates are consistent with other studies that conduct 3D hydrodynamic simulations of CEE, although with a red giant and less massive primary star, where they find accretion rates onto a MS secondary star of $10^{-3}-10^{1}~M_{\rm \odot}~{\rm yr}^{-1}$ \citep{Chamandy2018, Shiber2019, Lopez2022}. A MS star that accretes mass at a high rate might launch jets \citep{Shiber2016}, which can help in unbinding the envelope and perhaps even postpone the fast plunge-in (grazing envelope evolution; \citealt{Soker2015}, \citealt{ShiberSoker2018}, \citealt{Shiber2019}). However, we do not include jets in our simulation and disregard the jets' energy which can be approximated by taking a small factor $\eta$ of the accretion energy onto the secondary star, $E_{\rm jets}=\eta G M_2 \Delta M_2 / R_2 \simeq \eta \times 2\times 10^{47}~{\rm ergs}$. {{We also note that jets may act as a positive feedback that increases the accretion rate onto the companion, either by removing energy and decreasing the pressure in the vicinity of the accreting star, or by pushing mass towards the orbital plane where most of the accretion takes place. If this positive feedback is substantial, the increase in the accretion rate would result in higher jets' energy and the role of jets can become more important.}} 

During the later stages of the simulation, when the secondary star penetrates the envelope of the primary star (approximately at $t=30P_0$), the hydrodynamic and gravitational drag forces cause the secondary star to spiral further into the primary's envelope. Eventually, it merges with the helium (He) core of the primary star, which occurs within a timescale of approximately $15P_0$ or around 1 year. By this point, a total mass of approximately $\lesssim 1.3M_{\rm \odot}$ is ejected from the system (panel (c)), where some of it actually originated from the secondary star. This mass carries an energy of $6.5 \times 10^{46}~{\rm ergs}$, and z-angular momentum of $0.3 J_{\rm orb}^0$, where $J_{\rm orb}^0$, is the initial orbital angular momentum and equals $8.5\times 10^{53}~{\rm ergs~s}$. However, not all of this mass is necessarily unbound or remains unbound. In the subsequent subsection (subsection~\ref{ssec:outflows}), we will demonstrate that approximately half of this amount of mass becomes unbound during the merger process and ultimately flows out of the computational grid, contributing to the formation of a mergeburst transient event.

\begin{figure*}
    \centering
    \includegraphics[scale=0.6, trim={0 0.0cm 0 0cm},clip]{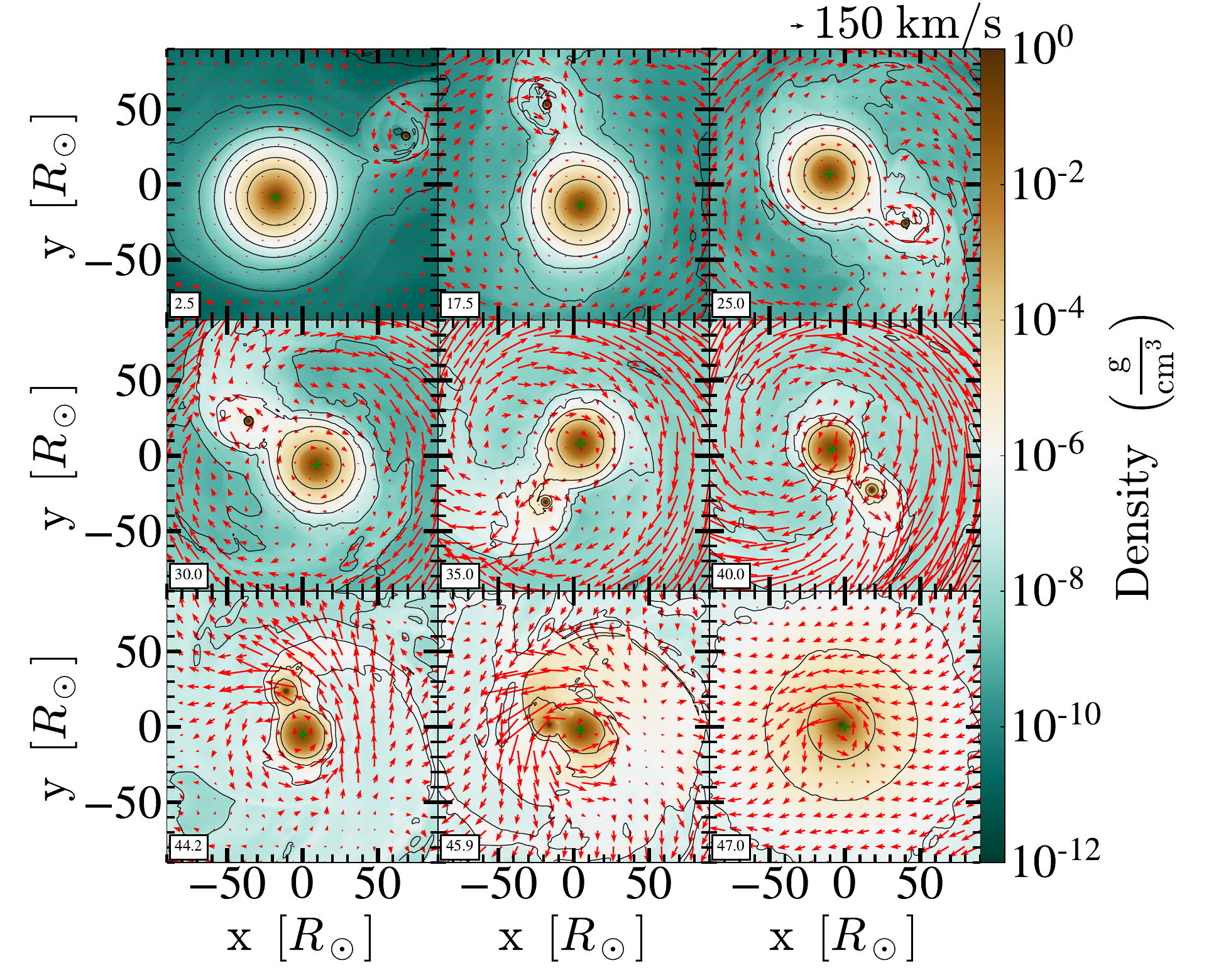}
    \caption{Equatorial density slices at nine different times  throughout the evolution up to the dynamical merger phase: $2.5P_0$, $17.5P_0$, $25.0P_0$, $30.0P_0$, $35.0P_0$, $40.0P_0$, $44.2P_0$, $45.9P_0$, and $47.0P_0$. We also plot the velocity field scaled by velocity magnitude (in red; key shown on the upper right corner) to highlight details of the flow. Velocities at the six first panels are measured in the frame rotating with momentary binary frequency, while in the three last panels they are measured in the frame rotating with the {\it initial} binary frequency. The panels are centered around the system's center of mass, and the green cross symbol denotes the location of the primary's star center of mass. {{The system rotates counter clock-wise}}}
    \label{fig:dens_evol}
\end{figure*}

\begin{figure*}
    \centering
    \includegraphics[scale=0.6, trim={0 0.0cm 0 0cm}, clip]{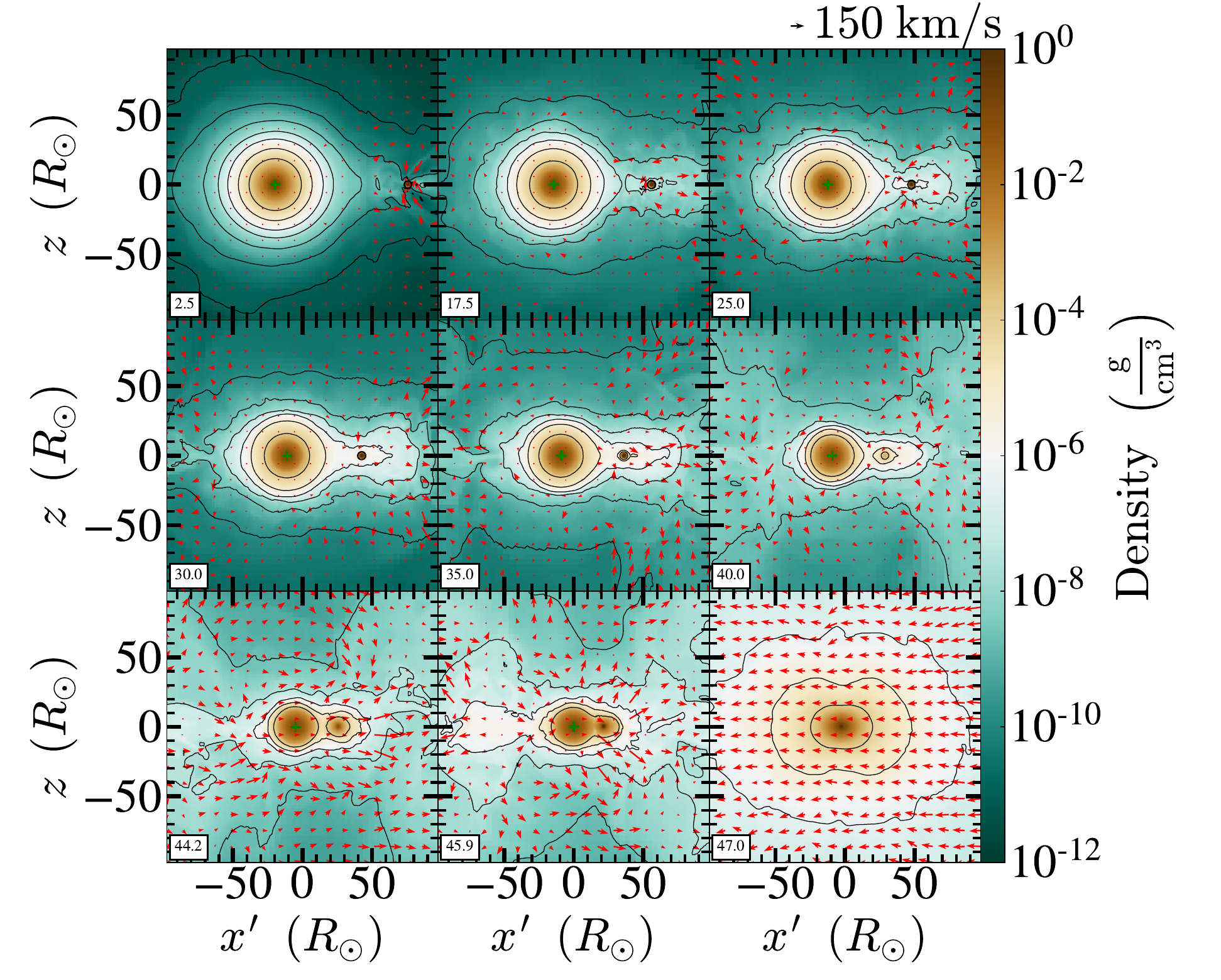}
    \caption{Meridional density slices at nine different times throughout the evolution up to the dynamical merger phase. Times and symbols are like in Figure~\ref{fig:dens_evol}}
    \label{fig:dens_evol_meri}
\end{figure*}

In Figures~\ref{fig:dens_evol}~and~\ref{fig:dens_evol_meri} we show density map slices along the equatorial (orbital) and meridional plane, respectively, at nine different times as denoted at the lower-left corner of each panel: $2.5P_0$, $17.5P_0$, $25.0P_0$, $30.0P_0$, $35.0P_0$, $40.0P_0$, $44.2P_0$, $45.9P_0$, and $47.0P_0$. We also plot the velocity field scaled by velocity magnitude (in red; key shown on the upper right corner) to highlight details of the flow\footnote{Movies of the simulation can be obtained via \href{https://lsu.box.com/s/nypv2te4cqgcno3sy2tsd03seei7mkrv}{this link}}. The plotted velocities at the six first panels (i.e., $t \leq 30P_0$) are measured with respect to a frame that rotates at the momentary orbital frequency (i.e., frame of reference that rotates with the binary). In the three last panels ($t > 30P_0$; closer to the merger time and afterwards), the plotted velocities are measured with respect to a frame that rotates at the initial orbital frequency (i.e., a frame of reference that rotates with the grid). {{The binary and as a consequence the grid itself both rotate at the counter clock-wise direction}}. Each panel is centered around the system's center of mass, and the green cross symbol denotes the location of the primary's star center of mass.

These plots illustrate the accumulation of mass around the secondary star and the system's overall mass loss through L2. Initially, mass flow originating from the primary star begins to envelop the secondary star, while mass escaping through L2 forms a spiral-shaped arm ($t=0-30P_0$). As the secondary star spirals inward and approaches the envelope of the primary star, a common envelope configuration is established ($t=30P_0$). Over time, hydrodynamical drag and gravitational forces act upon the system, driving the secondary star deeper into the envelope of the primary star, until the secondary tidally disrupts the primary's helium (He) core at $t=45.96P_0$. Subsequently, the secondary plunges into and mixes with the remnant helium core of the primary over the following couple of $P_0$. However, this final evolution is a consequence of the choices we had to make for numerical reasons, which resulted in a secondary core denser than the primary core. {{Since the central density of the primary star is $\sim 30$ times denser than the secondary's star central density, according to the {\sc MESA} models, in a more realistic evolution, it would be the primary core that disrupts the secondary}}. Therefore we are not able to say anything definite about the potential rejuvenation of the primary core.

Our finding that a substantial portion of the mass leaving the primary star is expelled through the L2 Lagrange point and is subsequently lost from the system instead of being accreted onto the secondary star could drastically affect the stability of the mass transfer. A conservative mass transfer in its simplest form dictates that a mass that passed from the more massive to the less massive star leads to a dynamical unstable mass transfer and to the formation of a CE. A more sophisticated analysis includes the response of each star to its mass loss or mass gain, respectively. In our specific case, the primary star possesses a predominantly radiative envelope that we approximate using a polytropic index of $n_1 = 4.1$. Consequently, its adiabatic mass-radius exponent, denoted as $\zeta_{\rm ad}$ and calculated as $(d\log R/d\log M)_{\rm ad}$, is approximately 2.82 and the primary radius will shrink when it looses mass. This needs to be compared with the mass-radius exponent associated with the Roche lobe, denoted as $\zeta{\rm_{RL}}$, which is estimated to be around $2.13 (M_1/M_2) - 1.67 \approx 6.5$ according to Tout et al. (1997) for conservative mass transfer. Therefore, based on this criterion, the RL will shrink faster than the primary star, and the mass transfer between the stars should be dynamically unstable.

However, our simulations reveal that the assumption that all the mass lost from the donor star is accreted by the secondary does not hold. This phenomenon can greatly enhance the stability of the mass transfer process, depending on the fraction of mass accreted, $f$, as well as on angular momentum losses through L2. A careful calculation of the stability of mass transfer for non-conservative cases shows that in the absence of angular momentum losses, $f < f_{\rm max} = 0.53$ would be dynamically stable. Angular momentum losses act to destabilize the mass transfer and thus effectively reduces $f_{\rm max}$ (for a detailed calculation we refer to appendix~\ref{app:mt_stability}). It is therefore plausible that a mass transfer on a thermal timescale is expected to occur under these conditions. Once the primary's envelope expands further and reaches the size of the orbit, the mass transfer will transition into a dynamically unstable regime. 

Moreover, even if the conditions are such that the mass transfer is dynamically unstable, it could be that the mass transfer rate will only slowly increase, implying that numerous orbits are still required to simulate before the actual merger will take place. It is not trivial to know a priori the amount of driving required to achieve a specific mass transfer rate nor to predict the subsequent behavior of the mass transfer rate once the driving is stopped. In fact we tried to stop the driving at several instances during the evolution, finding that the following evolution would be still too long to simulate. In practice we applied the driving mechanism continuously until the system merged.
Notably, the behavior observed while driving the system to contact (and beyond) suggests that the initial binary separation should have been chosen smaller, around $50 R_{\odot}$ instead of the $100 R_{\odot}$ estimated from the point at which the primary fills its RL. However, a smaller  initial separation can result in a lower final equatorial velocity, as demonstrated in Chatzopoulos et al. (2020) through Equation 5 and Figure 1.

\subsection{Mergerburst and outflow}
\label{ssec:outflows}
 
During the common envelope evolution, as the secondary star interacts with the envelope of the primary, there is an exchange of orbital and thermal energy. This process ultimately leads to the merger of the secondary star with the helium core of the primary. As the merger occurs, a powerful dynamical pulse is triggered that propagates through the envelope of the primary. Under certain conditions, this pulse can induce significant mass loss from the system. The gravitational energy released during the merger is primarily converted into kinetic energy, driving the high-speed outflow of gas from the primary's envelope. Additionally, a portion of the released energy is transformed into internal or thermal energy. As a result, the gas within the primary's envelope is expelled at high velocities, and a substantial fraction of it attains enough energy to surpass the gravitational potential barrier. Consequently, this gas becomes unbound from the system and is no longer gravitationally bound.
 
\begin{figure}
    \centering
    \includegraphics[scale=0.4]{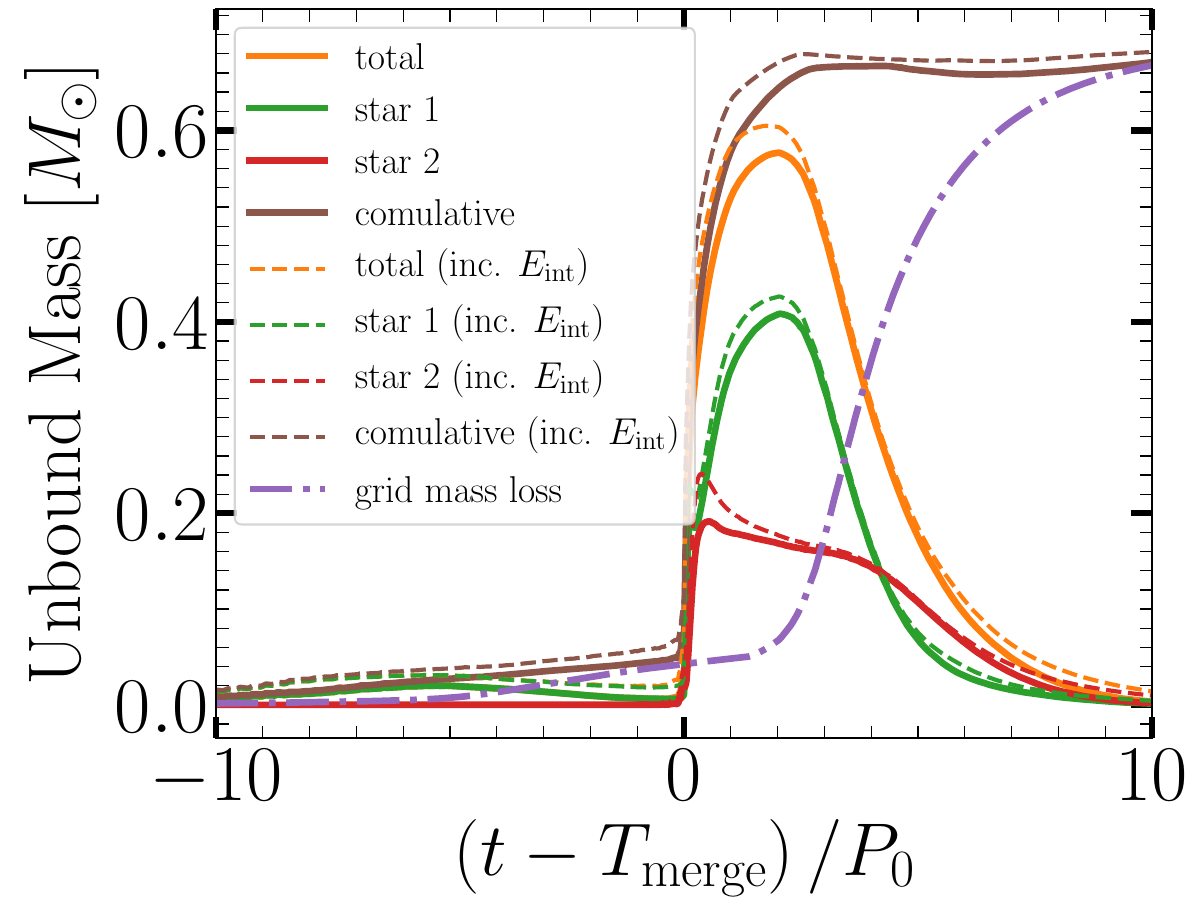}
    \caption{The unbound mass as a function of time in our simulation (solid lines). The mass that left the grid is also plotted (magenta dashed-dotted line). The cumulative unbound mass (brown lines) sums unbound gas in the grid plus the mass that left the grid, assuming that all the mass that left the grid, left it unbound and remains unbound afterwards. Including internal (thermal) energy only slightly increase the unbound mass (dashed lines)}
    \label{fig:unb_mass}
\end{figure}

Figure~\ref{fig:unb_mass} illustrates the quantity of mass that becomes unbound within the grid over time. We focus on the ten orbits preceding the merger time, $T_{\rm merge}=45.96P_0$, and the subsequent period, as only a small amount of mass becomes unbound during earlier times. We establish two criteria for identifying unbound mass. The first criterion considers gas as unbound when its kinetic energy plus gravitational energy, $E_k + E_g$, is positive (depicted by solid lines). The second criterion incorporates internal energy, where gas is considered unbound when its total energy, $E_k + E_g + E_{\rm int}$, is positive (represented by dashed lines). It is important to note that due to the finite-volume nature of our grid, the unbound mass eventually flows out of the grid, despite its relatively large size. Consequently, we include in the plot the quantity of mass that exits the grid (indicated by dash-dotted lines).

We observe that approximately $\sim 0.6M_{\rm \odot}$ of mass becomes unbound within the grid. The majority of this unbinding occurs at the moment of merger and during a time scale of one initial orbital period ($P_0=26~{\rm days}$). When accounting for internal energy in the energy balance, the amount of unbound mass is only slightly higher. Assuming that all the mass leaving the grid remains unbound, the maximum estimate of unbound mass is approximately $\sim 0.68~M_{\rm \odot}$ (indicated by brown-colored lines in Figure~\ref{fig:unb_mass}). This corresponds to approximately $\sim 0.07$ of the mass of the primary's envelope or $\sim 0.03$ of the total stellar mass of the system. Furthermore, a smaller phase of unbinding, approximately $\sim 0.02~M_{\rm \odot}$, occurs prior to the merger (beginning at $t=T_{\rm merge} - 5 P_0$ and decaying until $t=T_{\rm merge}$, allowing all the unbound gas to escape the grid before the merging time).

Interestingly, during the pre-merger unbinding phase, all the unbound mass originates from the primary star (depicted by green lines in Figure~\ref{fig:unb_mass}, labeled as 'star 1' in the legend). However, after the merger, when the majority of the unbinding occurs, only approximately $\sim 2/3$ of the unbound mass originates from the primary star, while the remaining $\sim 1/3$ originates from the secondary star (indicated by red lines, labeled as 'star 2' in the legend). This finding suggests that the common envelope interaction can unbind a portion of the spiraling-in secondary star's mass (approximately $0.2M_{\rm \odot}$ out of the $4M_{\rm \odot}$, which is 5\%), highlighting the need for future studies to consider this effect rather than relying solely on the simple $\alpha$ formalism often used. Such results, which cannot be captured when representing the secondary star as a point particle, warrant further investigation.

The fraction of the primary's envelope that becomes unbound (7\%) is relatively smaller compared to findings from previous studies on common envelope evolution of massive stars. However, in those studies they include some extra sources of energies that our simulation does not account such as recombination energy, radiation energy, and jets \citep{Lau2022a,Lau2022b,Moreno2022,Schreier2021}. 
Other previous works that do not include additional energy sources find ejection values more similar to ours (although for a low mass primary star; e.g., \citealt{Passy2012,Iaconi2017,Reichardt2019,Shiber2019}, and \citealt{sand2020}). Another possible explanation stems from the fact that our primary has not evolved yet to become a RSG and thus its envelope possesses much higher binding energy compared to the binding energy its envelope will posses when it will evolve to be a RSG star \citep{Klencki2021}. Other studies deliberately chose primary stars during its RSG phase and at maximal expansion to ease the envelope removal \citep{Lau2022a,Lau2022b,Moreno2022,ricker2019}.

\begin{figure}
    \centering
    \includegraphics[scale=0.4]{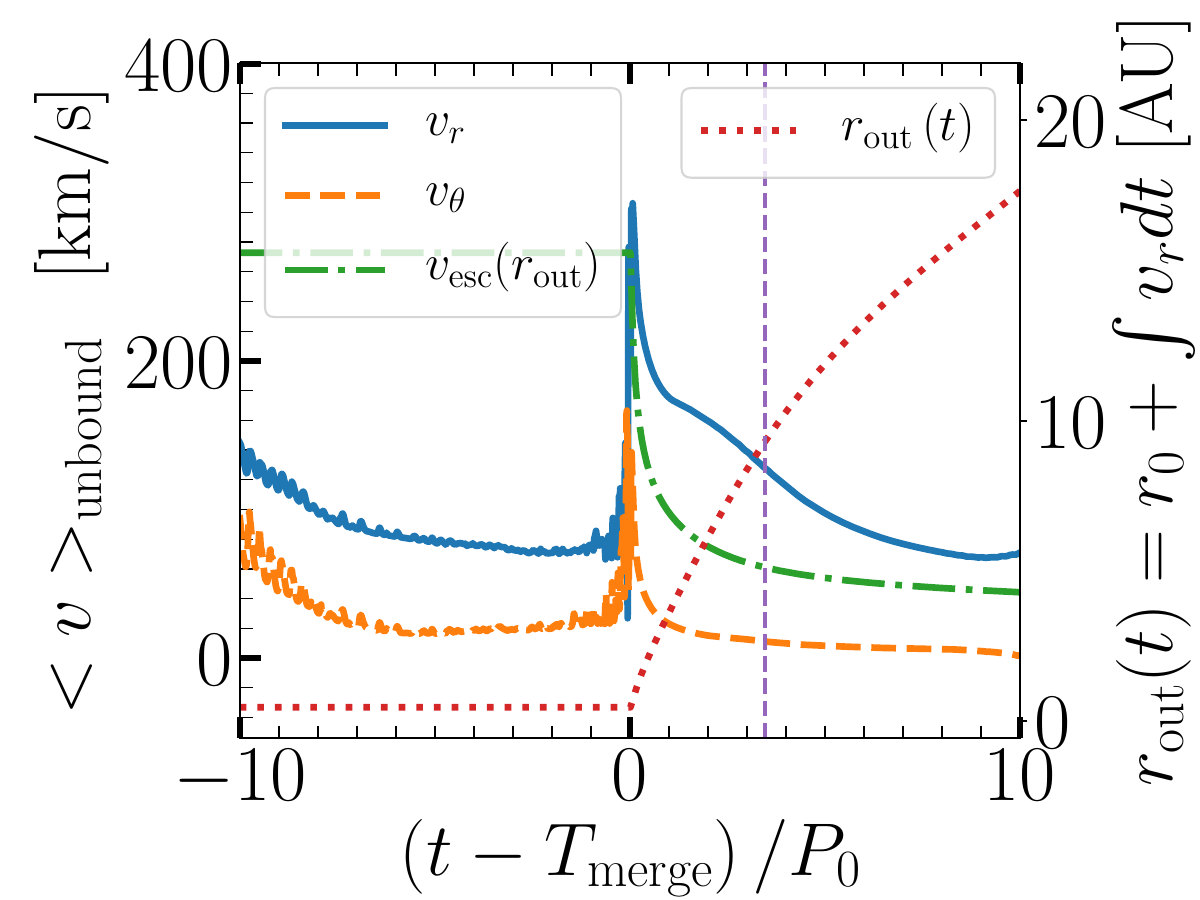}    
    \caption{Average unbound gas velocities and inferred radial distance as a function of time. The (solid) blue curve denotes spherical radial velocity, $v_r$, and the (dashed) orange curve azimuthal spherical velocity, $v_\theta$. The (dotted) red curve denotes the outflow average distance, $r_{\rm out}$, as calculated by integrating $v_r$, $r_{\rm out} = r_0 + \int v_rdt$ with $r_0$ being equal to the initial separation $100~R_{\rm \odot}$, while the (dashed) green curve denotes the escape velocity at this distance, $v_{\rm esc}(r_{\rm out})$. At each time the average velocity is a mass-averaged calculated only for unbound gas. The purple vertical line denotes the time in which the outflow arrives to the computational grid boundary
    }
    \label{fig:unb_mass_vel}
\end{figure}

\begin{figure*}
\centering
    \subfloat[$t=T_{\rm merge} + 0.26P_0$]{\includegraphics[width=6.0cm]{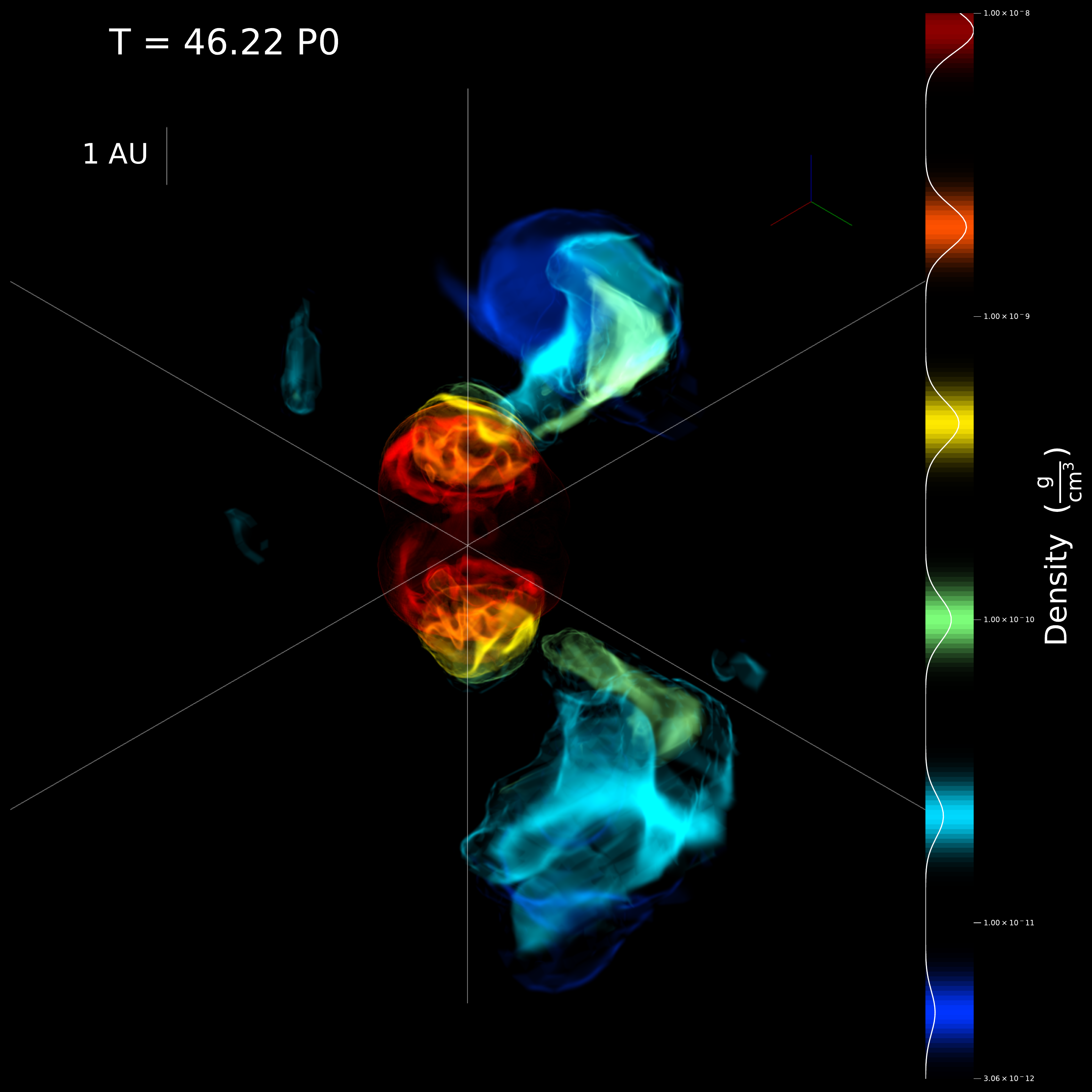}}%
    \subfloat[$t=T_{\rm merge} + 0.37P_0$]{\includegraphics[width=6.0cm]{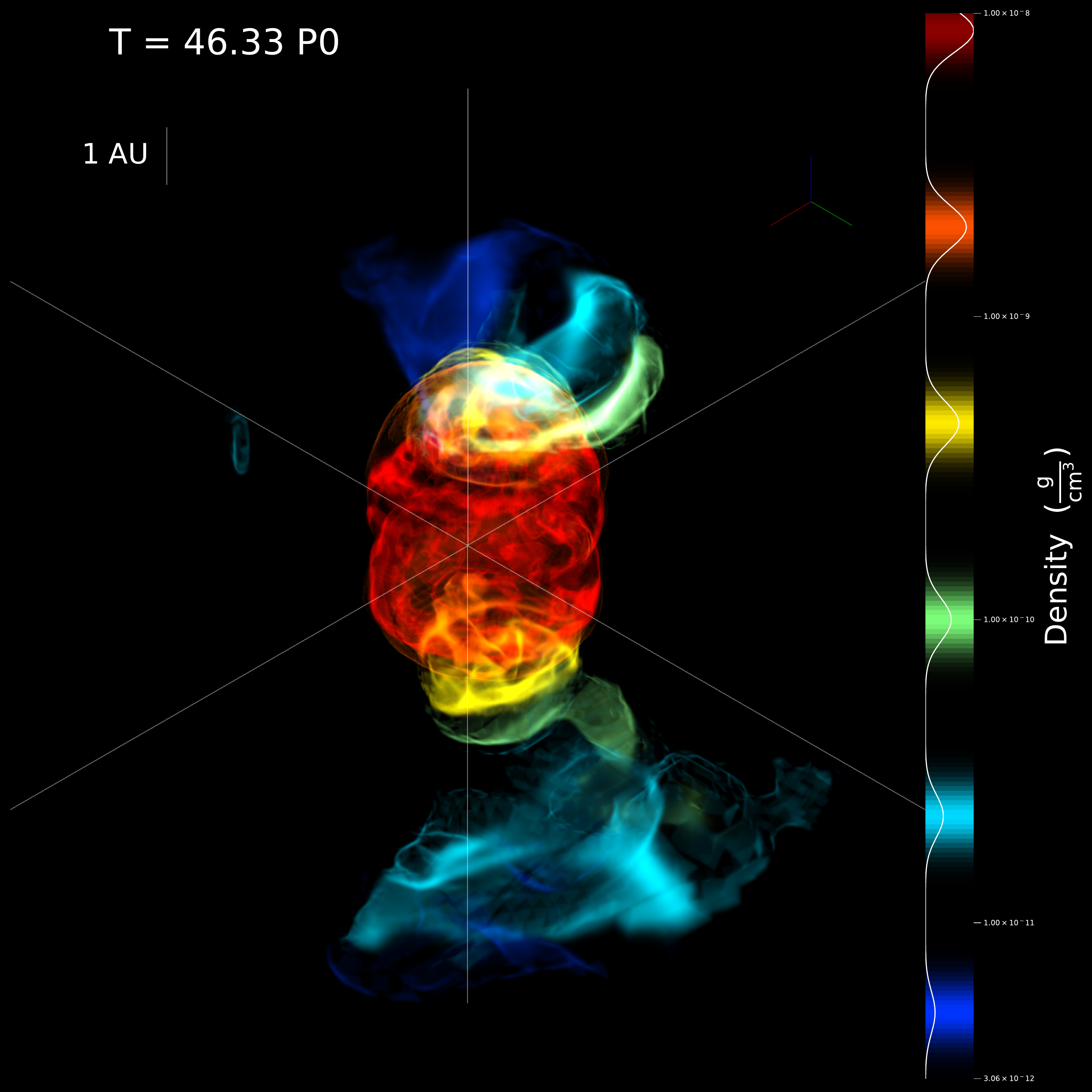}}
    \subfloat[$t=T_{\rm merge} + 0.54P_0$]{\includegraphics[width=6.0cm]{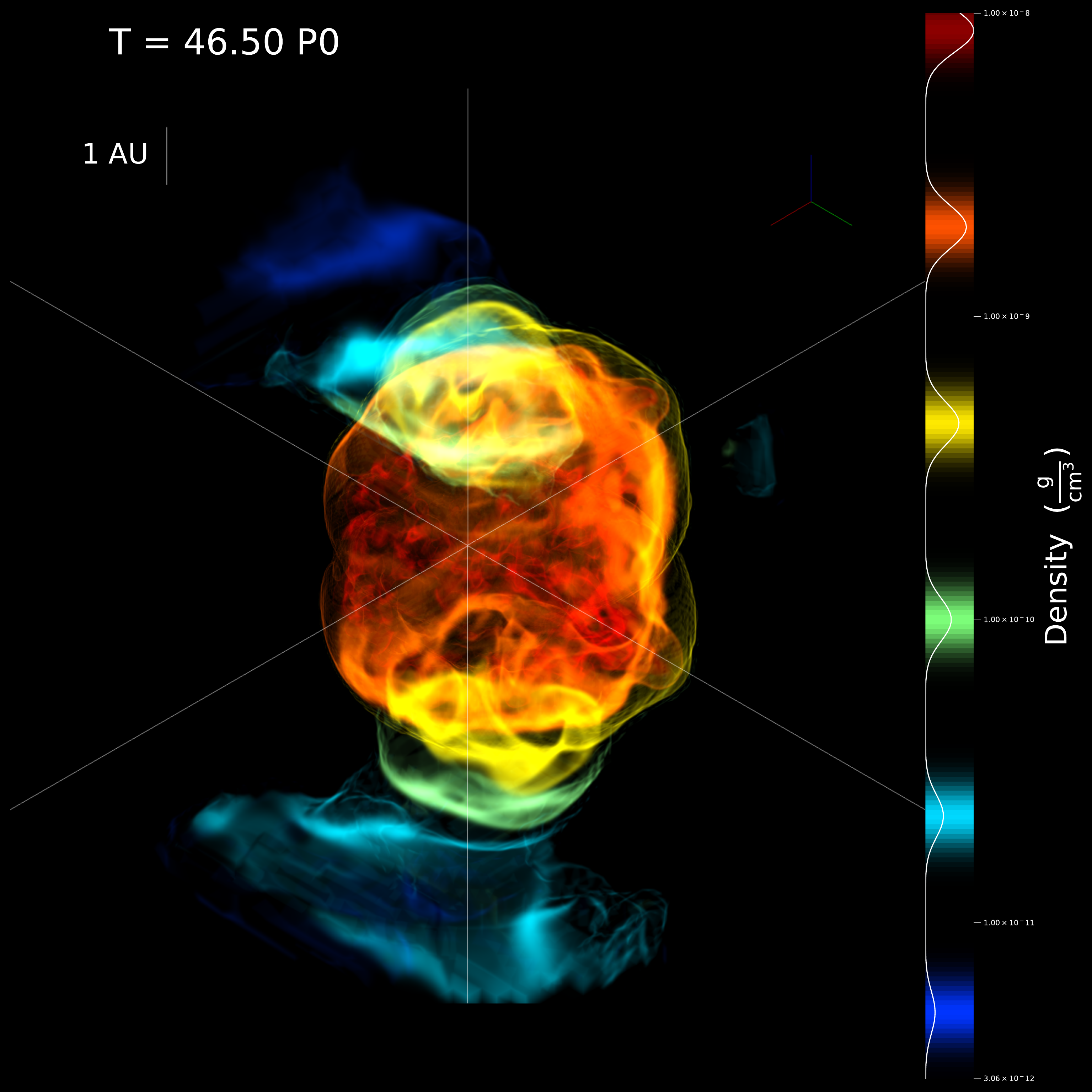}}    
    \qquad    
    \subfloat[$t=T_{\rm merge} + 0.79P_0$]{\includegraphics[width=6.0cm]{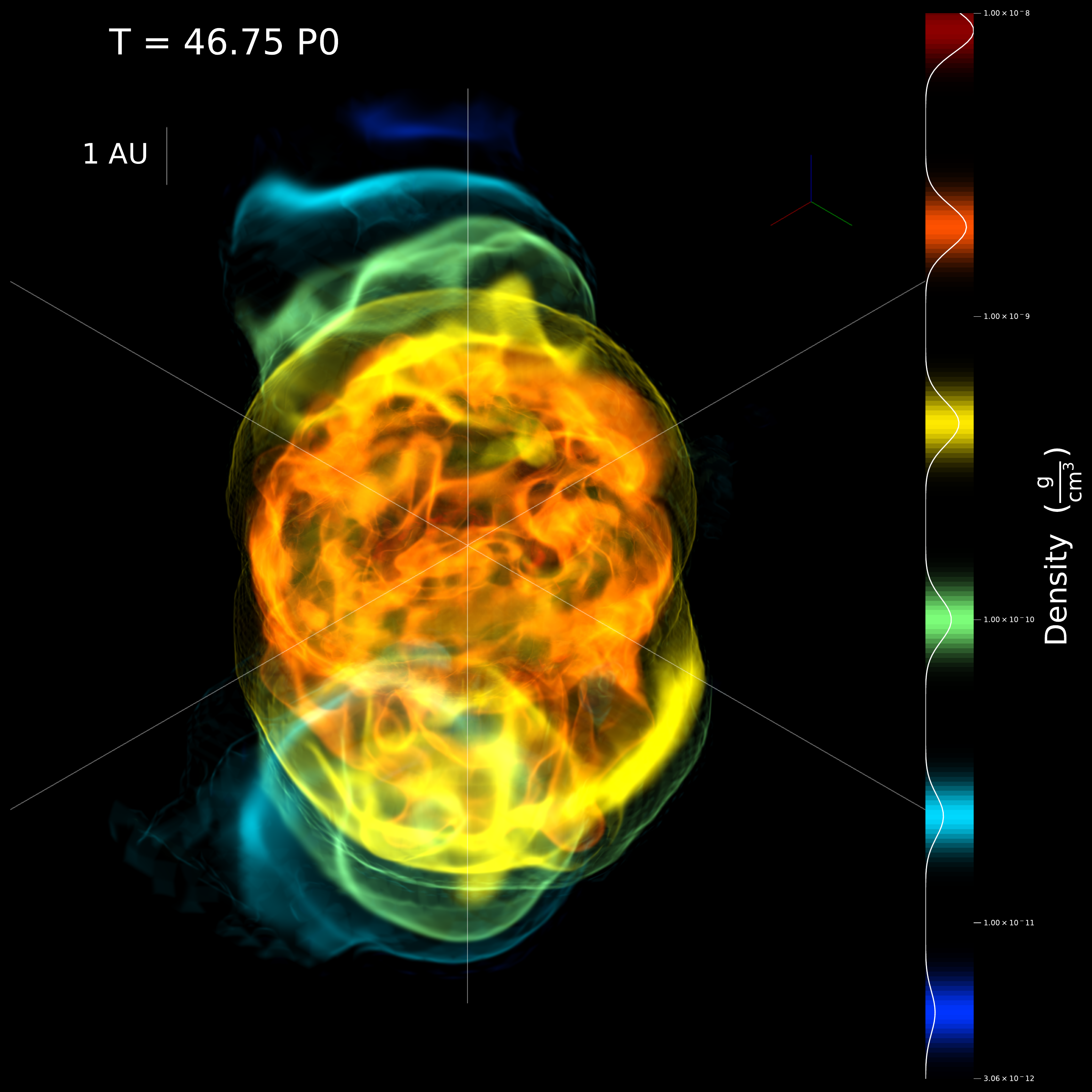}}%
    \subfloat[$t=T_{\rm merge} + 1.05P_0$]{\includegraphics[width=6.0cm]{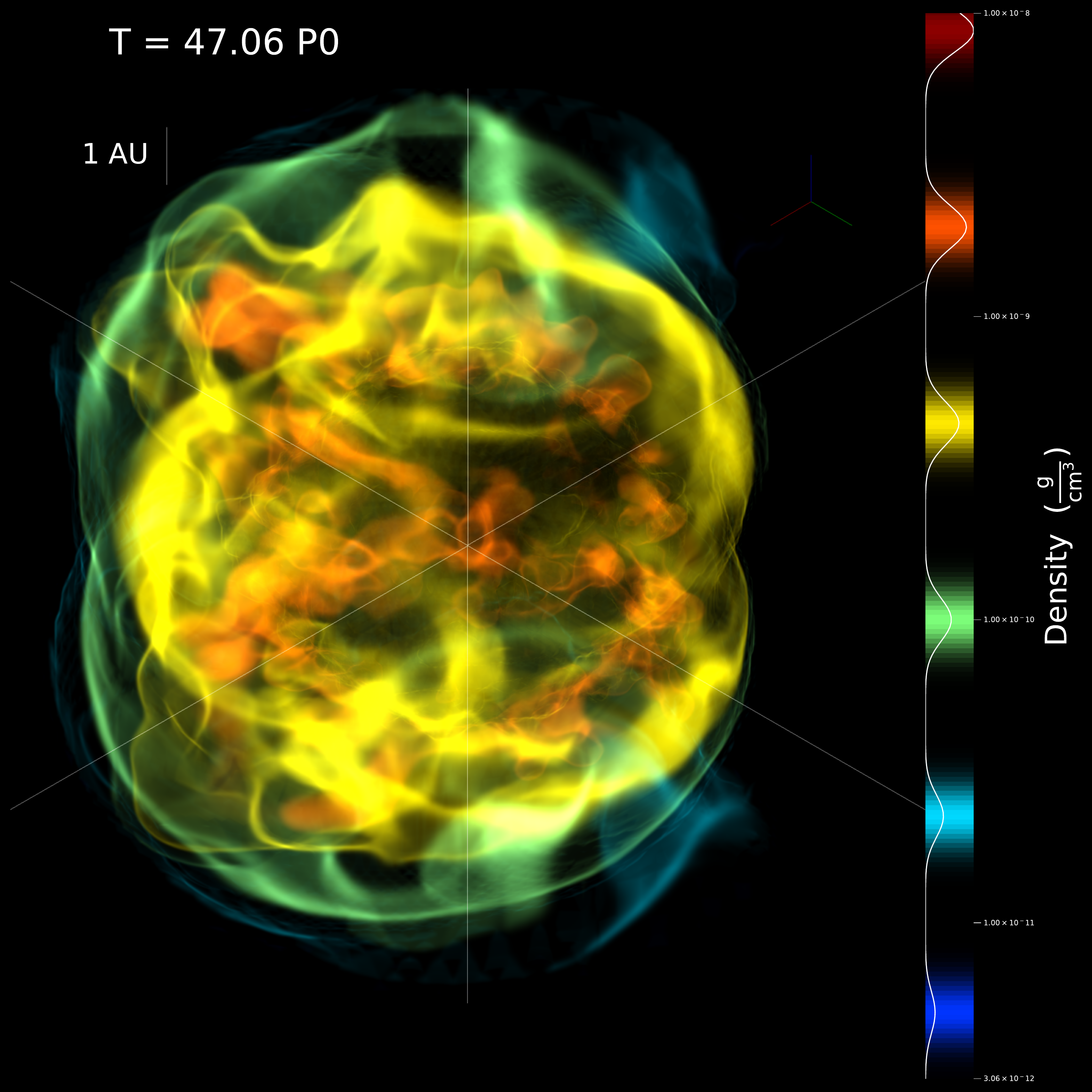}}
    \subfloat[$t=T_{\rm merge} + 1.54P_0$]{\includegraphics[width=6.0cm]{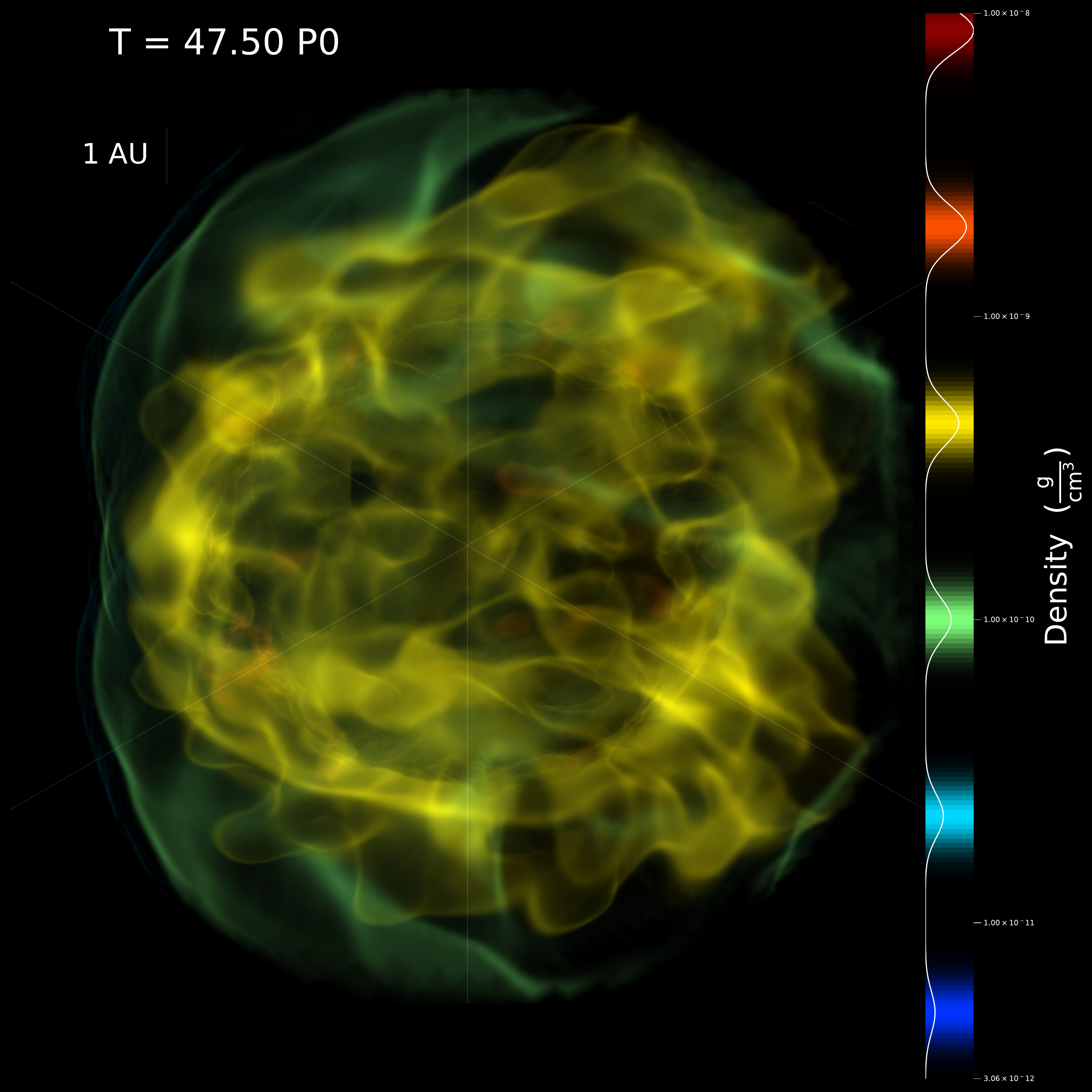}}    
     \caption{Outflow structure. Three-dimensional renderings of unbound gas density (i.e., gas with $E_k + E_g > 0$), at six different times: $0.26~P_0$ ($7~{\rm days}$), $0.37~P_0$ ($10~{\rm days}$), $0.54~P_0$ ($14~{\rm days}$), $0.79~P_0$ ($21~{\rm days}$), $1.05~P_0$ ($27~{\rm days}$), and $1.54~P_0$ ($40~{\rm days}$), after the stars merged. Each panel presents densities at the range of $10^{-8}-10^{-12}~{\rm g/cm^3}$ (red to dark blue, respectively, and according to the color map) as observed by a viewer at the corner of the grid looking towards the grid's center. These renderings extend up to an inner sphere of radius size $6.5~{\rm AU}$. {{A length scale of $1~{\rm AU}$ is also plotted. The positive z-axis is directed upward.}} }
    \label{fig:outflow_props_3d}
\end{figure*}

\begin{figure*}
\centering
        \includegraphics[scale=0.6]{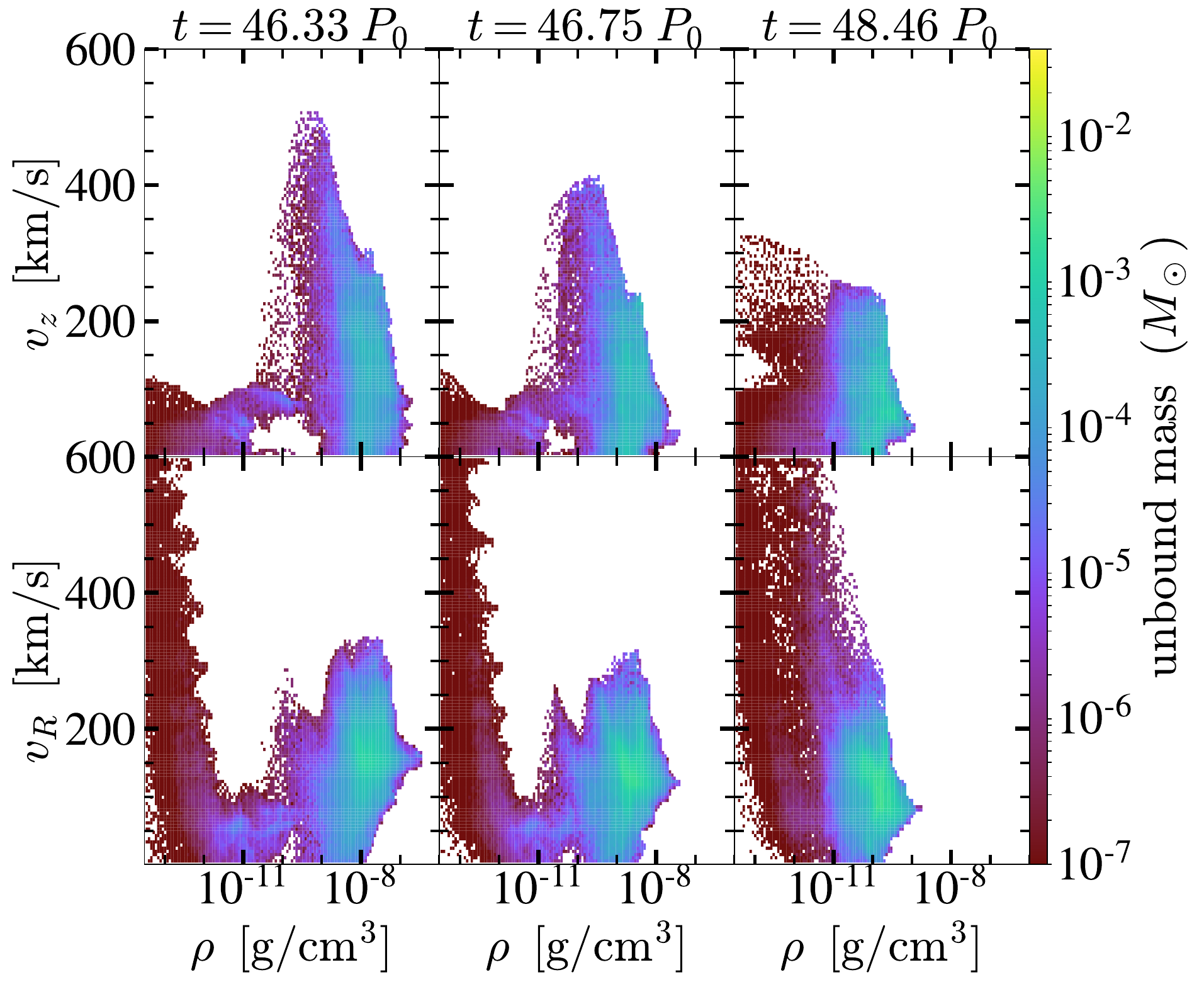}
     \caption{Outflow velocities at the z (upper panels) and cylindrical radius (bottom panels) directions, at three different times: $0.37~P_0$ ($10~{\rm days}$), $0.79~P_0$ ($21~{\rm days}$), and $2.5~P_0$ ($65~{\rm days}$) after the merger time. Each panel plots the amount of mass (in solar mass) the outflow, i.e, gas with $E_k + E_g >0$, has at a certain densities and velocities range. Following the merger, the mergerburst first expands but sustains its velocities ($t=T_{\rm merge} + 0.37P_0$, left panels), expands and decelerates ($t=T_{\rm merge} + 0.79P_0$, middle panels), and eventually reaches the grid boundary and escapes $t=T_{\rm merge} + 2.5P_0$, right panels. The fastest flow of the mergerburst is pointed toward the z-direction, as is reflected by the apparent bipolar structure}
    \label{fig:outflow_props}
\end{figure*}

In Figures~\ref{fig:unb_mass_vel}-\ref{fig:outflow_props} we explore the geometrical structure and kinematics of the unbound merger--driven outflow. Figure~\ref{fig:unb_mass_vel} shows the average (inertial) velocities of these unbound outflows as a function of time, where we focus at the same time-span as in Figure~\ref{fig:unb_mass}. We averaged by mass over gas with positive kinetic plus gravitational energy. These velocities can be compared with the nominal escape velocity from the system $v_{\rm esc}(r)=\sqrt{2G(M_1+M2)/r}\simeq 270~{\rm km/s}\cdot (r/100~R_{\rm \odot})^{-1/2}$, where $r$ is the distance from the system's center of mass.
Prior to the merging time, $T_{\rm merge}$, the amount of unbound mass is small and therefore does not contribute much to the mergerburst.
The unbound gas immediately after the merger escapes with radial velocities (solid blue line) which peak at $310~{\rm km/sec}$ but also spins at a tangential velocity of $140~{\rm km/sec}$ (dashed orange line). 

As the outflow propagates further out it decelerates due to the gravitational pull from the central, merged object. The tangential velocity decays faster than the radial velocity component. We assume for simplicity that this unbound mass originated from a distance of the initial orbital separation, $r_0 = a_0 = 100~R_{\rm \odot}$ and integrate its radial velocity from the merging time and onward to derive its radial distance, $r_{\rm out}$ as a function of time (red dotted line), $r_{\rm out}=r_0 + \int_{T_{\rm merge}}^t v_rdt$. We also calculate the escape velocity at the instantaneous radial distance of the outflow and show that the propagation speed of the outflow remains greater than the escape velocity at any time. A purple dashed vertical line at the post-merger time of $3.6P_0 = 93~{\rm days}$ denotes the time when the outflow arrives to the boundary of the grid. Note that the fastest flow reach the boundary earlier indicated by the peak in unbound mass (orange line in Figure~\ref{fig:unb_mass}) at $t\simeq 2P_0\simeq 52~{\rm days}$.

In Figure~\ref{fig:outflow_props_3d}, we plot three-dimensional renderings of the unbound gas density at six different times after the merger. In each panel we plot densities at the range of $10^{-8}-10^{-12}~{\rm g/cm^3}$ (colored red to dark blue, respectively, according to the color map). The point of view is of an observer at the corner of the grid, i.e, at $(2000~R_{\rm \odot}, 2000~R_{\rm \odot}, 2000~R_{\rm \odot})$, looking towards the grid's center.

Our simulation reveals that the outflow exhibits an asymmetric expansion pattern and possesses a complex inner structure. In the early stages (panels (a) and (b)), a distinctive two-ring structure emerges, with one ring located above and another below the equatorial plane (depicted in red). This structure expands over time and remains visible in later stages (panel (e), represented by orange). During the process of the secondary star's spiraling-in (prior to its merger), gas accumulates around it, causing the envelope to become inflated and concentrated primarily towards lower latitudes. Consequently, when the rapid outflow bursts out, it escapes and expands more rapidly towards the poles, resulting in the formation of a bipolar outflow structure. The bipolar and clumpy ring-like morphology of the outflow is reminiscent of circumstellar (CS) environments observed around some massive star systems (i.e., $\eta$--Carina, \citealt{2007ApJ...655..911S}) as well as supernova (SN) remnants like SN1987A \citep{1995ApJ...439..730P}. {{A clumpy ring structure has been also observed in several planetary nebulae (e.g., the necklace nebula; \citealt{Corradi2011}). However, in our simulation, the unbound outflow forms two clumpy rings, one above and another below the equatorial plane, which expand to higher latitudes rather than an equatorial knotted-ring.}}

Lastly, Figure~\ref{fig:outflow_props} presents a phase plot that displays the distribution of unbound mass (in solar mass units) within specific density and velocity ranges at three different time intervals following the merger: $0.37P_0$ (left column), $0.79P_0$ (middle column), and $2.5P_0$ (right column). The first row of plots illustrates the mass distribution in specific density and positive z-velocities bins, while the second row depicts the mass distribution in density and radial cylindrical velocities bins. These plots provide insights into the kinematics of the outflow. It is important to note that since only the positive z-direction is depicted, the mass shown in these plots (first row) represents approximately half of the total unbound mass within the grid.

During the early period up to $t=T_{\rm merge} + 0.37P_0$ (left column), the outflow expands steadily without decelerating. This expansion can be observed in the phase plots as a structure shifting to the left, to lower densities, without changing its shape. Subsequently, the expansion continues, but the outflow starts to decelerate (the structure shifts to lower densities and lower velocities, in the middle column  compared to left column). At this stage, the fastest velocities are still primarily in the z-direction. As the merger-burst reaches the boundary of the grid, less unbound mass remains within the grid. The fast outflow in the z-direction exits the grid first, resulting in a more uniform velocity distribution (right column). This can be seen in the phase plots of the right columns, where a similar shape is observed between the upper and lower panels. This analysis indicates that the fastest velocities are predominantly aligned with the z-direction.

\begin{figure*}
    \centering
    \includegraphics[scale=0.33]{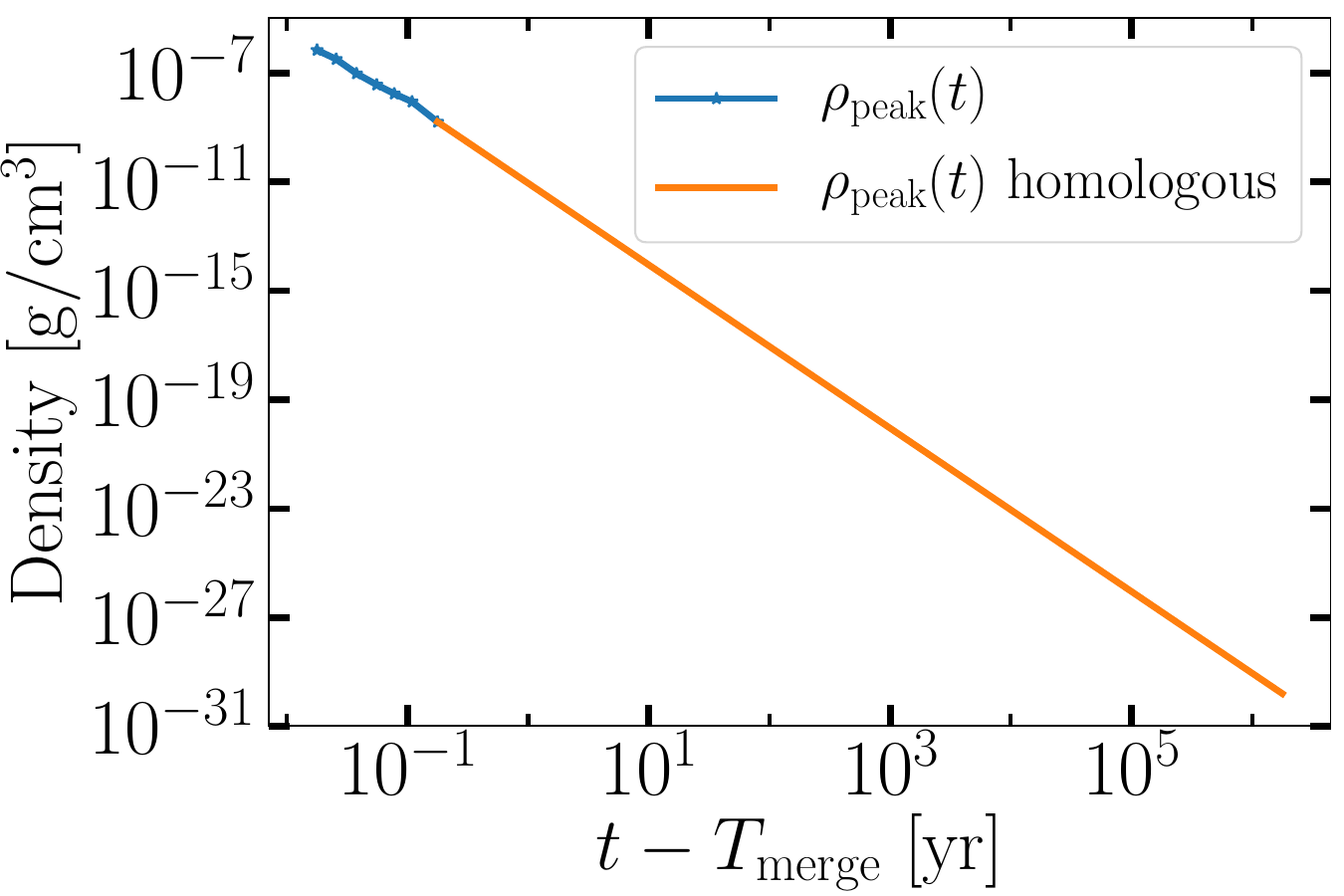}
    \includegraphics[scale=0.33]{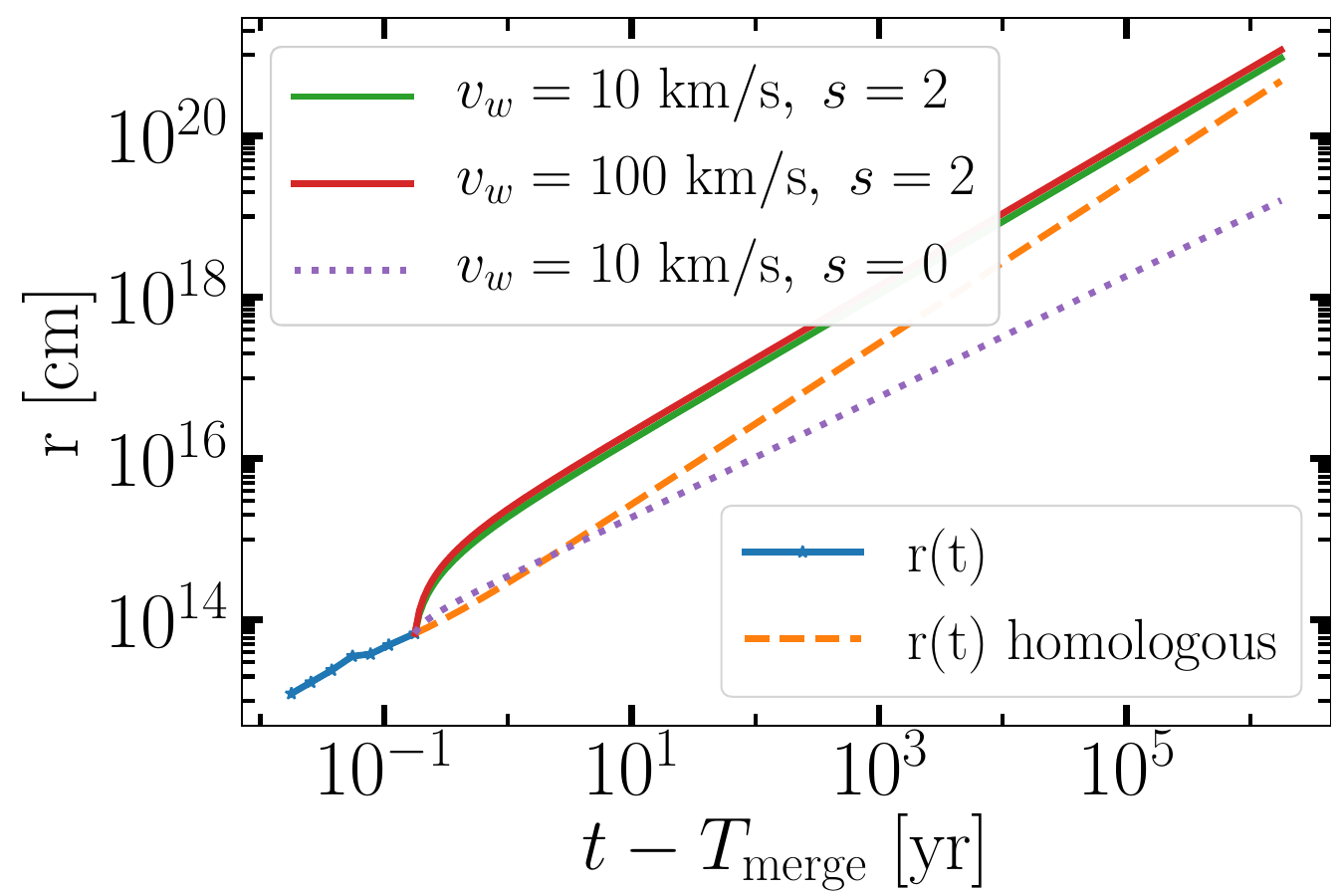}
    \caption{Extrapolation of the mergerburst to later stages post-merger. Left panel: Peak density. Right panel: Distance of the peak density from the post-merger object. The blue points are data points from the hydrodynamic simulation while the orange curves are extrapolation assuming homologous expansion. We also consider interaction of the mergerburst with winds blown by the massive primary star prior to the merger process with wind velocity $v_w$ and a profile that decreases as $r^{-s}$ (see details in the text)}
    \label{fig:out_expansion}
\end{figure*}

Based on these results we can extrapolate how a nebula produced by such a mergerburst will appear at later stages. In Figure~\ref{fig:out_expansion} we show the outflow peak density (left) and its distance from the merger (right) as a function of time. The blue points are data points from the hydrodynamic simulation while the orange curves are extrapolation assuming homologous expansion, i.e., $\rho_{\rm peak} \sim t^{-3}$, $r \sim t$ and disregarding interaction with the interstellar medium. 

In addition, we consider interaction of the outflow with previous winds blown by the massive primary star prior to the merger process. For that we use the equations by \cite{Chevalier1994}. They developed analytical formulas for the interaction of supernova-driven outflows with winds and we downscale them for mergerburst driven-outflows. We first fit the outflow density profile in the simulation domain to two power-laws. This yields a flat slope of $\delta = 1.8$ within the inner $6.7~{\rm AU}$, and a steep slope of $n = 11.9$ beyond this radius, which we round up to $n = 12$. The mass ($M$) and energy ($E$) of the outflow is the total mass and energy lost from the grid, i.e., $0.68~M_{\rm \odot}$ and $1.8 \times 10^{47}~{\rm ergs}$, respectively. As an estimation of the mass loss rate to winds we use the last value as prescribed by the \mesa\ model of our primary star, $\dot{M}=2\times 10^{-7}~M_{\rm \odot}~{\rm yr^{-1}}$. We assume circumstellar density that decreases as $r^{-s}$, where $s=2$, and plug two possible wind velocities of $10~{\rm km/sec}$ and $100~{\rm km/sec}$ to equation 2.7 from \cite{Chevalier1994} to derive the radius of peak densities shells, the green and red curves on the right panel, respectively. 

Lastly, we consider an interaction with a more powerful mass-loss of $\dot{M}=0.1~M_{\rm \odot}~{\rm yr^{-1}}$ and assuming a uniform wind profile, $s=0$. As in \cite{Chatzopoulos2012} (appendix B), we use a density scaling in the immediate vicinity of the primary star and therefore obtain $r_1=R_p=50~R_{\rm \odot}$. We plug wind velocity of $10~{\rm km/sec}$ and the right constants for $s=0$ and $n=12$ from table 1 of \cite{Chevalier1982} in equation B2 of \cite{Chatzopoulos2012} to obtain the forward shock location as an estimation for the radius of peak densities in this case. We plot this radius as a function of time in dotted purple line in the right panel of Figure~\ref{fig:out_expansion}.

From this figure we learn that if the merger will explode in a Core-collapse Supernova within the first few years after the merger, the nebula can affect the supernova lightcurve via interaction. At later times the nebula dissolves and would be difficult to detect nor to play any important role.

\subsection{Merger properties and further evolution}
\label{ssec:merger}

As mentioned in the Introduction (Section \ref{sec:intro}), observational evidence supports the presence of rapid surface velocities in a few giant and supergiant stars. For instance, studies on Betelgeuse have demonstrated \citep{Chatzopoulos2020, Sullivan2020} that a previous merger between a pre-main sequence massive star with a mass of approximately $15-17M_{\rm \odot}$ and a low-mass main-sequence companion with a mass of around $1-4M_{\rm \odot}$ could account for its implied high rotation rate. In this study, we aim to investigate whether the merger product in our simulation has the potential to evolve into a star similar to Betelgeuse. In this subsection, we explore the interaction between the secondary star and the primary star's envelope during the spiral-in process, which leads to the transfer of orbital angular momentum to the envelope. As a result, immediately following the merger, the envelope becomes inflated and rotates faster than its initial state. However, in order to determine whether this dynamical spin-up has any long-term effects on the significantly longer thermal and nuclear timescales that cannot be simulated hydrodynamically, we need to utilize a 1D representation of the post-merger object in a stellar evolution code. In Subsection \ref{sssec:merger_octo}, we analyze the post-merger structure within the hydrodynamic simulation, while in Subsection \ref{sssec:merger_mesa}, we investigate the long-term evolution by importing the post-merger structure into the {\sc MESA} stellar evolution code.

\subsubsection{Immediate dynamical properties}
\label{sssec:merger_octo}

\begin{figure*}
    \centering
    \includegraphics[scale=0.32]{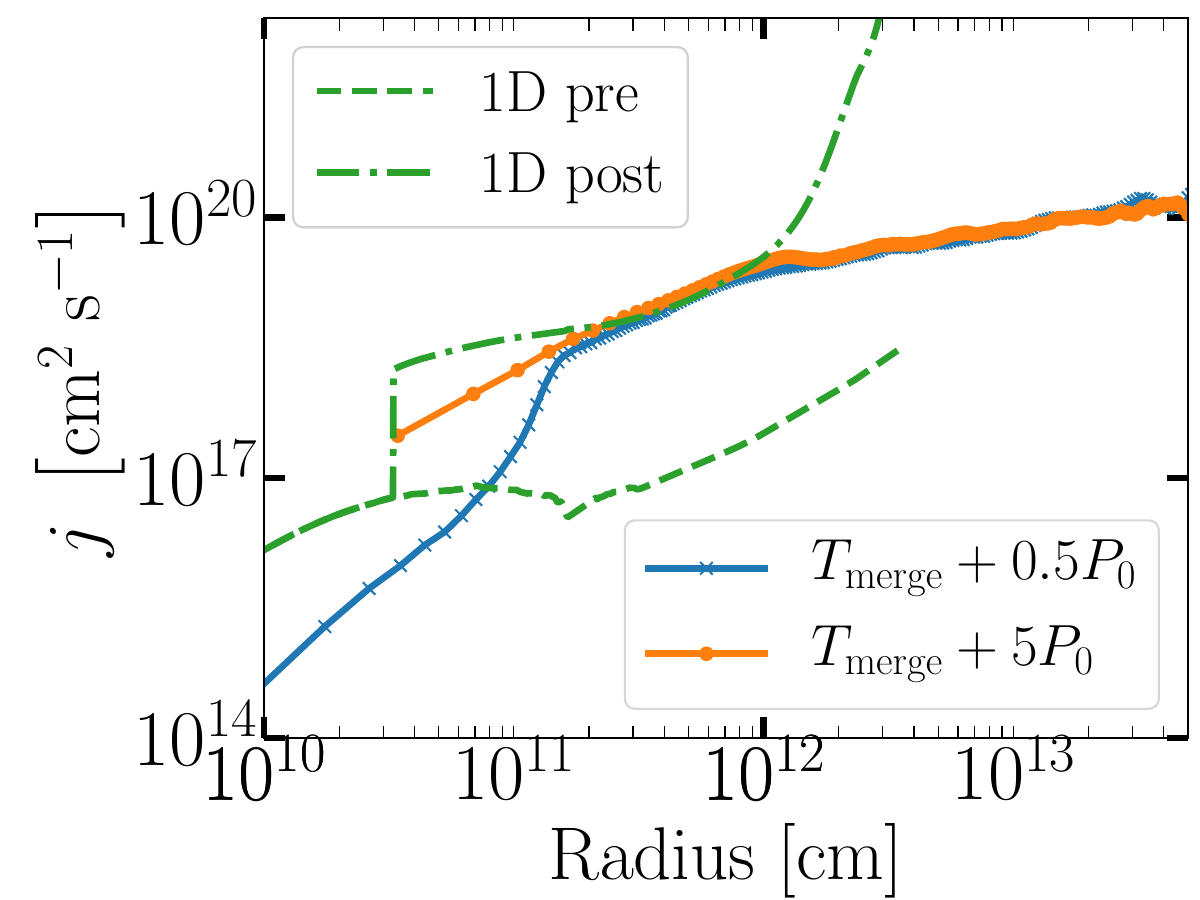}
    \includegraphics[scale=0.32]{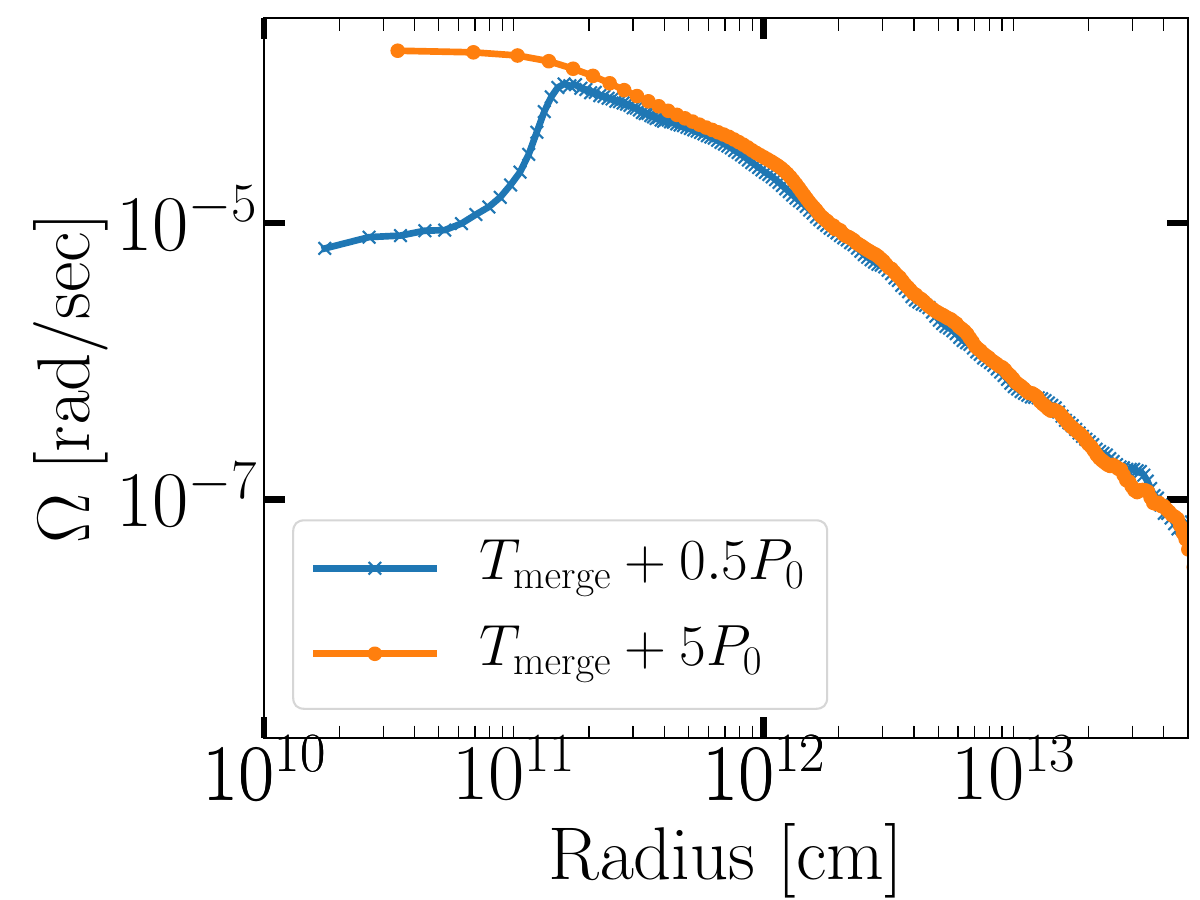}
    \includegraphics[scale=0.32]{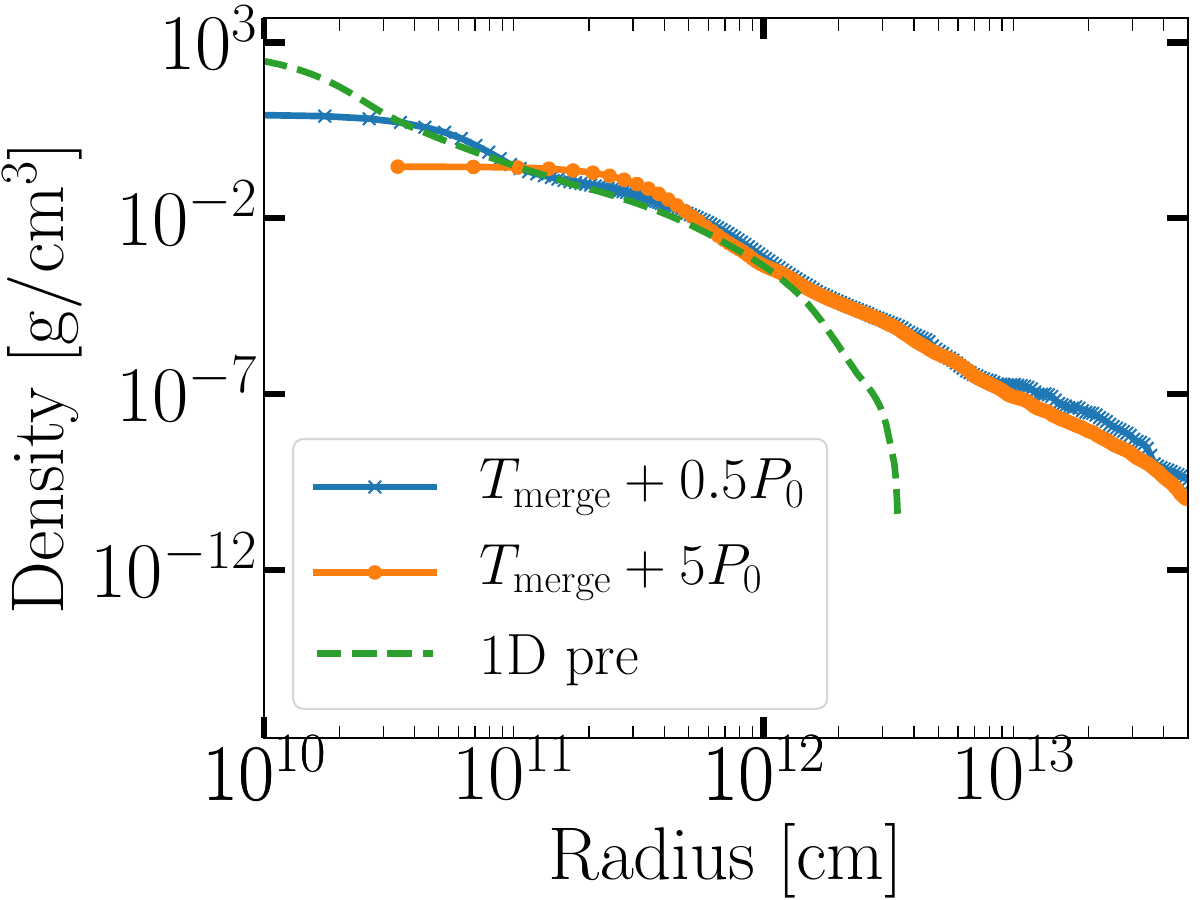}
    \includegraphics[scale=0.32]{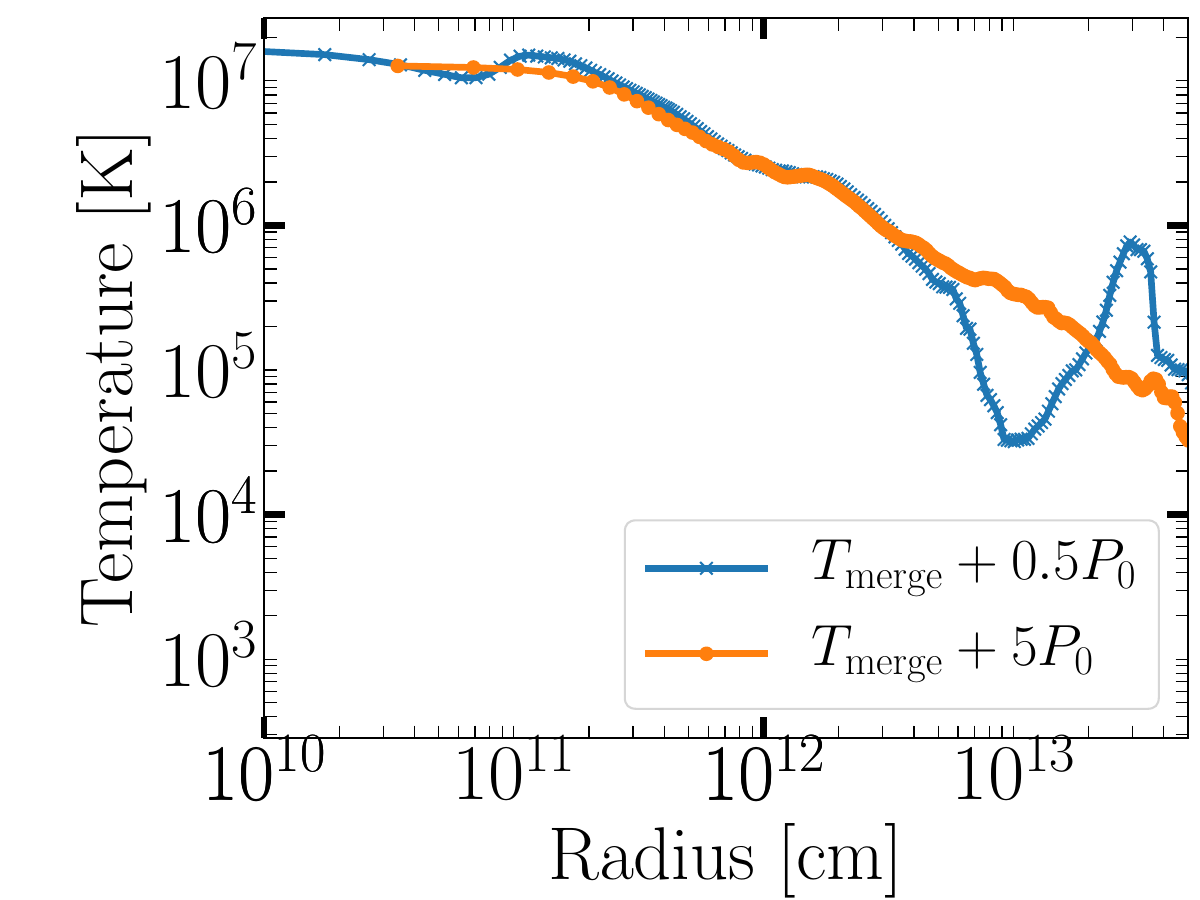}
    \caption{Post-merger star's structure azimuthally averaged over the equatorial plane at $0.5$ (blue curves) and $5$ (orange curves) orbits past merger. From the upper-left panel to the lower-right panel presented specific angular momentum; Angular velocity; Density; and Temperature. The density and specific angular momentum of the primary initial (MESA) model is also plotted (dashed green lines). We compare the specific angular momentum of the merger (as a result of our hydrodynamic simulation; upper-left panel, blue and orange lines) with the post-merger profile as calculated by Equations 13 and 14 in \cite{Chatzopoulos2020} and based on the pre-merger one-dimensional profile of the primary (green dashed-dotted line in the same panel)}
    \label{fig:prop_post}
\end{figure*}

Figure~\ref{fig:prop_post} illustrates the one-dimensional averaged structure of the merger. To obtain these profiles, we performed a mass average over the azimuthal angle around the point of maximum density, which we define as the center of the merger, and only considered data from the equatorial plane. The following quantities are plotted as a function of the distance from the merger's center (from upper left to lower right): specific angular momentum, angular velocity, density, and temperature. It is important to note that the specific angular momentum and angular velocity are computed with respect to the merger's center, meaning that the radius represents the distance from this point, and the velocities are corrected by subtracting the velocities of this point from the inertial velocities. Averaging only over the equatorial plane may lead to an overestimation of rotation values since the deposition of orbital angular momentum primarily occurs around the orbital plane. Consequently, it is expected that rotation would be slower at higher latitudes. However, as the post-merger object undergoes dynamic relaxation within a few orbits, the bound gas gradually becomes more spherically symmetric, causing the redistribution of angular momentum accordingly. Thus, only minor differences between latitudes are expected during this phase. In each panel, two profiles corresponding to two different times, namely $0.5P_0$ (13 days; blue solid crosses) and $5P_0$ (130 days; orange solid circles) after the merging time, are displayed. 

As the post-merger object undergoes dynamic relaxation, its internal structure gradually becomes smoother, as evident in the smoother temperature profile. The orange lines in the plots start from further out due to grid derefinement in the central regions following the merger. This de-refinement leads to a drop in central density but also facilitates the redistribution of angular momentum. To assess the effects of the merger on the post-merger object, we compare the specific angular momentum and density profiles with those of the original {\sc MESA} primary star's model (shown as green dashed lines, denoted as ``pre" in the legend). The specific angular momentum gain of the envelope resulting from the merger is clearly visible in the specific angular momentum plot. Moreover, the envelope becomes inflated, and the density contour of $10^{-9}~{\rm g/cm^3}$ expands from $50R_{\rm \odot}$ to $430~R_{\rm \odot}$. In the plot, we also include the specific angular momentum profile computed analytically (using Equations 13 and 14 from \citealt{Chatzopoulos2020}), shown as a dashed-dotted green line and denoted as "post" in the legend. This analytical profile was used in \cite{Chatzopoulos2020} to simulate the merger starting from the pre-merger 1D profile. Here, we include it for comparison purposes. It is worth noting that this profile rapidly increases at $10^{12}~{\rm cm}$ and further out, due to the steep density decline in these regions in the pre-merger primary profile. The specific angular momentum plot conclusively highlights the benefit of simulating the merger in a dynamical three-dimension simulation over a prescribed 1D model, which could not account important features of the merger such as envelope expansion.

It is worth mentioning though, that as a consequence of the continuous driving we apply on the system, the merger possesses less angular momentum than it would have without the driving. Specifically, the driving mechanisms absorbs almost 50\% of the initial orbital angular momentum. Outflows carried an additional $\sim 20\%$ of the angular momentum and the merger therefore acquires only 30\% of the orbital angular momentum that is initially available. This affects the specific angular momentum of the merger (and its angular velocity), which are probably underestimated and could be higher. In particular, this is true for the profiles at $t = T_{\rm merge} + 0.5P_0$ before viscosity torques and grid de-refinement introduce some numerical angular momentum increase (at the level of $\sim$5\%) 

\begin{figure}
    \centering
    \includegraphics[scale=0.4]{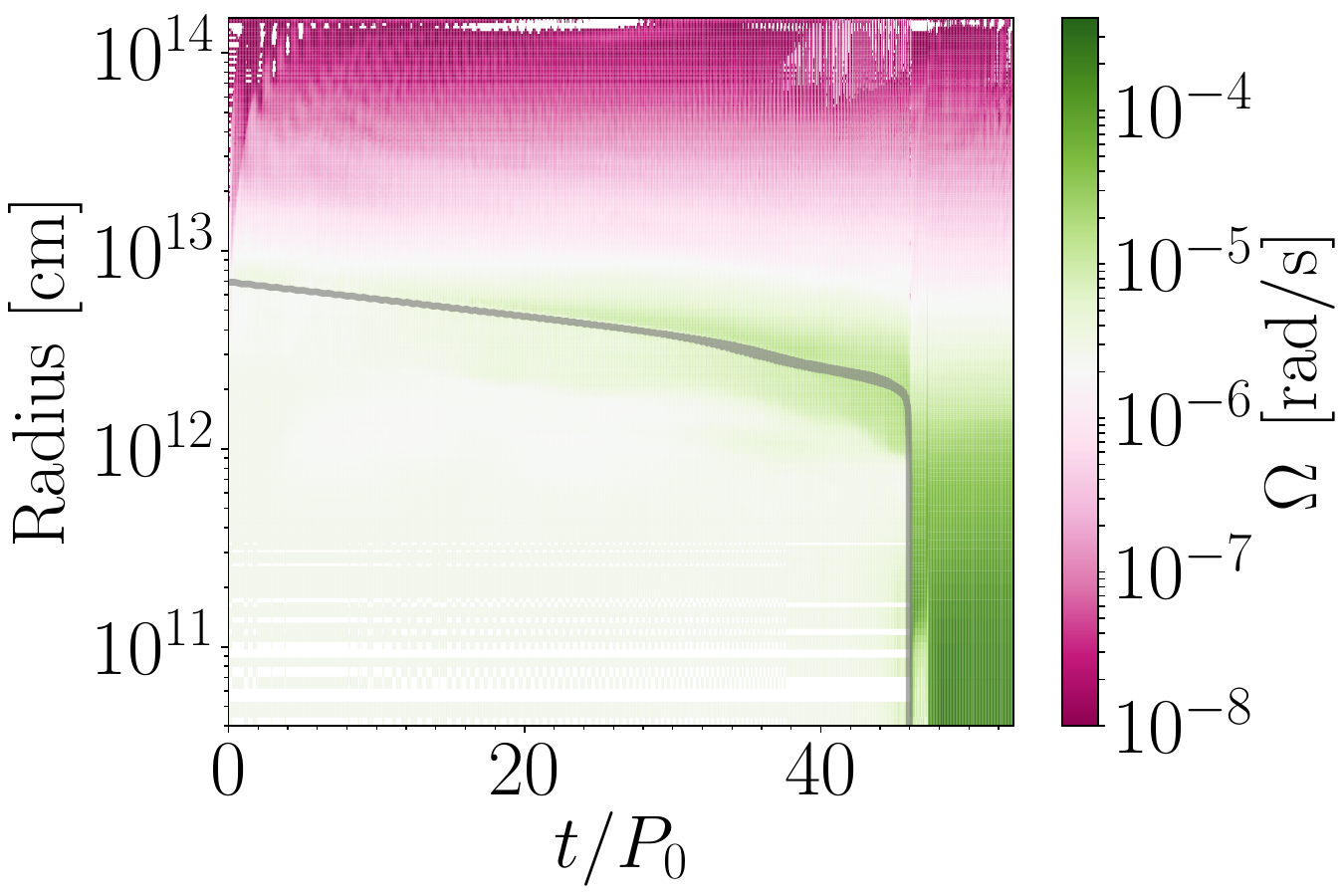}
    \caption{Illustration of the primary envelope spin-up during the merger process. The color scale corresponds to the angular velocity with respect to the primary's core position averaged at the equatorial plane. The color map is in units of ${\rm rad/s}$. Time (in units of initial periods, $P_0$) is plotted on the x-axis. Distance from the primary core (in units of ${\rm cm}$) is plotted on the y-axis. The orbital separation as a function of time is overplotted as a gray curve}
    \label{fig:spin-up}
\end{figure}

To emphasize the dynamic spin-up of the envelope, we present the angular velocity profile of the envelope over time in Figure~\ref{fig:spin-up}. Prior to the merger, the angular velocity is calculated relative to the point of maximal density, which belongs to the primary star and is determined using the diagnostics scheme described in Section~\ref{sec:numerics}. Following the merger, the angular velocity is calculated relative to the center of the post-merger star. Additionally, we include the orbital separation as a function of time, represented by a gray line.
This figure reveals that the envelope in regions near the secondary star undergoes a spin-up process during the slow spiraling-in phase. As the binary system approaches contact and enters the fast spiraling-in phase, the rate of angular momentum gain increases. Subsequently, during grid de-refinements, the angular momentum redistributes within the system. As a result, the inner region of approximately $50~R_{\rm \odot}$ exhibits a significant increase in rotational velocities, exceeding their initial rotation by more than an order of magnitude.

\subsubsection{Long term nuclear evolution}
\label{sssec:merger_mesa}

To study the long-term evolution of the merger on a nuclear timescale, we import the averaged 1D structure of the merger into the {\sc MESA} stellar evolution code. This is achieved by performing a mass-weighted spherical averaging procedure and importing the system 5 initial orbits after the merger event. The resulting profiles of entropy, specific angular momentum, and composition serve as inputs for {\sc MESA}. However, due to insufficient resolution in the inner regions of the hydrodynamical simulation, we replace the inner $9.6 M_{\rm \odot}$ of the entropy profile with a core structure from a similar star evolved in {\sc MESA} (representing the primary's pre-merger/original input structure) that matches in terms of age, mass, and radius. Assuming that the chemical composition remains unchanged by the merger, we also import the abundance profile from the {\sc MESA} pre-merger model. The specific angular momentum profile is directly imported from the \octo\ model using cylindrical shell averages. After spherically averaging the bound mass of the post-merger object, the resulting mass of the post-merger is $18.6 M_{\rm \odot}$

\begin{figure*}
    \centering
    \includegraphics[scale=0.5]{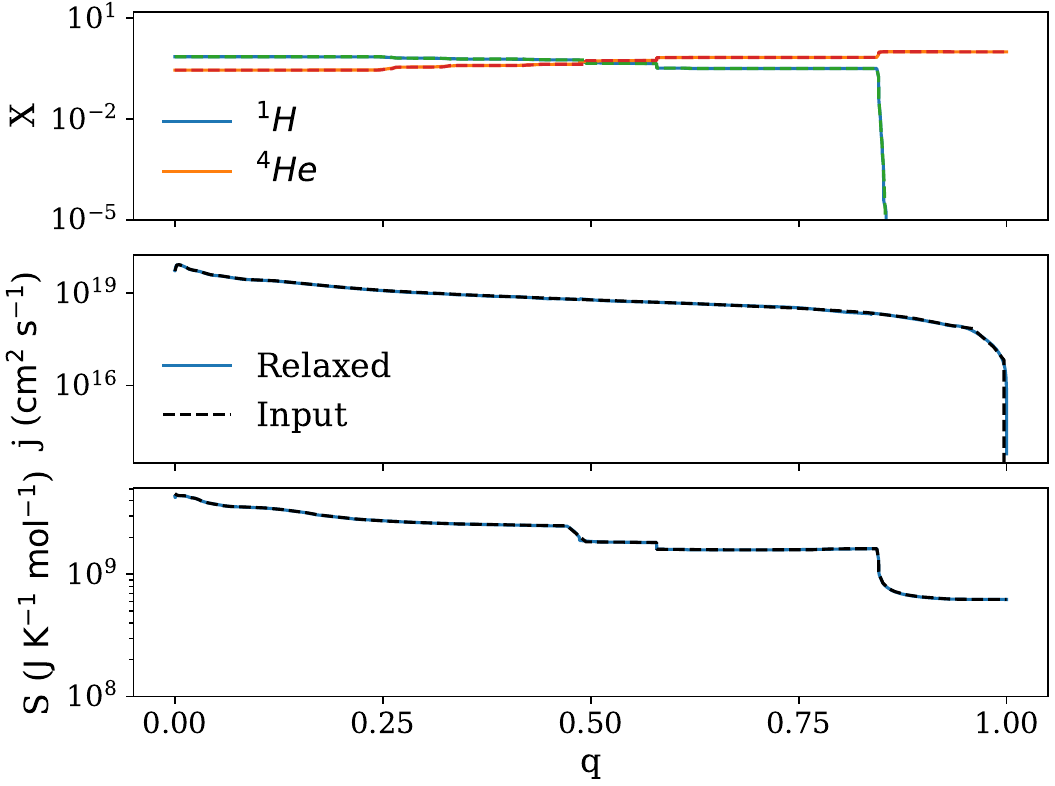}
    \includegraphics[scale=0.5]{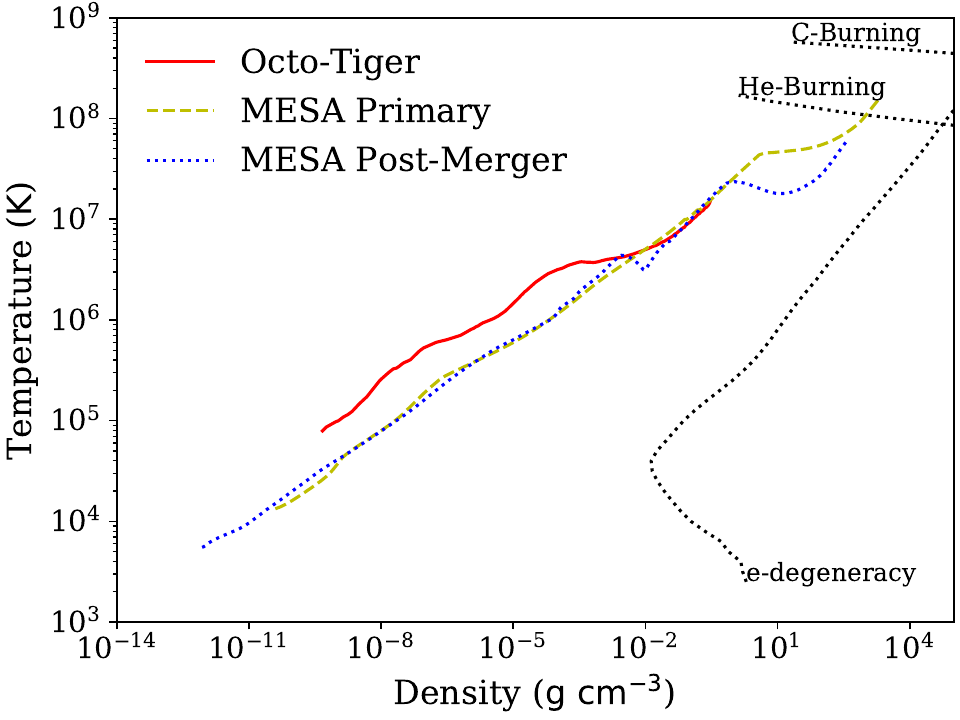}
    \caption{A comparison of the inputs from \octo\ to the relaxed output of {\sc MESA}. The left panels show the input and relaxed profiles of the H and He composition (top), angular momentum (middle), and entropy (bottom). {{The x-axis "q" is the normalized exterior mass coordinate, namely, $q=0$ is the star's surface and $q=1$ is the star's center. The interface in the entropy between the {\sc MESA} and \octo\ initial models occurs at $q=0.5$}}. The panel on the right shows the Temperature-Density diagram for the \octo\ post-merger, the primary star from {\sc MESA}, and the relaxed output from {\sc MESA}.}
    \label{fig:mesa_input}
\end{figure*}

Once we have obtained the necessary profiles, we use the external model relaxation procedures in {\sc MESA} to create a single star with a structure similar to our post-merger object. Figure \ref{fig:mesa_input} illustrates the comparison between our input data and the resulting relaxed structure. It can be observed that all three profiles relax to a state that closely matches the input structure. However, it is important to note that the same entropy profile in {\sc MESA} will yield a different temperature and density profile compared to \octo\ due to the differences in equation of state (EoS) and the requirement of hydrostatic equilibrium (HSE) in {\sc MESA}. Additionally, the inclusion of radiation effects and tabulated opacity in {\sc MESA} leads to a larger relaxed structure compared to the post-merger structure observed in \octo. The disparities in the density and temperature profiles are evident in Figure~\ref{fig:mesa_input}. It should be mentioned that due to the aforementioned core resolution issue, the \octo\ core does not have a sufficient temperature for helium burning. Conversely, the {\sc MESA} profile suggests that the primary should be undergoing helium burning. It is important to acknowledge that although the post-merger object also possesses a core that is currently too cool for helium burning, the structure has not yet fully settled after the relaxation procedure, and the star will contract, resulting in a hotter core.

After relaxing the post-merger into {\sc MESA}, we allow the star to dynamically settle and then evolve over nuclear timescales. During the evolution we use the \texttt{Dutch} wind prescription, which uses the \texttt{de Jegar} prescription for effective temperatures less than 8000K \citep{deJager88}, and the \texttt{Vink} wind prescription for higher effective temperatures \citep{Vink01}. Typically, for non-rotating models, a standard value of the wind scaling factor is 0.8. This could be lower for rotating models so we use a value of 0.4. \citet{Heger00} give an analytical expression for enhanced mass loss due to rotation, which is utilized by default in {\sc MESA} version r21.12.1, of the following form:
\begin{equation}
\label{eq:enhanced_mdot}
    \dot{M} = \dot{M}_0\left[\frac{1}{1-v/v_{crit}}\right]^\xi
\end{equation}
Where $\dot{M}_0$ is the mass loss without rotation, $v$ is the surface rotation, $v_{crit}$ is the critical surface rotation, and $\xi$ is the power factor (the default value is 0.43).
Following some of the parameters used for the {\sc MESA} test suite options, we use the \texttt{Cox} MLT option \citep{Cox68}. We include the Ledoux criterion and use a mixing length parameter of 1.6. We also include the effects of semiconvection and thermohaline mixing, but do not focus on overshooting.

Some of the most important set of parameters for this study are the effects of rotationally induced mixing for both chemical mixing and angular momentum diffusion. Because the surface of the star is spun up due to the coalescence of the secondary into the envelope of the primary, it is important that not too much angular momentum is diffused downward toward the core. The types of rotationally induced mixing we use are Solberg Hoiland (SH), Secular Shear Instability (SSI), Eddington-Sweet Circulation (ES), and Goldreich-Schubert-Fricke Instability (GSF) which are all described in \citet{Heger00}. We also use Spruit-Taylor dynamo action (ST) of \citet{Heger05} which includes the effects of magnetic fields, but we note that there is a lot of uncertainty regarding the strength and effect of magnetic fields for giant stars. Finally, the overall coefficient which is multiplied by the sum of the diffusion due to all of the above mixing effects is set to 1/30 according to the recommendation of \citet{Heger00}.

When evolving the post-merger in \mesa\, we vary the efficiency of rotationally induced diffusion. The reason for this is these effects are inherently 3-dimensional, and their use in \mesa\ is based on the work of \citet{Heger00}, who approximated 1D diffusion coefficients for each process. In addition, we acknowledge that in the case of a star merger, most of the angular momentum deposition occurs over the equatorial plane and non homogeneously throughout the primary star's envelope. The efficiency factors we vary are used in order to account for the uncertainty of the effectiveness of diffusion for each of these mechanisms. This method of simulating mixing is known as the diffusion approximation and \citet{Paxton2013} note that there is another method they refer to as the diffusion-advection approach. While these two methods have nearly identical chemical mixing, the transport of angular momentum can vary significantly. More details regarding this other method of angular momentum diffusion can be found in \citet{Maeder98} and \citet{Zahn92}. A summary of our model parameters, the results, and Betelgeuse observations are shown in Table \ref{tab:bg}.
\begin{table*}[!htb]
    \centering
    \begin{tabular}{ccccccccccc}
    \hline
    \hline
        Model & ST & SH & GSF & ES & SSI & $v_{\rm surf}$ (km/s) & $\epsilon_C$ & $\epsilon_N$ & $\epsilon_O$ & $t_{cool}$ (years)\\ 
        \hline
        1 & 1 & 1 & 1 & 1 & 1 & 0.036 & 8.48 & 8.62 & 8.90 & 14700 \\ 
        2 & 0 & 1 & 1 & 1 & 1 & 0.29 & 8.46 & 8.74 & 8.89 & 17400 \\ 
        3 & 1 & 1 & 1 & 0 & 1 & 0.041 & 8.48 & 8.63 & 8.90 & 14900 \\ 
        4 & 0.1 & 1 & 1 & 0.1 & 1 & 0.063 & 8.48 & 8.65 & 8.90 & 14800 \\ 
        5 & 0.01 & 1 & 1 & 0.01 & 1 & 0.12 & 8.48 & 8.64 & 8.90 & 14600  \\ 
        6 & 0.001 & 1 & 1 & 0.001 & 1 & 0.23 & 8.48 & 8.65 & 8.90 & 14500 \\ 
        7 & $10^{-7}$ & 1 & 1 & $10^{-7}$ & 1 & 1.9 & 8.48 & 8.65 & 8.90 & 25600\\
        8 & 0 & 1 & 1 & 0 & 1 & 5.8 & 8.32 & 8.55 & 8.74 & 22200 \\
        BG &  &  &  &  &  & 5-15 & 8.25-8.55 & 8.45-8.75 & 8.65-8.95
    \end{tabular}
    \caption{A summary of our model parameters and results for eight different models. BG refers to observed surface values of Betelgeuse taken from \cite{2018A&A...609A..67K} and \cite{1984ApJ...284..223L}. Columns 2 through 6 are the efficiency factors used for each diffusion mechanism. $v_{surf}$ is the equatorial surface velocity as calculated in {\sc MESA} averaged over the time spent in the HR diagram box. $\epsilon_i$ is the surface value of element $i$ calculated with Equation \ref{eq:surf_abund}. {{$t_{cool}$ is the number of years it takes for the star to cool from 5000K to 4000K (yellow to red in color)}}.}
    \label{tab:bg}
\end{table*}

The post-merger evolution is found to be primarily influenced by the viscosity coefficients associated with the ST and ES mechanisms. These mechanisms play a dominant role in the outer regions of the post-merger star's envelope, facilitating the efficient transport of excess angular momentum towards the inner regions. As a result, they significantly reduce the equatorial rotation rate of the star. However, we have reasons to consider lower efficiencies for the ST and ES processes. This is motivated by the non-spherically symmetric deposition of angular momentum during the 3D common envelope phase. The high rotation is mainly concentrated around the equator and not evenly distributed in the azimuthal layers of the model, thereby reducing the effectiveness of the ES mechanism. In addition, as we discussed earlier, the artificial driving by angular momentum extraction during the pre-merger phase in \octo\ may lead to an under-estimate of the actual angular momentum that would have been deposited due to drag forces driving the in-spiraling phase alone.
These uncertainties justify our choice of lower efficiencies for the ES and ST mechanisms. However, more comprehensive and computationally intensive 3D numerical simulations are necessary to accurately quantify these effects over time.

\begin{figure}
    \centering
    \includegraphics[scale=0.5]{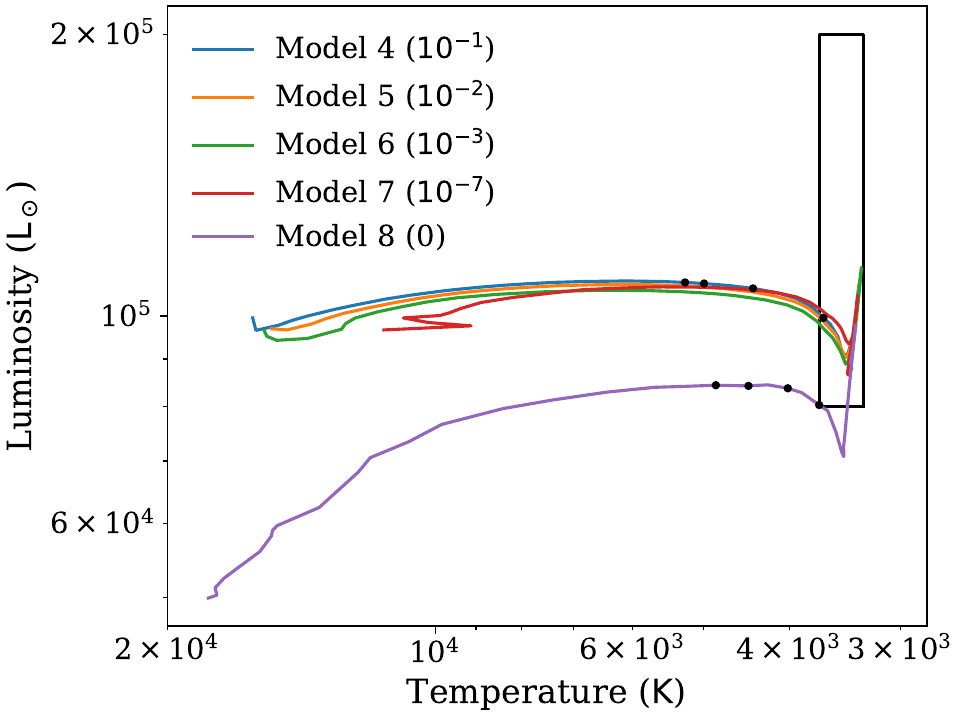}
    \caption{{{HR diagram for Models 4-8, which use smaller diffusion coefficients for the ST and ES mechanisms. The black dots are placed approximately 10,000 years apart starting roughly 30,000 years before the model enters the observed effective temperature range for Betelgeuse. The values in the parentheses are the coefficients for ES and ST mechanisms. The box represents the observed range of effective temperature and luminosity of Betelgeuse.}} }
    \label{fig:mesa_hr}
\end{figure}

After the relaxation, the post-merger first goes through a contraction phase as it settles from the initial structure of the merger. This contraction phase lasts until the core becomes hot enough for He-burning, on the order of 10\textsuperscript{4} years. {{The evolution we are interested in and that we show in Figure~\ref{fig:mesa_hr} is during the He-burning phase and later, which takes place for another 10\textsuperscript{6} years and is indicated by the solid lines in Figure \ref{fig:mesa_hr}. }} 

During these 10\textsuperscript{6} years, the star evolves into the box representing the observed surface temperature and luminosity range of Betelgeuse. {{We additionally plot black dots placed 10,000 years apart and starting roughly 30,000 years before the model enters this box, for models 4 and 8 (see Table~\ref{tab:bg}). This illustrates the star's color evolution, as there are some indications, according to historical records, that Betelgeuse has evolved rapidly in color   \citep{Neuhauser2022}. We also list the time, $t_{cool}$, which takes for each model to change its effective temperature from 5000K to 4000K in Table~\ref{tab:bg}. We find a slower color evolution of several 10,000 years compared to the suggested fast evolution of two millennia by \cite{Neuhauser2022}.}} 

\begin{figure*}
    \centering
    \includegraphics[scale=0.75]{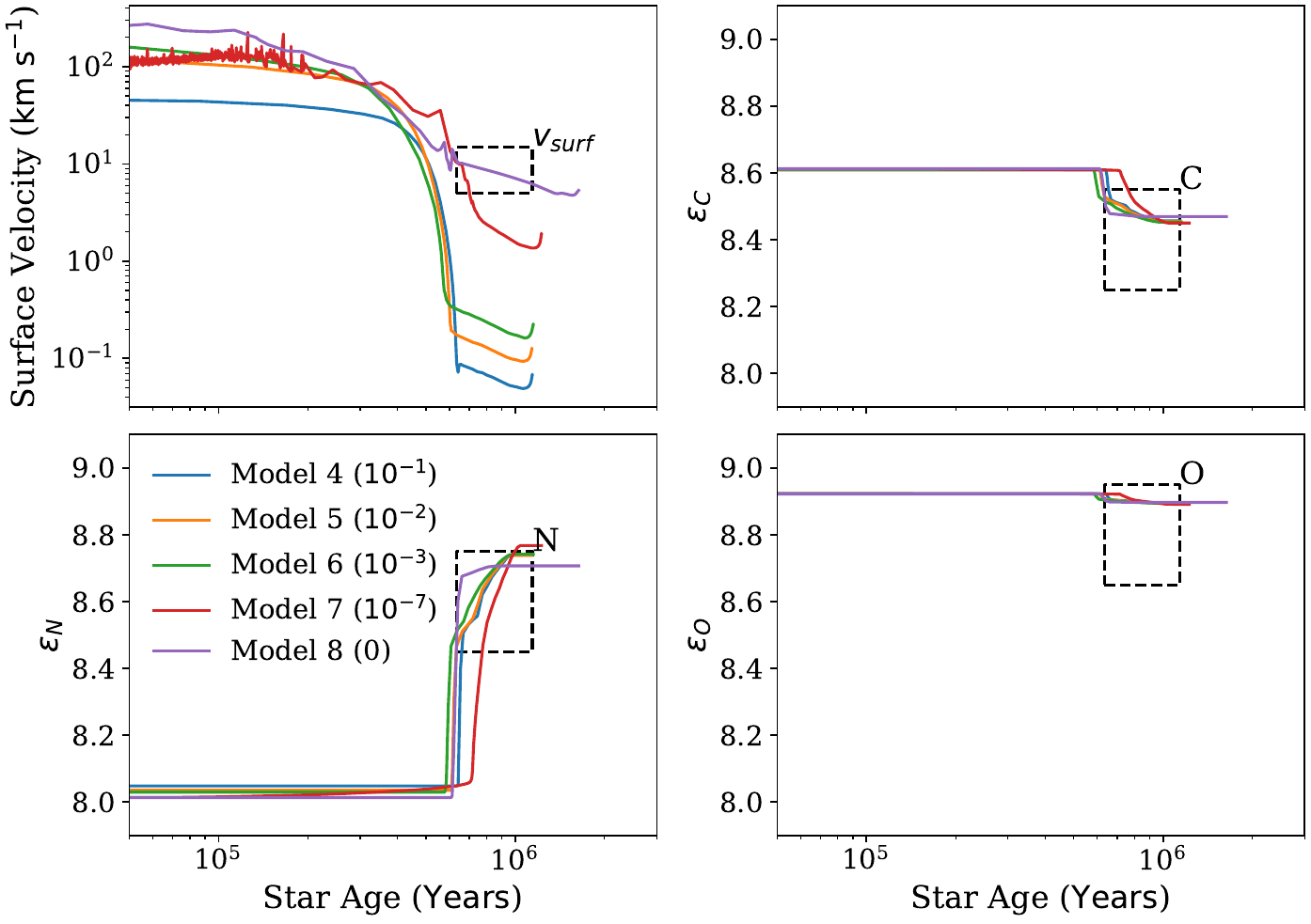}
    \caption{Evolution of the post merger object for five {\sc MESA} runs with different rotation mixing coefficients: $0$ (magenta), $10^{-7}$ (red), $10^{-3}$ (green), $10^{-2}$ (orange), and $10^{-1}$ (blue). Plotted are (from top left to bottom right): surface equatorial rotational velocity in km/s, Carbon, Nitrogen, and Oxygen solar scaled surface abundances. The dashed black square frames the time when the models are within the $3\sigma$ error bars of the observed luminosity and effective temperature of Betelgeuse (on the x-axis) and the observed Betelgeuse ranges (on the y-axis).  }
    \label{fig:mesa_sum}
\end{figure*}

We further analyze the surface composition and rotation rate of our models. The surface composition is calculated using the standard equation (see, for example \citealt{2009ARA&A..47..481A}):
\begin{equation}
\label{eq:surf_abund}
    \epsilon_i = \log(X_i/X_H\mu_i) + 12
\end{equation}
In order to compare our results to the observations of Betelgeuse, we plot the evolution of the surface composition and velocity and use a box that represents the target values from observations. The top and bottom edges of the box represent the range of acceptable values from observations taken from \citet{1984ApJ...284..223L}, while the left and right edges represent the time that the model spends with a surface temperature and luminosity matching Betelgeuse. This analysis is demonstrated in Figure \ref{fig:mesa_sum}. 

We can see from the upper left panel that the surface rotational velocity decreases rapidly by nearly three orders of magnitude. The primary reason for such a decrease in surface rotation is the diffusion of angular momentum towards the core. We note that the strongest contributors are specifically the ST and ES diffusion coefficients with ST being about an order of magnitude stronger as demonstrated in Table \ref{tab:bg} by comparing Models 2 and 3.

After identifying the two biggest contributors to the loss of surface rotation, we ran Models 4-8 in order to explore the effects of decreasing the efficiency of ST and ES diffusion. The results of this experiment are shown in Figure \ref{fig:mesa_sum}. We note that while the range of efficiency factors we explored do not have a significant effect on the surface composition, the less efficient diffusion of angular momentum is critical to the evolution of the surface rotation. We find for our models, we do not see a sufficiently fast rotation rate unless the efficiency is reduced down to $\sim$ 10\textsuperscript{-7} or more of the default value of 1.0. We also note that as the efficiency is reduced and angular momentum cannot diffuse towards the core as rapidly, the model becomes more unstable during the contraction phase.

\section{Summary and Discussion}
\label{sec:summary}

{{This paper presents an attempt to model the complete evolution of a binary star merger, spanning from the onset of the interaction to the common envelope phase, tidal disruption phase, and subsequent aftermath}}. Our study focuses on the merger between a 16 $M_{\odot}$ primary star and a 4 $M_{\odot}$ secondary star during the post-main sequence evolution of the primary towards the RSG branch. The common envelope phase occurs when the primary's expanding envelope comes into contact with the secondary, leading to the decay of the secondary's orbit within the primary's extended envelope. The secondary star merges with the He core of the primary, resulting in a mergeburst transient phase and the ejection of approximately 0.6 $M_{\odot}$ of material in a bipolar outflow.

To conduct this study, we employed various methods and numerical techniques, including 3D hydrodynamical simulations with the \octo\ code and nuclear evolution modeling with the MESA code. The transition between the two codes posed challenges due to differences in assumptions, such as the equation of state, treatment of nuclear burning, and inclusion of radiation effects. The 3D \octo\ simulation required significant resolution (to ensure that the structures of both the primary and secondary stars are adequately resolved for dynamical stability) and a large simulation box (to allow for the tracking of the merger-induced mass-loss in order to investigate the properties of the unbound material). Finally, to facilitate a reasonable computation time, we drove the merger by removing angular momentum from the orbit at a low rate of 1\% per (initial) orbital period.

Despite the limitations in our approach, we successfully achieved our main goal of studying the post-merger evolution and comparing it to observations of stars with peculiar envelope properties, such as Betelgeuse. Key findings of our study include the following:
\begin{itemize}
\item During the common envelope phase, which lasts for 340 days after contact at 50 $R_{\odot}$, the in-spiral of the secondary into the primary's envelope results in significant angular momentum deposition along the equatorial plane (Figure~\ref{fig:spin-up}).
\item Due to resolution limitations, we couldn't investigate the details of the secondary's tidal disruption, stream-core interaction, and the potential rejuvenation of the primary. However, based on the mass ratio studied (q = 0.25), we assume a ``quiet merger" scenario \citep{Chatzopoulos2020}, where the post-merger star continues to evolve toward the RSG phase.
\item The dynamical merger leads to the ejection of approximately 0.6 $M_{\odot}$ of material at velocities of 200-300 km/s, characteristic of mergeburst events (see Figures~\ref{fig:unb_mass}~and~\ref{fig:unb_mass_vel}, respectively). Interestingly, 1/3 of this unbound gas originated from the spiralling-in secondary star (Figure~\ref{fig:unb_mass}). The unbound gas exhibits a distinct bipolar geometry with most of its mass concentrated in two clumpy rings (Figure~\ref{fig:outflow_props_3d}).
\item If the post-merger star explodes as a Type IIP supernova within the next 50,000 to 500,000 years, the circumstellar material formed by the previous merger would be located at a distance of $1-100$ pc, depending on the specific modeling of wind interaction (Figure~\ref{fig:out_expansion}). The densities of this material are unlikely to significantly affect the radiative properties of the supernova.
\item The long-term nuclear evolution of the post-merger star is consistent with the well-known RSG star Betelgeuse, as evidenced by the observed surface enhancement of $^{14}$N (see Figure~\ref{fig:mesa_sum}) and its location in the HR diagram (see Figure~\ref{fig:mesa_hr}).
\item The long-term surface rotation rate of the post-merger star depends on the efficiency of angular momentum transport mechanisms in the 1D evolution calculation (Figure~\ref{fig:mesa_sum}). Rapidly spinning post-mergers are found when the efficiency of meridional circulation (Eddington-Sweet mechanism) is low and magnetically-driven angular momentum transport (Spruit-Taylor mechanism) is minimal (see Table~\ref{tab:bg}). We argue that the anisotropic deposition of angular momentum concentrated along the equatorial plane in a merger scenario suggests low efficiency of inward angular momentum transport. Other factors influencing the surface rotation rate include the angular momentum carried away by mass loss during the merger and post-merger evolution, as well as the underestimation of the total angular momentum deposited in the primary envelope due to our merger driving method. Additionally, the limited size of the primary's envelope expansion up to a radius of 50 $R_{\odot}$ restricts the number of secondary orbits during the common envelope phase and, consequently, the spin-up potential. Previous studies have shown that contact at higher radii (100-500 $R_{\odot}$) can lead to higher post-merger spin-up rates \citep{Chatzopoulos2020}.
\end{itemize}

Our results shed light on the post-merger evolution and provide valuable insights into the properties of merger events and their impact on stellar evolution. However, further improvements are necessary, including refining the codes and techniques used, investigating the details of the tidal disruption and stream-core interaction, and addressing uncertainties related to angular momentum transport and mass loss during the merger and post-merger phases.

Moving forward, our future work will focus on enhancing the \octo\ code, incorporating a more suitable equation of state and radiation effects, and exploring the interaction between supernovae and merger-driven circumstellar environments. By delving deeper into the physics of stellar mergers, we aim to advance our understanding of massive star evolution, the properties of supernova progenitors, and the role of mergers in shaping the astrophysical transient landscape.

\section*{Acknowledgments}
We are grateful to Athira Menon and Craig Wheeler for discussions on Betelgeuse and SN1987A-like progenitors. We also thank Geoffrey Clayton, Dominic Marcello, and Orsola De Marco for useful feedback and discussions. The research of EC was supported by the Department of Energy Early Career Award DE-SC0021228. The numerical work was carried out using the computational resources (QueenBee2) of the Louisiana Optical Network Initiative (LONI) and Louisiana State University’s High Performance Computing (LSU HPC). Our use of BigRed3 at Indiana University was supported by Lilly Endowment, Inc., through its support for the Indiana University Pervasive Technology Institute. This work also required the use and integration of a Python package for astronomy, yt (http://yt-project.org, \citealt{Turk2011}).

\section*{Supplementary materials}
{{
\octo\ is available on GitHub\footnote{\url{https://github.com/STEllAR-GROUP/octotiger}} and was built using the following build chain\footnote{\url{https://github.com/STEllAR-GROUP/OctoTigerBuildChain}}. On Queen-Bee and BigRed \octo\ version \cite{dominic_marcello_2021_4432574} was used. 
All necessary files to reproduce the Betelgeuse post-merger evolution in MESA are available on Zenodo\footnote{\url{https://doi.org/10.5281/zenodo.8422498}}.
}}

\software{\octo\ \citep{Marcello2021}, MESA-r21.12.1 \citep{Paxton2011,Paxton2013,Paxton2015,Paxton2018,Paxton2019}, Python (available from python.org), Matplotlib \citep{matplotlib}, Numpy \citep{numpy}, yt \citep{Turk2011}}

\clearpage
\bibliography{bibliography}

\appendix

\section{Stability tests}
\label{app:pri_stability}

To ensure that the binary interaction is the sole cause of the deformation and of the spin-up of the primary's envelope, we have performed stability tests for our primary star. This includes running the stationary star test as discussed in \cite{Marcello2021} but instead of simulating a $n=1.5$ polytrope, we have simulated a $n=4.1$ polytrope, the polytropic index which have been used to build the primary star, through the SCF, in the binary simulation. Additionally, we scale the star mass and star radius according to the primary star mass and radius, i.e., $M_1=15.5~M_{\rm \odot}$, and $R_1=50~R_{\rm \odot}$, respectively.
The resultant stellar structure resembles the primary star of the binary simulation except that it does not incorporate tidal deformations induced by the secondary star and thus it has spherical symmetry. As in \cite{Marcello2021}, the star occupies half of the computational domain in each direction. We have carried-out two uniform grid simulations, one of $128^{3}$ cells and a second with twice the resolution, i.e., $256^{3}$ cells. As stability measures, we observe the central density, $\rho_c$, behavior, as well as whether a density contour of $\rho_{\rm con}=10^{-5}~{\rm g/cm^3}$ remains at the same radius as it starts at. We have run the simulations for many dynamical times, $t_{\rm dyn}$, where the dynamical time equals to the free-fall time and therefore 
\begin{equation}
    t_{\rm dyn}=(GM_1 / R_1^3)^{-1/2}=1.7~{\rm days}.
    \label{tdyn}
\end{equation} 

First, we find that right at the beginning of the simulation, perturbations develop inside the star. These perturbations stem from the specific adiabatic index of $n=4.1$ that results in a steep density profile at the star's center, which our code cannot adequately resolve. A secondary contribution comes from evolving a polytrope, which assumes a polytropic relation at initialization, with an ideal gas equation of state. The disturbance appears in both resolutions, although its effect is stronger in the lower resolution run. The perturbations quickly propagate outwards, moving out of the star, then propagating out of the grid and decay (at a time span of approximately $4t_{\rm dyn}\simeq 6.6~{\rm days}$). This rapid transient causes the outer, lower-density, layers of the primary star to slightly expand initially, but does not affect the stability of the star. To further confirm that the polytropic index induces this transient, we ran a similar simulation of an $n=3.1$ polytrope, where we have observed that none of these perturbations are formed. A similar transient incurs to the primary's star in our binary simulation as well, and as a result its outer shells slightly expand.

\begin{figure*}
    \centering
    \includegraphics[scale=0.33]{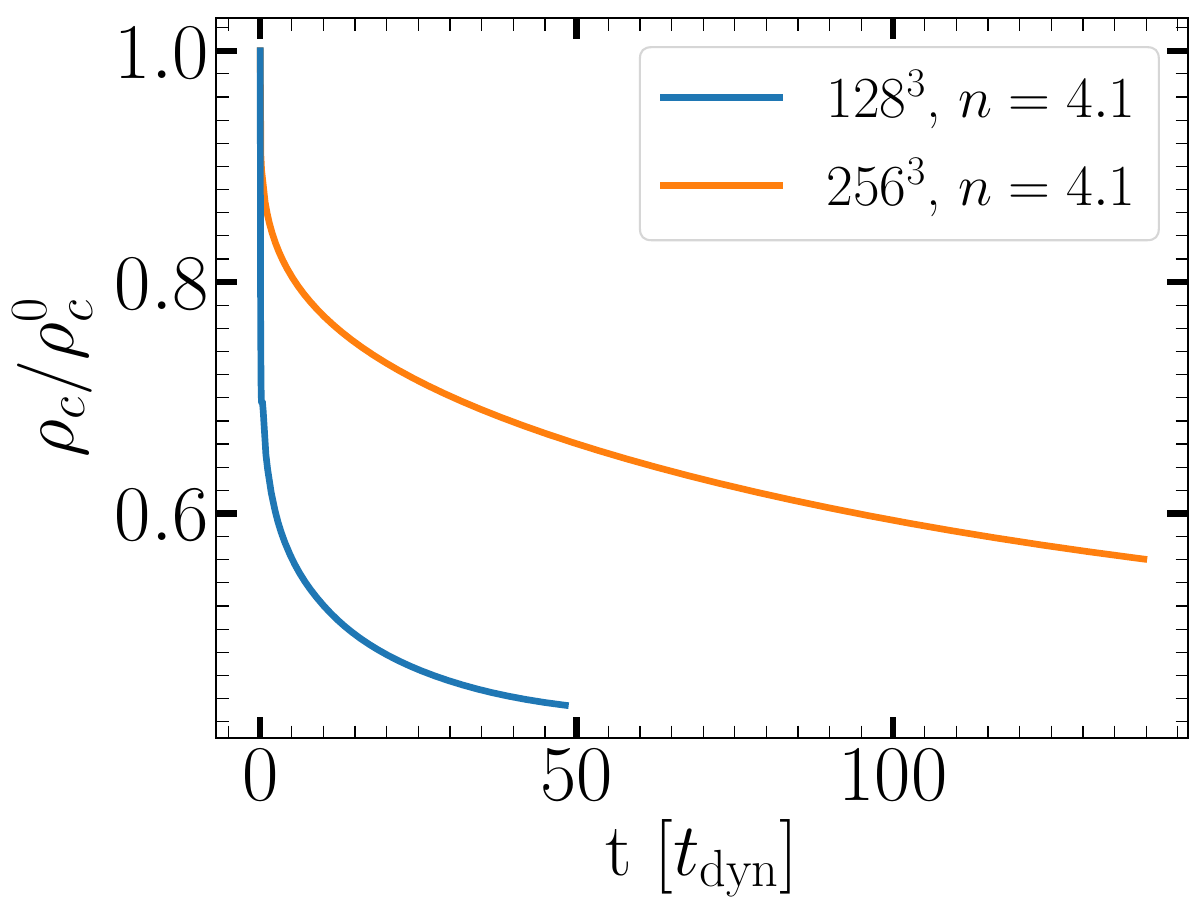}
    \includegraphics[scale=0.33]{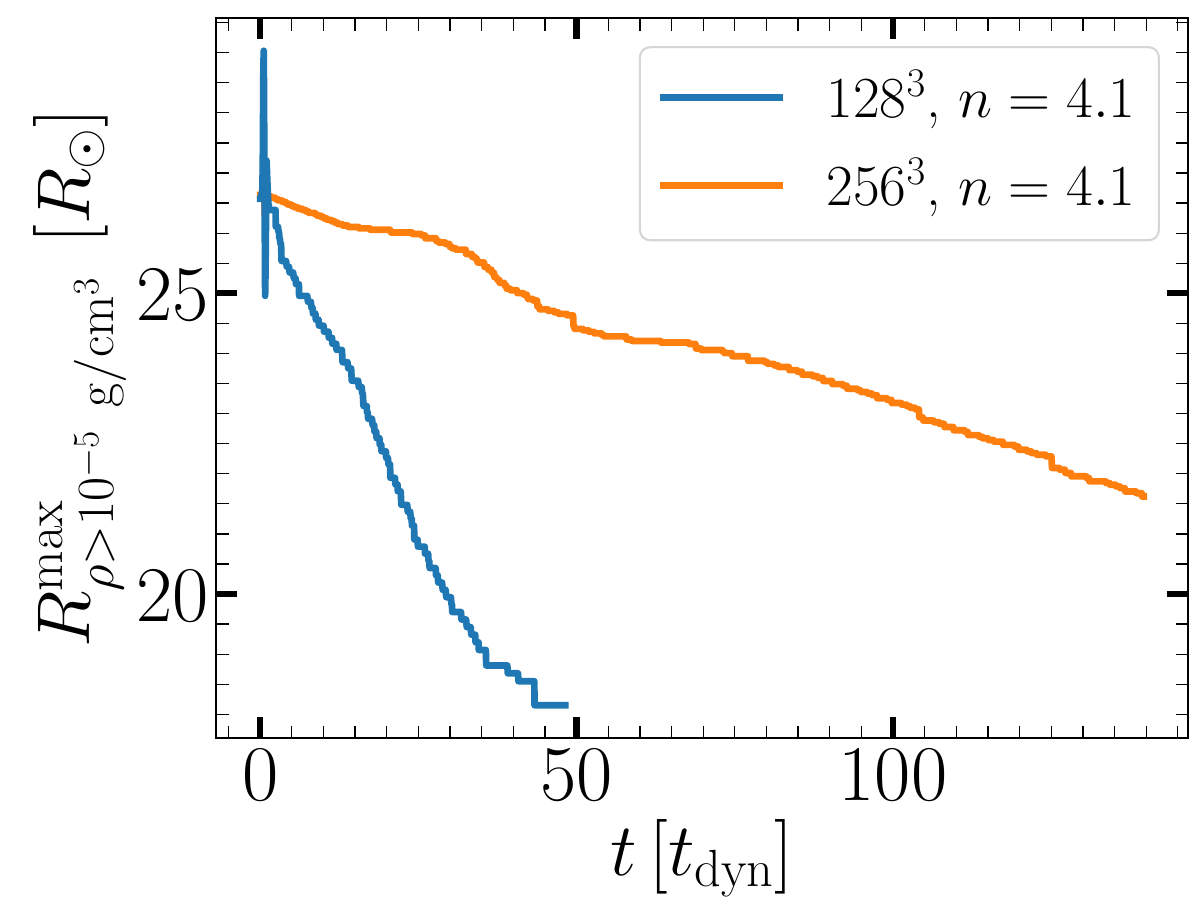}
    \caption{Stability tests for our primary star. We plot the central density divided by the initial central density (left) and the location of a $10^{-5}~{\rm g/cm^3}$ density contour (right) as a function of time of two uniform-grid single-star simulations of an $n=4.1$ polytrope that differ in resolution as denoted in the legend }
    \label{fig:stability_res}
\end{figure*}

In Figure~\ref{fig:stability_res} we plot the central density divided by the initial central density (left) and the location of a $10^{-5}~{\rm g/cm^3}$ density contour (right) as a function of time of the two simulations. 
We find that adequate resolution is crucial in maintaining the $n=4.1$ star stable. In both runs the central density initially falls, however the decrease is much more notable in the lower resolution. Furthermore, in the lower resolution run, the star experiences a collapse, as seen by the contraction of the $10^{-5}~{\rm g/cm^3}$ density contour. After 50 dynamical times the structure of the star greatly deviates from the initial structure and it instead collapses to a shape of $n=1.5$ polytrope. In the higher resolution run the collapse is much more moderate and after $140$ dynamical times the star remains fairly stable with the density contour collapsing by less than a 20\%. We conclude that a minimal resolution of 128 cells across the star's diameter (as in the high resolution run) is required in order to ensure the stability of the star long enough. In the binary simulation presented in this paper we resolve the primary star diameter into $192$ cells, guaranteeing the stability of the star throughout the complete evolution of the simulation. 

\section{Stability of mass transfer}
\label{app:mt_stability}

The standard discussion of the stability of mass transfer under conservative assumptions indicates that mass transfer from the more massive component (primary, labeled 1) is dynamically unstable. This well-known result needs modification when only a fraction of the mass transfer from the primary is accreted by the secondary, while most of it escapes through $L_2$. Considering only the orbital angular momentum 
$J_{\rm orb}$, and assuming that the fraction accreted by the secondary is $f$, we have $\dot M_2=-f\dot M_1$, and $\dot M= \dot M_1 + \dot M_2 = (1-f)\dot M_1$. With the standard argument using the logarithmic derivatives, it is easy to show that the binary separation $a$ obeys the equation
\begin{equation}
  \frac{\dot a}{a} = \frac{2\dot J_{\rm orb}}{J_{\rm orb}}  -\frac{2\dot M_1}{M_1}\left(1-\frac{f}{q} - \frac{1-f}{2(1+q)}\right)\, ,
  \label{eq:adot}
\end{equation}
where the first term on the r.h.s. stands for angular momentum loss (AML), and $q=M_2/M_1$ (in the case considered here $q=M_{\rm accretor}/M_{\rm donor}$). Since $\dot M_1<0$, in the absence of driving, the binary will expand if the quantity in parenthesis is positive. That is if
\begin{equation}
    f\leq f_{\rm max}=\frac{q(1+2q)}{2+q}\, .
\end{equation}
In our case $q=1/4$, and the binary will expand if $f\leq 1/6$. Further analysis requires assumptions about the specific angular momentum carried away by the mass loss, and the rate of external driving. Let us suppose that the specific angular momentum carried by mass loss is $j_{L2}$, and that the system is driven at a rate $\dot J_{\rm sys}$. We may take the total $\dot J_{\rm orb}$ to be the sum of systemic and consequential terms:
\begin{equation}
  \dot J_{\rm orb} = \dot J_{\rm sys} + 
  \dot M_1 (1-f)j_{L2}\, .
  \label{eq:jdot}
\end{equation}
As a result of mass transfer the radius of the donor and the effective Roche-lobe radius will vary according to
\begin{equation}
    \frac{\dot R_1}{R_1} = \zeta_1\frac{\dot M_1}{M_1}
    \quad {\rm and} \quad
    \frac{\dot R_{L1}}{R_{L1}} = \zeta_{L1}\frac{\dot M_1}{M_1}\, ,   
\end{equation}
respectively. Using for simplicity the Pacy\'nski expression for the Roche lobe, we have
\begin{equation}
    \frac{\dot R_{L1}}{R_{L1}} = \frac{\dot a}{a} + \frac{1}{3}\frac{\dot M_1}{M_1} -
    \frac{1}{3}\frac{\dot M}{M} \, .  
    \label{eq:pacynski}
\end{equation}
Note that changes in binary separation and hence Roche-lobe sizes can occur in the absence of mass transfer, due to systemic AML. The Roche-lobe exponent $\zeta_{L1}$ describes changes due solely to mass transfer.  Without loss of generality, we can take $j_{L2}=\Omega a_{L2}^2$, and use Equations~\ref{eq:adot}~and~\ref{eq:jdot} in \ref{eq:pacynski}, to obtain after some algebra
\begin{equation}
     \frac{\dot R_{L1}}{R_{L1}} = \frac{2\dot J_{\rm sys}}{J_{\rm orb}} +
     \frac{\dot M_1}{M_1}
     \left(2(1-f)\frac{1+q}{q} \left(\frac{a_{L2}}{a}\right)^2 - \frac{2+q}{q(1+q)}(f_{\rm max}-f)
     +\frac{1}{3}\left(1-\frac{1-f}{1+q}\right)\right)\, ,
\end{equation}
where the expression in parenthesis in the second term on the r.h.s. stands for $\zeta_{L1}$. If we assume that he outflow carries the specific AM of $L_2$, then $a_{L2} = r(q) a$, where $r(q)$ is a slowly varying function of $q$, taking values around $\sqrt{1.5}$ \citep{Pribulla1998}. In reality, depending on flow details, the gas leaving $L_2$ may carry a fraction $\epsilon$ of the specific AM at $L_2$. After some rearrangement, we may then write 
\begin{equation}
    \zeta_{L1} = -\frac{5}{3} +\frac{2}{q} +
    (1-f)\left(3\epsilon\frac{1+q}{q}-\frac{2+4q/3}{q(1+q)}\right)\, .
\end{equation}
For the standard case  $f=1$, in which all mass transferred is accreted, $\zeta_{L1}$ reduces to the well-known expression with $1/q$ being the mass ratio of donor to accretor. If mass transfer is stable, its equilibrium value is given by

\begin{equation}
    \left(\frac{\dot M_1}{M_1}\right)_{\rm eq} =  
    \frac{2{\dot J}_{\rm sys}/J_{\rm {orb}}}
    {\zeta_1-\zeta_{L1}} \, ,
\end{equation}
and the condition for stability is $\zeta_1-\zeta_{L1}>0$.
In the absence of driving, any initial mass transfer will decrease to zero if $\zeta_1-\zeta_{L1}>0$, or grow exponentially if $\zeta_1-\zeta_{L1}<0$. In our simulation, mass transfer is started by driving, and we find $f\approx 0.25$. With $\zeta_1=2.82$, and $\zeta_{L1}\approx 6.333 + 0.75(15\epsilon -112/15)$, one can see that $\epsilon > 0.185$ is required for mass transfer to grow in the absence of driving. In our simulation we find $\epsilon\approx 1$ throughout, therefore $\zeta_{L1}>\zeta_1$, for any value of $f$, and the mass transfer is unstable. However,
simulating the evolution with no driving would be unreasonably expensive (see text).

\end{document}